\documentclass[10pt,conference]{IEEEtran}

\usepackage[acronyms,nonumberlist,nopostdot,nomain,nogroupskip]{glossaries}
\usepackage[numbers,sort&compress]{natbib}
\usepackage{xspace}
\usepackage{orcidlink}

\AtBeginDocument{%
  \providecommand\BibTeX{{%
    \normalfont B\kern-0.5em{\scshape i\kern-0.25em b}\kern-0.8em\TeX}}}

\usepackage[utf8]{inputenc}
\usepackage{xcolor}
\usepackage{amsmath}
\usepackage{amssymb}

\usepackage[acronyms,nonumberlist,nopostdot,nomain,nogroupskip]{glossaries}
\usepackage{booktabs}
\usepackage{lipsum}
\usepackage{tabularx}
\usepackage[noend]{algorithmic}
\usepackage{algorithm, setspace}
\usepackage{tikz}
\usepackage{pgfplots}
\pgfplotsset{compat=newest}
\pgfplotsset{plot coordinates/math parser=false}
\newlength\fheight
\newlength\fwidth
\usetikzlibrary{plotmarks,patterns,decorations.pathreplacing,backgrounds,calc,arrows,arrows.meta,spy,matrix}
\usepgfplotslibrary{patchplots,groupplots,colorbrewer}
\usepackage{tikzscale}
\usepackage{colortbl}
\usepackage{array}
\newif\ifexttikz
\exttikzfalse
\ifexttikz
	\usetikzlibrary{external}
\fi

\usepackage{multirow}

\renewcommand{\figurename}{Fig.}
\usepackage[font=footnotesize]{subcaption}
\usepackage[font=footnotesize]{caption}

\usepackage{mathtools}

\newcommand{\eq}[1]{Eq.~\eqref{#1}}

\newcommand{\beq}{\begin{equation}}
\newcommand{\eeq}{\end{equation}}
\newcommand{\bit}{\begin{itemize}}
\newcommand{\eit}{\end{itemize}}

\tikzstyle{startstop} = [rectangle, rounded corners, minimum width=2cm, minimum height=0.5cm,text centered, draw=black]
\tikzstyle{io} = [trapezium, trapezium left angle=70, trapezium right angle=110, minimum width=3cm, minimum height=1cm, text centered, draw=black]
\tikzstyle{process} = [rectangle, minimum width=2cm, minimum height=0.5cm, text centered, draw=black, alignb=center]
\tikzstyle{decision} = [ellipse, minimum width=2cm, minimum height=1cm, text centered, draw=black]
\tikzstyle{arrow} = [thick,<->,>=stealth]
\tikzstyle{line} = [thick,>=stealth]
\tikzstyle{darrow} = [thick,<->,>=stealth,dashed]
\tikzstyle{sarrow} = [thick,->,>=stealth]
\tikzstyle{larrow} = [line width=0.1mm,dashdotted,->,>=stealth]

\makeatletter
\def\grd@save@target#1{%
  \def\grd@target{#1}}
\def\grd@save@start#1{%
  \def\grd@start{#1}}
\tikzset{
  grid with coordinates/.style={
    to path={%
      \pgfextra{%
        \edef\grd@@target{(\tikztotarget)}%
        \tikz@scan@one@point\grd@save@target\grd@@target\relax
        \edef\grd@@start{(\tikztostart)}%
        \tikz@scan@one@point\grd@save@start\grd@@start\relax
        \draw[minor help lines] (\tikztostart) grid (\tikztotarget);
        \draw[major help lines] (\tikztostart) grid (\tikztotarget);
        \grd@start
        \pgfmathsetmacro{\grd@xa}{\the\pgf@x/1cm}
        \pgfmathsetmacro{\grd@ya}{\the\pgf@y/1cm}
        \grd@target
        \pgfmathsetmacro{\grd@xb}{\the\pgf@x/1cm}
        \pgfmathsetmacro{\grd@yb}{\the\pgf@y/1cm}
        \pgfmathsetmacro{\grd@xc}{\grd@xa + \pgfkeysvalueof{/tikz/grid with coordinates/major step x}}
        \pgfmathsetmacro{\grd@yc}{\grd@ya + \pgfkeysvalueof{/tikz/grid with coordinates/major step y}}
        \foreach \x in {\grd@xa,\grd@xc,...,\grd@xb}
        \node[anchor=north] at (\x,\grd@ya) {\pgfmathprintnumber{\x}};
        \foreach \y in {\grd@ya,\grd@yc,...,\grd@yb}
        \node[anchor=east] at (\grd@xa,\y) {\pgfmathprintnumber{\y}};
      }
    }
  },
  minor help lines/.style={
    help lines,
    gray,
    line cap =round,
    xstep=\pgfkeysvalueof{/tikz/grid with coordinates/minor step x},
    ystep=\pgfkeysvalueof{/tikz/grid with coordinates/minor step y}
  },
  major help lines/.style={
    help lines,
    line cap =round,
    line width=\pgfkeysvalueof{/tikz/grid with coordinates/major line width},
    xstep=\pgfkeysvalueof{/tikz/grid with coordinates/major step x},
    ystep=\pgfkeysvalueof{/tikz/grid with coordinates/major step y}
  },
  grid with coordinates/.cd,
  minor step x/.initial=.5,
  minor step y/.initial=.2,
  major step x/.initial=1,
  major step y/.initial=1,
  major line width/.initial=1pt,
}
\makeatother
\newacronym{6g}{6G}{sixth generation}
\newacronym{5g}{5G}{fifth generation}
\newacronym{snr}{SNR}{signal-to-noise ratio}
\newacronym{bw}{BW}{bandwidth}
\newacronym{mod}{Mod}{Modulation}
\newacronym[plural=\gls{cnn}s,firstplural=convolutional neural networks (CNNs)]{cnn}{CNN}{convolutional neural network}

\newacronym{iq}{I/Q}{in phase/quadrature}
\newacronym{mbc}{MBC}{modulation and bandwidth classification}
\newacronym{ml}{ML}{machine learning}
\newacronym{phy}{PHY}{physical layer}
\newacronym{cvl}{CVL}{convolutional layer}
\newacronym[plural=\gls{dnn}s,firstplural=deep neural networks (DNNs)]{dnn}{DNN}{deep neural network}
\newacronym{mmwave}{mmWave}{millimeter wave}
\newacronym{dsp}{DSP}{digital signal processing}
\newacronym{dsa}{DSA}{dynamic spectrum access}
\newacronym{ism}{ISM}{industrial, scientific and medical}
\newacronym{csi}{CSI}{channel state information}
\newacronym{fcc}{FCC}{Federal Communication Commission}
\newacronym{rfp}{RFP}{radio fingerprinting}
\newacronym[plural=\gls{sdr}s,firstplural=long training fields (SDRs)]{sdr}{SDR}{software-defined radio}

\newacronym{pus}{PUs}{primary users}
\newacronym{sus}{SUs}{secondary users}
\newacronym{iot}{IoT}{Internet of Things}
\newacronym{mimo}{MIMO}{multi-input, multi-output}
\newacronym{mum}{MU-MIMO}{multi-user \gls{mimo}}
\newacronym{sum}{SU-MIMO}{single-user \gls{mimo}}
\newacronym{iui}{IUI}{inter-user interference}
\newacronym{isi}{ISI}{inter-stream interference}

\newacronym{wlan}{WLAN}{Wireless LAN}
\newacronym{wlans}{WLANs}{Wireless Local Area Networks}
\newacronym{rlnc}{RLNC}{Random Linear Network Coding}
\newacronym{drx}{DRX}{Discontinuous Reception}
\newacronym{cpu}{CPU}{Central Processing Unit}
\newacronym{soc}{SoC}{system-on-chip}
\newacronym{dcm}{DCM}{distributed cooperative \gls{mimo}}
\newacronym{comp}{CoMP}{Coordinated Multi-Point}
\newacronym{ap}{AP}{access point}
\newacronym{sta}{STA}{station}
\newacronym{dl}{DL}{downlink}

\newacronym{mcs}{MCS}{modulation and coding scheme}
\newacronym{cfr}{CFR}{channel frequency response}
\newacronym{ndp}{NDP}{null data packet}
\newacronym[plural=\gls{ltf}s,firstplural=long training fields (LTFs)]{ltf}{LTF}{long training field}
\newacronym{vht}{VHT}{very high throughput}
\newacronym{ofdm}{OFDM}{orthogonal frequency-division multiplexing}
\newacronym{cfo}{CFO}{carrier frequency offset}
\newacronym{sfo}{SFO}{sampling frequency offset}
\newacronym{pdd}{PDD}{packet detection delay}
\newacronym{ppo}{PPO}{phase-locked loop offset}
\newacronym{pa}{PA}{phase ambiguity}
\newacronym{sbc}{SBC}{single board computer}

\newacronym[plural=\gls{cm}s,firstplural=confusion matrices (CMs)]{cm}{CM}{confusion matrix}

\newacronym{id}{ID}{identifier}
\newacronym{aoa}{AoA}{angle of arrival}
\newacronym{ul}{UL}{uplink}

\newacronym{svd}{SVD}{singular value decomposition}

\newacronym[plural=\gls{pdf}s,firstplural=probability density functions (PDFs)]{pdf}{PDF}{probability density function}

\newcommand{\FW}{\texttt{DeepCSI}\xspace}

\newcommand{\MYfooter}{\smash{
\hfil\parbox[t][\height][t]{\textwidth}{\centering
\thepage}\hfil\hbox{}}}

\makeatletter

\def\ps@IEEEtitlepagestyle{%
\def\@oddhead{\parbox[t][\height][t]{\textwidth}{\centering
To be presented at IEEE ICDCS, Bologna, Italy, July 10-13, 2022.\\
}\hfil\hbox{}}%
\def\@evenhead{\scriptsize\thepage \hfil \leftmark\mbox{}}%
\def\@evenfoot{\MYfooter}}

\makeatother

\begin{document}

\title{\textit{\FW}: Rethinking Wi-Fi Radio Fingerprinting Through MU-MIMO CSI Feedback Deep Learning\vspace{-0.3cm}}
\author{\IEEEauthorblockN{Francesca Meneghello$^*$\orcidlink{0000-0002-9905-0360}, Michele Rossi$^{* \dag}$\orcidlink{0000-0003-1121-324X}, and Francesco Restuccia$^{\mathsection}$\orcidlink{0000-0002-9498-2302}\vspace{-0.4cm}}\\
 \IEEEauthorblockA{$^*$ Department of Information Engineering, University of Padova, Italy}
 \IEEEauthorblockA{$^{\dag}$ Department of Mathematics, University of Padova, Italy}
 \IEEEauthorblockA{$^{\mathsection}$ Institute for the Wireless Internet of Things, Northeastern University, United States}
 \IEEEauthorblockA{e-mail: \texttt{meneghello@dei.unipd.it}, \texttt{michele.rossi@unipd.it}, \texttt{frestuc@northeastern.edu}}
 \vspace{-0.9cm}}

\pagestyle{plain}

\maketitle

\begin{abstract} We present \FW, a novel approach to \mbox{Wi-Fi} radio fingerprinting (RFP) which leverages standard-compliant beamforming feedback matrices to authenticate MU-MIMO \mbox{Wi-Fi} devices on the move. 
By capturing unique imperfections in off-the-shelf radio circuitry, RFP techniques can identify wireless devices directly at the physical layer, allowing low-latency low-energy cryptography-free authentication. However, existing \mbox{Wi-Fi} RFP techniques are based on software-defined radio (SDRs), which may ultimately prevent their widespread adoption. Moreover, it is unclear whether existing strategies can work in the presence of MU-MIMO transmitters -- a key technology in modern \mbox{Wi-Fi} standards. Conversely from prior work, \FW does not require SDR technologies and can be run on any low-cost \mbox{Wi-Fi} device to authenticate MU-MIMO transmitters. Our key intuition is that imperfections in the transmitter's radio circuitry percolate onto the beamforming feedback matrix, and thus RFP can be performed without explicit channel state information (CSI) computation. \FW is robust to inter-stream and inter-user interference being the beamforming feedback not affected by those phenomena. We extensively evaluate the performance of \FW through a massive data collection campaign performed \textit{in the wild} with off-the-shelf equipment, where 10 MU-MIMO \mbox{Wi-Fi} radios emit signals in different positions. Experimental results indicate that \FW correctly identifies the transmitter with an accuracy of up to 98\%. The identification accuracy remains above 82\% when the device moves within the environment. To allow replicability and provide a performance benchmark, we pledge to share the 800~GB datasets -- collected in static and, for the first time, dynamic conditions -- and the code database with the community.\vspace{-0.1cm}
\end{abstract}

\vspace{-0.1cm}
\IEEEpeerreviewmaketitle
\glsresetall
\section{Introduction and Motivation}\label{sec:intro}\vspace{-0.1cm}
The sheer expansion of \gls{iot} is rapidly saturating unlicensed spectrum bands \cite{SpectrumCrunch}. With the global mobile data traffic projected to reach 164 exabytes per month in 2025 \cite{EricssonMobility2020}, spectrum congestion will soon decrease data throughput to intolerable levels. To alleviate the issue, the \gls{fcc} has recently released 150~$\mathrm{MHz}$ additional bandwidth in the 3.5~$\mathrm{GHz}$ spectrum band \cite{FCC3p5GHz}, as well as 1.2~$\mathrm{GHz}$ in the 6~$\mathrm{GHz}$ band (5.925--7.125), the latter providing opportunities to use up to 320 $\mathrm{MHz}$ channels to expand capacity and increase network performance \cite{FCC6GHz}.

The release of these spectrum bands for unlicensed use implies that previously licensed users (also known as incumbents), unlicensed \mbox{Wi-Fi} devices \cite{WiFi6E} and 5G cellular networks \cite{5G-6GHz-Band} will need to coexist in the same spectrum bands. This will necessarily require the enactment of strict, fine-grained \gls{dsa} rules \cite{WiFi5GCoexist}, which will require spectrum administrators to continuously monitor \textit{which} unlicensed \mbox{Wi-Fi} device is using the spectrum, and \emph{when} the device is using it. To this end, cryptography-based techniques are substantially unfeasible in this context, since a spectrum observer should possess the private keys exchanged among all the nodes in the network, which is unrealistic.

On the other hand, \gls{rfp} has attracted significant attention as reliable and effective spectrum-level authentication technique~\cite{zheng2019fid,16-Peng-ieeeiotj2018,17-Xie-ieeeiotj2018,18-Xing-ieeecomlet2018,sankhe2019oracle,12_vo2016fingerprinting,al2020exposing}. \gls{rfp} leverages naturally-occurring circuitry imperfections to compute a unique ``fingerprint'' of the device directly at the waveform level \cite{johnson1966physical}. Although \gls{rfp} for \gls{phy} \mbox{Wi-Fi} authentication has been explored, existing approaches require \gls{sdr} devices to extract \gls{rfp} features. This may ultimately prevent widespread adoption, since \glspl{sdr} require expert knowledge and are usually more expensive than off-the-shelf devices. Moreover, existing work has tackled \mbox{Wi-Fi} fingerprinting up to the legacy 802.11a/g/b standards, which do not support \gls{mimo} techniques. However, newer \mbox{Wi-Fi} releases such as 802.11ac/ax and the upcoming 802.11be will heavily rely on \gls{mum} techniques to deliver significantly higher throughput than previous standards \cite{ong2011ieee,khorov2018tutorial,deng2020ieee}. Thus, it is still unknown whether existing \gls{rfp} strategies can be applied in the significantly more complex \gls{mum} scenario, where \gls{iui} and \gls{isi} can significant decrease the quality of the fingerprint itself.\vspace{-0.3cm}
\begin{figure}[h]
    \centering
    \includegraphics[width=1\columnwidth]{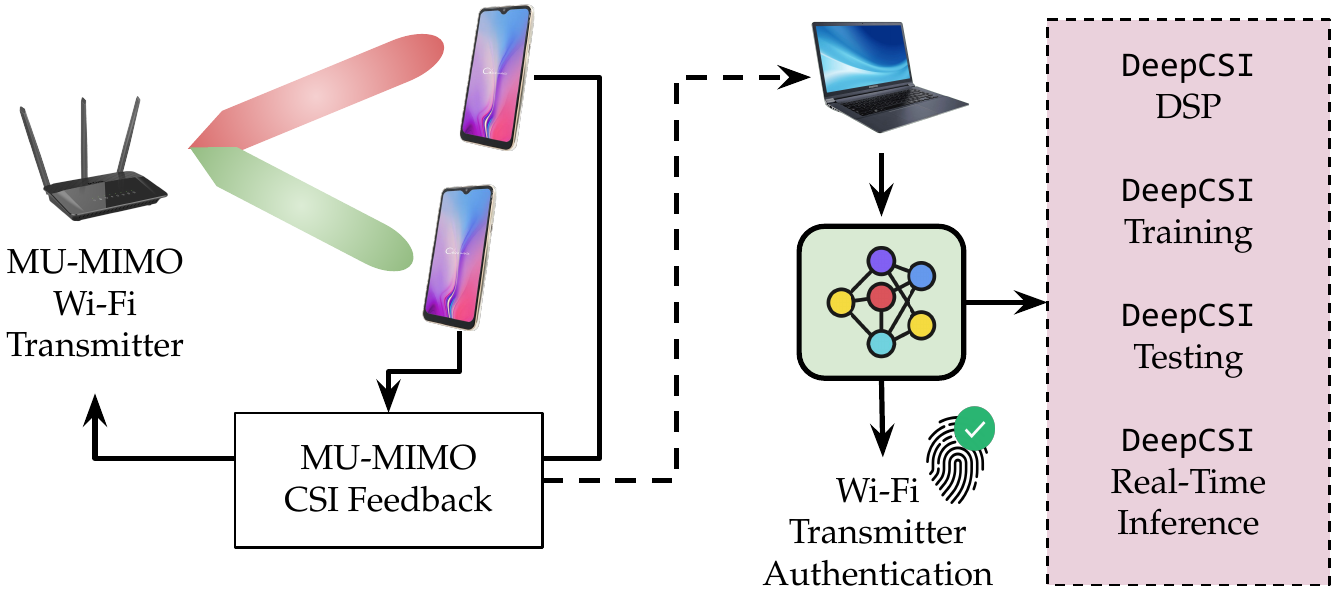}
    \setlength\abovecaptionskip{-.3cm}
    \setlength\belowcaptionskip{-.3cm}
    \caption{Main operations of \FW. The beamformer's fingerprint can be independently extracted from the feedback of any beamformee.}
    \label{fig:pipeline}
\end{figure}

To fill the current research gap, in this paper we propose \FW, a brand-new technique for \gls{rfp} of \mbox{Wi-Fi} devices which is summarized in \figurename~\ref{fig:pipeline}. The core intuition behind \FW is that the circuitry imperfections in the transmitter's radio interface will percolate onto the \gls{mum} \gls{csi} feedback sent by the receiver to the transmitter to perform beamforming. By demodulating this \gls{phy}-level information and performing deep learning techniques on a processed version of the feedback, an observer can fingerprint the transmitter without the need of \gls{sdr} capabilities. Note that the observer can leverage the feedback from any beamformee associated with the target beamformer to compute the beamformer's fingerprint.
The key advantage is that our technique is not affected by \gls{isi} nor by \gls{iui} -- see Section~\ref{subsec:challenges}. As pointed our earlier, the effect of \gls{iui} and \gls{isi} may prevent the correct devices' authentication. We stress that previous work only considered non-\gls{mimo} transmissions where \gls{iui}/\gls{isi} are not present and, in turn, the plain \gls{csi} information suffices to perform \gls{rfp}. On the other hand, the core concern of performing \gls{rfp} without direct \gls{csi} access is that it is unknown whether the imperfections will actually percolate onto the beamforming feedback matrix.
In this context, it is crucial to evaluate the \gls{phy} fingerprinting techniques as a function of different channels and different transmitter-receiver positions, since these can significantly undermine the fingerprint \cite{al2020exposing}. 

To summarize, this paper makes the following contributions:

$\bullet$ We propose \FW, the first approach to perform \gls{rfp} of MU-MIMO Wi-Fi devices. \FW uses deep learning of the standard-compliant beamforming matrices to learn the device-unique imperfections located in the \gls{csi} and authenticate MU-MIMO \mbox{Wi-Fi} devices directly at the \gls{phy} layer. The core intuition is that imperfections in the transmitter's radio circuitry are also present in the beamforming feedback matrix that is transmitted in clear text. Thus, conversely from prior work, explicit \gls{csi} computation through \gls{sdr} technologies are not needed and \FW can be run on any low-cost \mbox{Wi-Fi} device. Through \FW an observer can leverage the beamforming feedback matrix from any beamformee -- \textit{one at a time} -- associated with the beamformer to be authenticated. Given the small memory footprint, the trained learning algorithm can be run to perform the online inference on low-cost Wi-Fi devices, e.g., laptops, without the need for powerful facilities. 

$\bullet$ We extensively evaluate the performance of \FW through a massive data collection campaign performed \textit{in the wild} with off-the-shelf equipment, where 10 \mbox{Wi-Fi} radios emit MU-MIMO signals to multiple receivers located at different positions (and thus, with different beam patterns). Experimental results indicate that \FW is able to correctly identify the transmitter with an accuracy above 98\%, which shows that \gls{rfp} of MU-MIMO devices can be performed leveraging the \gls{csi} beamforming feedback matrices. We evaluate the impact of the feedback quantization error on the performance -- where quantization is applied for transmission efficiency reasons as per the Wi-Fi standards~\cite{IEEE-80211ac, IEEE-80211ax} -- observing an accuracy increase of up to 63\% when changing the feedback \gls{phy} parameters. We show that \FW achieves at least 17\% more accuracy than methods based on \gls{csi} phase cleaning, since the latter partially remove the imperfections due to the hardware circuitry. Finally, we evaluate the beamformer identification accuracy on the move where \FW achieves an accuracy above 82\%. \textbf{We pledge to share our code and 800~GB datasets with the community, which will allow replicability and provide a performance benchmark to other researchers in the field~\cite{dataset-code}}.

\section{Background, Related Work, and Challenges}\label{sec:related}

Thanks to their capability of identifying transmitters without the need of computation-hungry cryptography techniques, \gls{rfp} techniques have received a significant amount of attention from the research community
\cite{12_vo2016fingerprinting,16-Peng-ieeeiotj2018,17-Xie-ieeeiotj2018,18-Xing-ieeecomlet2018,1_xu2016device}. While early work has demonstrated the feasibility of \gls{rfp}, it has focused on the extraction of complex hand-tailored features, which do not scale well with the device population, or work in ad hoc propagation settings only. Among the first works on \mbox{Wi-Fi}-specific \gls{rfp}, Vo \textit{et al.} \cite{12_vo2016fingerprinting} propose \gls{rfp} techniques that extract features from the scrambling seed, the level of frequency offset and transients between symbols. However, the models achieve accuracy up to 50\% on 100 devices. The authors in~\cite{16-Peng-ieeeiotj2018}, instead, demonstrated that up to 54 ZigBee devices can be fingerprinted with about 95\% accuracy through PSK transients. More recently, Zheng \textit{et al.} \cite{zheng2019fid} studied and evaluated in a testbed of 33 devices a model-based approach to summarize imperfections in the modulation, timing, frequency and power amplifier noise. It is not clear, however, whether the approach in \cite{zheng2019fid} generalizes to different channel environments.

In stark contrast with early work, recent \gls{rfp} papers have leveraged deep learning techniques to fingerprint wireless devices \cite{restuccia2019deepradioid,sankhe2019oracle,Riyaz-ieeecommag2018,merchant2018deep,das2018deep}. A key advantage of deep learning techniques is that they are able to perform feature extraction and classification at the same time, thus avoiding manual extraction of device-distinguishing features. For example, Das \emph{et al.} \cite{das2018deep} and Merchant \textit{et al.} \cite{merchant2018deep} \glspl{dnn} achieve more than 90\% accuracy with a population of 7 ZigBee devices and 30 LoRa devices. To further increase accuracy, \cite{sankhe2019oracle,Riyaz-ieeecommag2018} proposed the introduction of artificial impairments at the transmitter's side. However, without compensation, this approach inevitably increases the bit error rate (BER). The usage of complex-valued \glspl{cnn} has been explored by Gopalakrishnan \textit{et al.} \cite{gopalakrishnan2019robust}, while in \cite{restuccia2019deepradioid} and \cite{doro2021can} the authors propose the usage of finite impulse response (FIR) filters to compensate for the adverse action of the wireless channel on the fingerprinting accuracy. The key limitation of existing work is that it is entirely based on \glspl{sdr}, which is very specialized, expensive equipment that is not widely available in common \mbox{Wi-Fi} networks. Moreover, to the best of our knowledge, no prior work has tackled the issue of assessing whether \gls{rfp} is feasible in \gls{mum} \mbox{Wi-Fi} networks. In this work, we address both issues at once by presenting \FW, a framework that (i) can be run on any off-the-shelf \mbox{Wi-Fi}-compliant device, and (ii) can accurately fingerprint \gls{mum} devices. We evaluate the performance of \FW in static and -- for the first time -- dynamic conditions, assessing the robustness of the learned fingerprint to changing transmission channel characteristics.\vspace{-0.1cm}

\subsection{Challenges of MU-MIMO Fingerprinting}\label{subsec:challenges}

Performing \gls{rfp} of devices operating in \gls{dl} \gls{mum} mode is significantly more challenging than \gls{rfp} of devices operating with omnidirectional antennas. First, transmissions are inevitably impaired by imperfect beamforming weights that do not accurately compensate the wireless channel. Secondly, (i) inter-stream interference (ISI) occurs between streams transmitted to the same receiver, and (ii) inter-user interference (IUI) affects streams directed to different receivers. The time-varying behavior of both \gls{isi} and \gls{iui} complicates the identification of the device-specific imperfections. Moreover, it has been shown in prior work that the \gls{rfp} process may be adversely impacted by the presence of the wireless channel \cite{sankhe2019oracle,al2020exposing}. This reasoning led us to design a different approach for extracting effective radio fingerprints. Specifically, we use the beamforming feedback matrix $\mathbf{\Tilde{V}}$ described in Section~\ref{subsec:beamforming_feedback}. The matrix $\mathbf{\Tilde{V}}$ is estimated based on the \gls{vht}-\glspl{ltf} of the \gls{ndp} that is sent in broadcast mode without being beamformed. Moreover, the \gls{vht}-\glspl{ltf} are sent over the different antennas in subsequent time slots of 4~$\mu$s each. Therefore, the \gls{ndp} and, in turn, $\mathbf{\Tilde{V}}$, \textit{are not affected by \gls{iui} nor by \gls{isi}}. 
However, since the feedback matrix is \textit{quantized} before transmission, quantization errors are inevitable. In Section \ref{sec:results}, we analyze the effect of the quantization error and investigate the generalization capability of our \gls{rfp} approach to multiple channels and beamformee positions, and to beamformer's mobility.

\section{The \FW Framework}\label{sec:deepcsi}
Henceforth, we will adopt the following notation for mathematical expressions.
The superscripts $T$ and $\dag$ respectively denote the transpose and the Hermitian of a matrix, i.e., the complex conjugate transpose. 
By $\angle{\mathbf{C}}$, we refer to the matrix whose elements are the phases of the corresponding elements in the complex-valued matrix $\mathbf{C}$.
diag$(c_1, \dots, c_j)$ indicates the diagonal matrix with elements $(c_1, \dots, c_j)$ on the main diagonal. 
The $(c_1, c_2)$ entry of matrix $\mathbf{C}$ is denoted by $\left[\mathbf{\mathbf{C}}\right]_{c_1, c_2}$. 
Finally, $\mathbf{I}_{c}$ refers to a $c \times c$ identity matrix while $\mathbf{I}_{c\times d}$ is a $c\times d$ matrix with ones on the main diagonal and zeros elsewhere.\vspace{-0.1cm}

\subsection{Preliminaries on MU-MIMO in Wi-Fi}\label{sec:mu-mimo}

In the following, we will consider Wi-Fi devices operating with the IEEE 802.11ac (Wi-Fi 5) standard and 802.11ax (WiFi 6 and 6E)~\cite{IEEE-80211ac, IEEE-80211ax}. These devices operate on the 2.4~GHz, 5~GHz and 6~GHz frequency bands with channels having up to 160~MHz of bandwidth. In Wi-Fi, data is transmitted via \gls{ofdm} by dividing the selected channel into $K$ partially overlapping and orthogonal sub-channels, spaced apart by $1/T$.
The input bits are grouped into \gls{ofdm} samples, $a_k$, and symbols, $\mathbf{a} = [a_{-K/2}, \dots, a_{K/2-1}]$, collecting $K$ samples each. After being digitally modulated, the $K$ samples of one \gls{ofdm} symbol are simultaneously transmitted though the $K$ \gls{ofdm} sub-channels, occupying the channel for $T$ seconds. 

\noindent Up-converted to the carrier $f_c$, the transmitted signal is
\beq \label{eq:tx_signal}
	 s_{\rm tx}(t) = e^{j2\pi f_c t} \sum_{k=-K/2}^{K/2-1} a_{k} e^{j2\pi kt/T}.
\eeq

\begin{figure}[h]
    \centering
    \includegraphics[width=0.95\columnwidth]{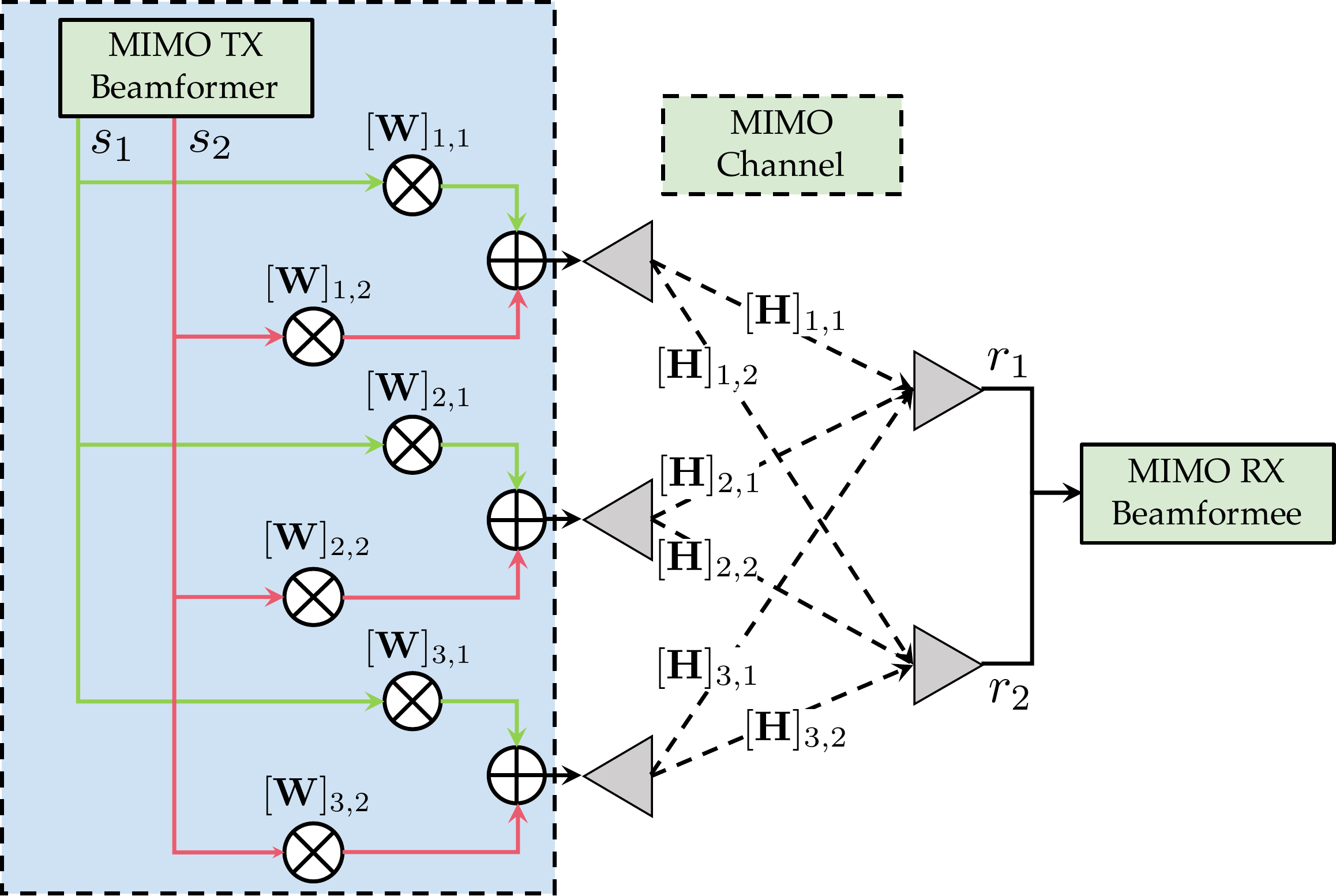}
    \setlength\belowcaptionskip{-0.3cm}
    \caption{$3 \times 2$ MIMO system. The grey triangles represent the antennas. ${s_1, s_2}$ and ${r_1, r_2}$ stand for the transmitted and received signals respectively. $\mathbf{W}$ is the steering matrix containing the weights to shape the beams. $\mathbf{H}$ is the \gls{cfr}.}
    \label{fig:beamforming}
\end{figure}

To improve the \gls{snr}, the transmitter can use \emph{beamforming} to focus the power toward the intended receiver. The beamforming may also compensate the effect of the wireless channel from the transmitter (\textit{beamformer}) to the receiver (\textit{beamformee}). When both devices in the communication link are equipped with antenna arrays (\gls{mimo} system), each pair of transmitter and receiver antennas forms a physical channel that can be exploited for wireless communication. This spatial diversity allows shaping multiple beams, referred to as \textit{spatial streams}, to transmit different signals to the beamformee, in a parallel fashion. To this end, the signals are combined at each transmitter antenna through steering weights, $\mathbf{W}$, derived from the \gls{cfr} matrix $\mathbf{H}$. The \gls{cfr} needs to be estimated for every \gls{ofdm} sub-channel over each pair of transmitter (TX) and receiver (RX) antennas, thus obtaining a $K \times M \times N$ matrix, where $M$ and $N$ are respectively the number of TX and RX antennas. 
In \figurename~\ref{fig:beamforming} we report an example of beamforming for a $3 \times 2$ \gls{mimo} system. At the beamformee side, the original signals are retrieved from their combination exploiting the fact that, ideally, $[\mathbf{H}]_{\bar{\ell}, \bar{i}} [\mathbf{W}]_{\ell, i} = 0$ when $\bar{\ell} \neq \ell$ or $\bar{i} \neq i$.\vspace{-0.3cm}

A meaningful model for the \gls{cfr} $\mathbf{H}$ in indoor spaces is obtained by considering the proprieties of the wireless propagation. After being irradiated by the transmitter antenna \mbox{$m \in \{0, \dots, M-1\}$}, the signal is reflected by objects in the environment and, in turn, $P$ different copies of $s_{\rm tx}(t)$ are collected at the receiver antenna \mbox{$n \in \{0, \dots, N-1\}$}. Each received signal is characterized by an attenuation $A_p$ and a delay $\tau_p$ that depends on the length of the path followed by the transmitted wave.
Thus, the $(k, m, n)$ element of $\mathbf{H}$ is 
\beq \label{eq:h_subcarrier}
	 [\mathbf{H}]_{k, m, n} =\sum_{p=0}^{P-1} A_{m, n, p} e^{-j2\pi (f_c + k/T)\tau_{m, n, p}}.
\eeq
By knowing $\mathbf{H}$, the beamformer can generate the steering matrix $\mathbf{W}$ to maximize the power sent toward the beamformee or simultaneously send parallel data streams to multiple beamformees. These communication modes are respectively referred to as \gls{sum} and \gls{mum}.
While IEEE 802.11n only supports \gls{sum} mode, in 802.11ac and above \gls{mum} can be enabled in the \gls{dl} direction, i.e., at the \gls{ap} side~\cite{IEEE-80211ac}. In 802.11ax \gls{mum} can be also enabled in the \gls{ul} \cite{IEEE-80211ax}. \vspace{-0.3cm}

\subsection{Compressed Beamforming Feedback}\label{subsec:beamforming_feedback}

In IEEE 802.11ac/ax, \gls{dl} \gls{mum} is enabled by the \textit{pre-coding} and the \textit{channel sounding} procedures~\cite{IEEE-80211ac}. Pre-coding linearly combines the signals to be simultaneously transmitted to the different beamformees. This procedure shapes the beams focusing the power in the correct directions. The combination weights are \mbox{antenna-specific} and are computed based on channel sounding performed through a \gls{ndp}, transmitted without beamforming. After receiving the \gls{ndp}, each beamformee estimates $\mathbf{H}$ based on a \gls{vht}-\gls{ltf} for each spatial stream. Next, the beamformee feeds back the matrix to the beamformer in the form of a \textit{compressed beamforming feedback}, which is computed for each sub-channel $k$ as follows.

Let $\mathbf{H}_k$ be the $M \times N$ sub-matrix of $\mathbf{H}$ containing the \gls{cfr} samples (see \eq{eq:h_subcarrier}) related to sub-channel $k$. $\mathbf{H}_k$ is first decomposed via \gls{svd}:
\beq
    \mathbf{H}_k^T = \mathbf{U}_k\mathbf{S}_k\mathbf{Z}_k^\dag,
\eeq
where $\textbf{U}_k$ and $\mathbf{Z}_k$ are, respectively, $N \times N$ and $M \times M$ unitary matrices, while $\mathbf{S}_k$ is an $N\times M$ diagonal matrix collecting the singular values. Next, the first $N_{\rm SS} \le N$ columns of $\mathbf{Z}_k$ are extracted to form the complex-valued beamforming matrix $\mathbf{V}_k$ that is used by the beamformer to compute the pre-coding weights for the $N_{\rm SS}$ spatial streams directed to the beamformee. Note that the beamformee can be served with at maximum $N_{\rm SS}=N$ spatial streams (see Chapter 13 of~\cite{perahia_stacey_2008}).
Thus, the beamformee is required to send back $\mathbf{V}_k$ to the beamformer. To do that efficiently, instead of sending the complete matrix, the beamformee derives and transmits its \textit{compressed representation}. 
Specifically, the feedback is a number of angles obtained by converting $\mathbf{V}_k$ into polar coordinates. 
The transformation is based on the procedure in Algorithm~\ref{alg:beamf_feedback}, where $\mathbf{D}_{k,i}$ and $\mathbf{G}_{k,\ell,i}$ are defined as
\beq
    \mathbf{D}_{k,i} =
	 \begin{bmatrix}
	\mathbf{I}_{i-1} & 0 & \multicolumn{2}{c}{\dots} & 0 \\
	0 & e^{j\phi_{k,i,i}} & 0 & \dots & \multirow{2}{*}{\vdots} \\
	\multirow{2}{*}{\vdots} & 0 & \ddots & 0 &  \\
	 & \vdots & 0 & e^{j\phi_{k,M-1,i}} & 0 \\
	0 & \multicolumn{2}{c}{\dots} & 0 & 1
	 \end{bmatrix},\label{eq:d_matrix}
\eeq
\beq
    \mathbf{G}_{k,\ell,i} =
	 \begin{bmatrix}
	\mathbf{I}_{i-1} & 0 & \multicolumn{2}{c}{\dots} & 0 \\
	0 & \cos{\psi_{k,\ell,i}} & 0 & \sin{\psi_{k,\ell,i}} & \multirow{2}{*}{\vdots} \\
	\multirow{2}{*}{\vdots} & 0 & \mathbf{I}_{\ell-i-1} & 0 &  \\
	 & -\sin{\psi_{k,\ell,i}} & 0 & \cos{\psi_{k,\ell,i}} & 0 \\
	0 & \multicolumn{2}{c}{\dots} & 0 & \mathbf{I}_{M-\ell}
	 \end{bmatrix}.\label{eq:g_matrix}
\eeq
The obtained matrices allows rewriting $\mathbf{V}_k$ as
\beq
    \mathbf{V}_k = \mathbf{\Tilde{V}}_k \mathbf{\Tilde{D}}_k,
\eeq
with
\beq
    \mathbf{\Tilde{V}}_k = \prod_{i=1}^{\min(N_{\rm SS}, M-1)} \Bigg( \mathbf{D}_{k,i} \prod_{l=i+1}^{M}\mathbf{G}_{k,l,i}^T\Bigg) \mathbf{I}_{M\times N_{\rm SS}}, \label{eq:v_matrix}
\eeq
where the products represent matrix multiplications. Note that, by construction, the last row of the complex-valued $\mathbf{\Tilde{V}}_k$ matrix, i.e., the feedback for the $M$-th transmitter antenna, consists of non-negative real numbers.
Next, the $K \times M \times N_{\rm SS}$ beamforming matrix $\mathbf{\Tilde{V}}$ is obtained by stacking the $\mathbf{\Tilde{V}}_k$ matrices for $k \in \{-K/2, \dots, K/2-1\}$. Thanks to this transformation, the beamformee is only required to transmit the $\phi$ and $\psi$ angles from which the $\mathbf{\Tilde{V}}_k$ matrices can be reconstructed.
The beamforming performance is equivalent at the beamformee when using $\mathbf{V}_k$ or $\mathbf{\Tilde{V}}_k$ to construct the steering matrix $\mathbf{W}$ and, in turn, the feedback for $\mathbf{\Tilde{D}}_k$ is not sent~\cite{perahia_stacey_2008}.\vspace{-0.1cm}
\begin{algorithm}[h]
\caption{$\mathbf{V}_k$ matrix decomposition}\label{alg:beamf_feedback}
\begin{algorithmic} 
\small
\setstretch{1.2}
\STATE \textbf{Input:} $\mathbf{V}_k$
\STATE \textbf{Output:} $\mathbf{D}_{k,i}$ and $\mathbf{G}_{k,\ell,i}$ for $i \in \{1, \dots, \min(N_{\rm SS},M-1)\}$, $\ell \in \{i+1, \dots, M\}$
\smallskip
\smallskip
\STATE $\mathbf{\Tilde{D}}_k = {\rm diag}(e^{j \angle \left[\mathbf{V}_k\right]_{M,1}}, \dots, e^{j \angle \left[\mathbf{V}_k\right]_{M,N_{\rm SS}}})$ 
\STATE $\mathbf{\Omega}_k = \mathbf{V}_k\mathbf{\Tilde{D}}_k^\dag$
\FOR{$i \leftarrow 1$ to $\min (N_{\rm SS}, M-1)$}
    \STATE $\phi_{k,\ell,i} = \angle \left[\mathbf{\Omega}_k\right]_{\ell, i}$ with $\ell={i, \dots, M-1}$
    \STATE compute $\mathbf{D}_{k,i}$ through \eq{eq:d_matrix}
    \STATE $\mathbf{\Omega}_k \leftarrow \mathbf{D}_{k,i}^\dag \mathbf{\Omega}_k$
    \FOR{$\ell \leftarrow i+1$ to $M$}
        \STATE $\psi_{k,\ell,i} = \arccos \left( \frac{[\mathbf{\Omega}_k]_{i, i}}{\sqrt{[\mathbf{\Omega}_k]_{i, i}^2 + [\mathbf{\Omega}_k]_{\ell, i}^2}} \right)$ 
        \STATE compute $\mathbf{G}_{k,\ell,i}$ through \eq{eq:g_matrix}
		\STATE $\mathbf{\Omega}_k \leftarrow \mathbf{G}_{k,\ell,i} \mathbf{\Omega}_k$
	\ENDFOR
\ENDFOR 
\end{algorithmic}
\end{algorithm}\vspace{-0.1cm}

The angles are quantized for transmission using $b_{\phi} \in \{7, 9\}$ bits for $\phi$ and $b_{\psi} = b_{\phi}-2$ bits for $\psi$.
Next, the quantized values are packed into the \gls{vht} compressed beamforming frame and transmitted without encryption, thus allowing any device that can access the wireless channel to capture the information sent by the beamformee to the beamformer.
The $b_{\phi}$ and $b_{\psi}$ values can be read in the \gls{vht} \gls{mimo} control field of the frame, together with other information including the number of columns ($N_{\rm SS}$) and rows ($M$) in the beamforming matrix and the channel bandwidth. At the beamformer, the $\phi$ and $\psi$ angles are retrieved from their quantized versions \mbox{$q_{\phi} = \{0, \dots, 2^{b_{\phi}}-1\}$} and \mbox{$q_{\psi} = \{0, \dots, 2^{b_{\psi}}-1\}$} using
\beq
    \left[\phi, \psi \right] =\left[ \pi \left( \frac{1}{2^{b_{\phi}}} + \frac{q_{\phi}}{2^{b_{\phi}-1}} \right), \pi \left( \frac{1}{2^{b_{\psi}+2}} + \frac{q_{\psi}}{2^{b_{\psi}+1}} \right)\right]
\eeq

\subsection{\FW Workflow and Learning Architecture}\label{subsec:processing}

\begin{figure}[t]
    \centering
    \includegraphics[width=0.95\columnwidth]{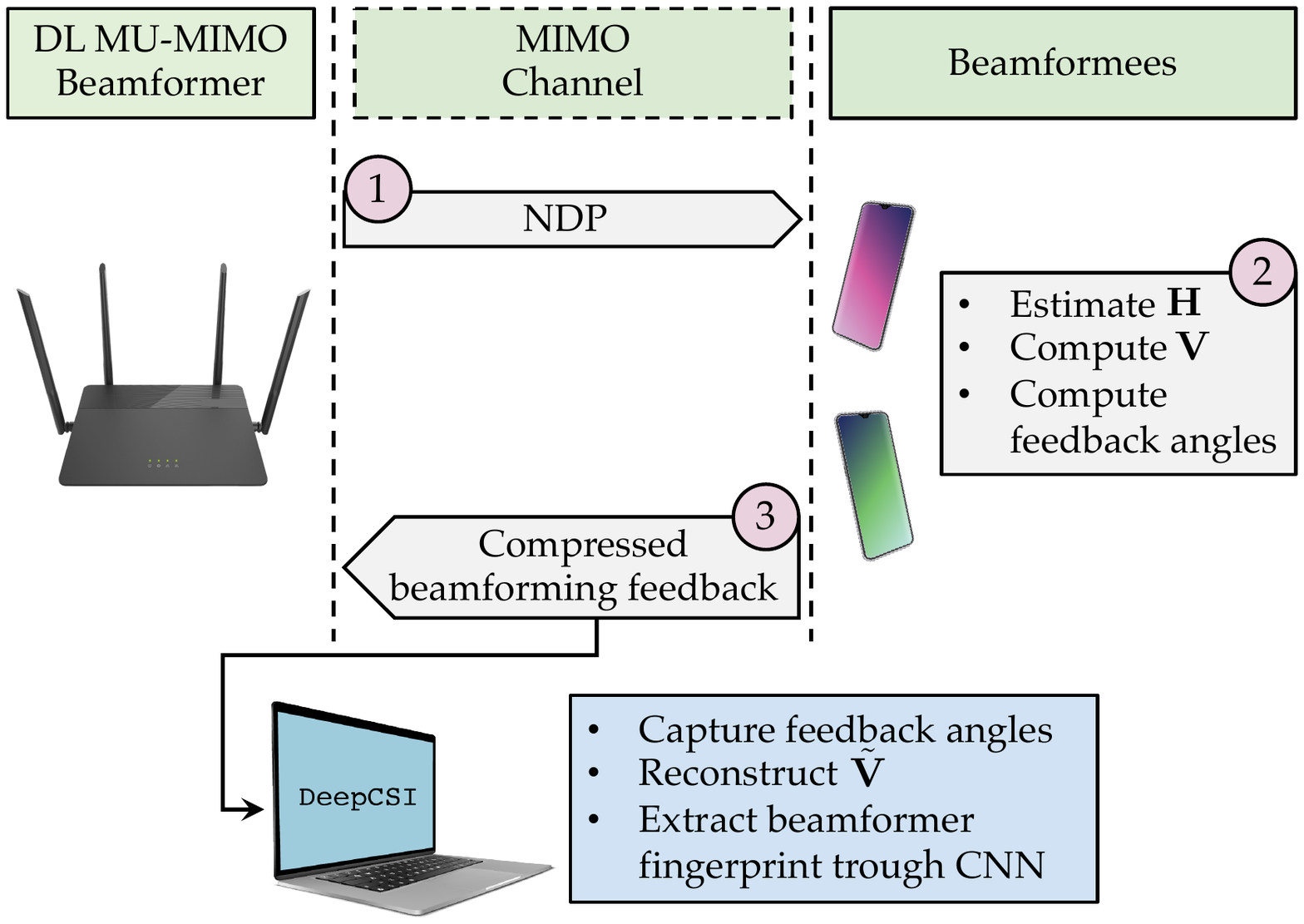}
    \setlength\belowcaptionskip{-.6cm}    
    \caption{\FW workflow. The compressed beamforming feedback computed by any of the beamformees as the final step of the sounding protocol is leveraged by \FW to obtain a fingerprint of the beamformer.}
    \label{fig:data_proc}
\end{figure}

\begin{figure}[t]
    \centering
    \includegraphics[width=0.85\columnwidth]{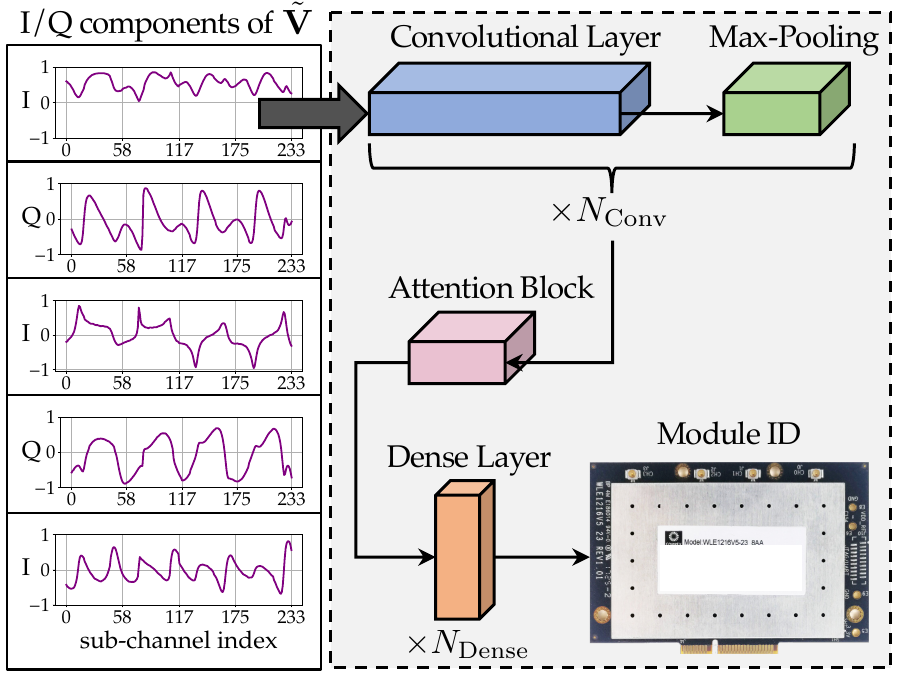}
    \setlength\belowcaptionskip{-.6cm}  
    \caption{\FW learning algorithm. The I and Q components of $\mathbf{\Tilde{V}}$ serves as input for a neural network classifier that computes the beamformer fingerprint and returns, as output, the estimated Wi-Fi module ID.}
    \label{fig:neural_network}
\end{figure}

\figurename~\ref{fig:data_proc} summarizes how \FW leverages the sounding protocol mechanism described in Section~\ref{subsec:beamforming_feedback} to obtain a fingerprint of the IEEE 802.11ac/ax \gls{ap} (beamformer). The sounding is triggered by the beamformer before sending data in \gls{dl} \gls{mum} mode to the beamformees, and concludes with the transmission of the feedback angles. \FW exploits the fact that the angles can be easily collected by any Wi-Fi compliant device by setting the \mbox{Wi-Fi} interface in monitor mode and using a network analyzer toolkit, e.g., Wireshark~\cite{orebaugh2006wireshark}, to capture the packet containing the feedback. Notice that \FW does not require the monitor device to be authenticated with the target \gls{ap}. Once obtained the feedback angles, \FW reconstructs $\mathbf{\Tilde{V}}$ through \eq{eq:v_matrix}. Next, the beamforming feedback matrix is used as input for the \gls{dnn} classifier depicted in~\figurename~\ref{fig:neural_network} to extract the \gls{rfp} of the beamformer. Once trained, the \gls{dnn} can be deployed and utilized in real time for device authentication at the \gls{phy} level.  The observer can leverage the feedback from {\it any beamformee} associated with the target beamformer to compute a beamformer's fingerprint. In turn, \FW is independent of the number of terminals connected to the \gls{ap}. Moreover, different fingerprints can be obtained for the same beamformer and can be indifferently used to authenticate the device.

The elements of the feedback matrix are fed to the \gls{dnn} as follows. The I/Q components of the beamforming feedback are stacked into an $N_{\rm row} \times N_{\rm col} \times N_{\rm ch}$ matrix, where \mbox{$N_{\rm col} \le K$} identifies the number of selected \gls{ofdm} sub-channels, \mbox{$N_{\rm row}\leq N_{\rm SS}$} and \mbox{$N_{\rm ch} < 2M$} refer to the columns and rows of $\mathbf{\Tilde{V}}$ used for fingerprinting and the 2-factor is for the I/Q components. Note that the feedback for the last transmitting antenna consists of the sole I information as, by construction, the last row of each $\mathbf{\Tilde{V}}_k$ (\eq{eq:v_matrix}) is composed of non-negative real values~\cite{IEEE-80211ac}. 
The learning architecture is inspired from \cite{OShea-ieeejstsp2018} and consists of a series of $N_{\rm conv}$ convolutional layers followed by \texttt{selu} activation function~\cite{klambauer2017self}, and by a \mbox{max-pooling} layer. The output of the previous block (in blue and green in \figurename~\ref{fig:neural_network}) is forward through an attention block and -- after being flattened -- is processed by $N_{\rm dense}$ dense layers with \texttt{selu} activation function. A final dense layer with \texttt{softmax} activation is used for classification. 
Alpha-dropout layers are interposed between the dense layers.
The attention block is inspired by the spatial attention module in \cite{Woo_2018_ECCV}. First, the maximum and the average feature maps are obtained by computing respectively the maximum and the mean of the input feature maps over the channel dimension. Next, the two maps are concatenated and forwarded through a convolutional layer with \texttt{sigmoid} activation function that outputs the weights to attend the input feature maps. Specifically, the attention operation consists in multiplying the input by the computed weights. A skip connection is also implemented by summing the output of the attention block with its input before passing the result to the subsequent dense layers. Thanks to the attention block, the algorithm learns where the most relevant information is located within the feature maps. This allows the network to focus on the relevant regions obtaining a more effective fingerprint.

We performed hyper-parameter evaluation in Section~\ref{sec:results}, and established through experiments that a good set of hyper parameters is $N_{\rm conv}=$~5 with 128 filters each, and $N_{\rm dense}=$~2 dense layers with 128 and 64 neurons each. This architecture yields a \gls{dnn} containing 489,301 trainable parameters, which is relatively small compared to state-of-the-art DNNs. The \FW learning algorithm is trained in an offline fashion by back propagating the \mbox{cross-entropy} loss between the module \gls{id} predicted by the classifier and the actual one. 

\section{Experimental Setup}\label{sec:setup}

\begin{figure}[t]
    \centering
    \includegraphics[width=0.85\columnwidth]{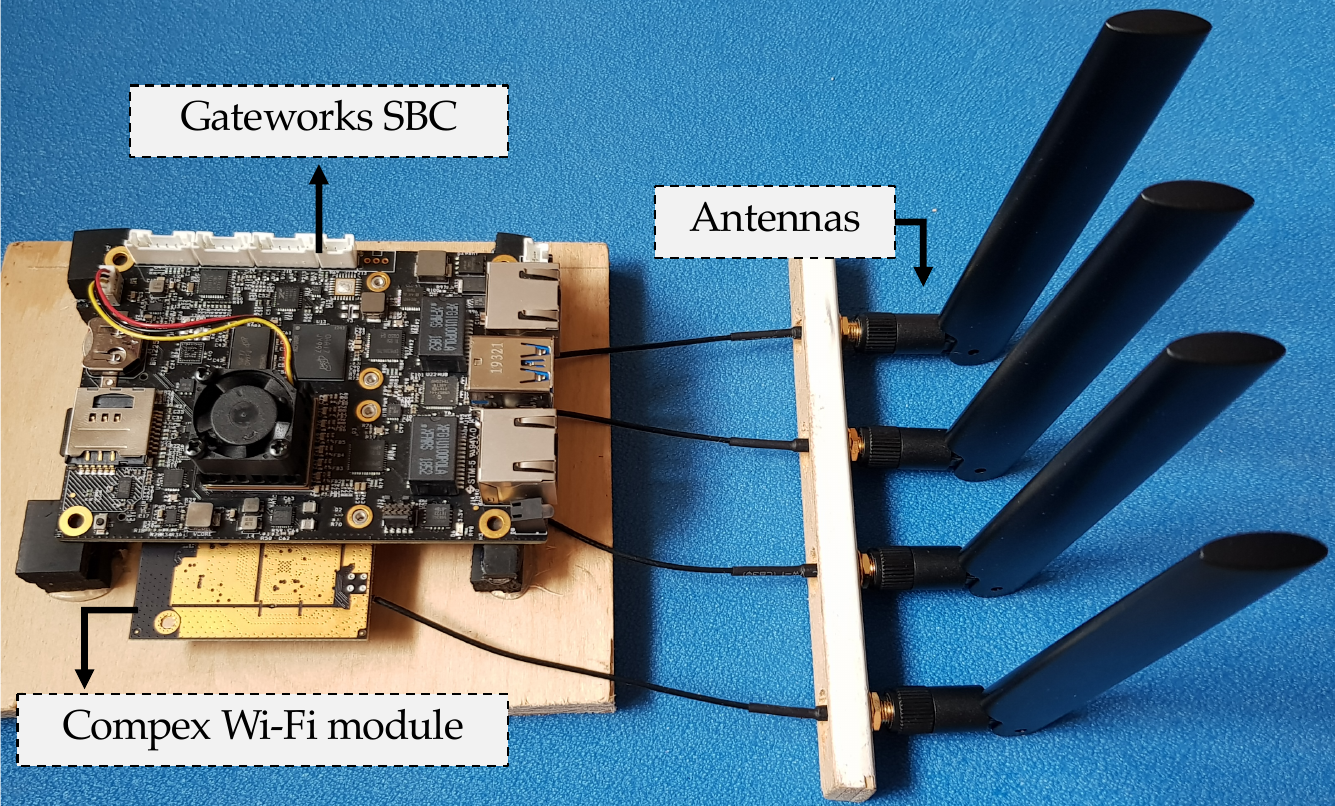}
    \setlength\belowcaptionskip{-.6cm}  
    \caption{\gls{dl} \gls{mum} enabled \mbox{Wi-Fi} \gls{ap} (beamformer). The Compex WLE$1216$v$5$-$23$ \mbox{Wi-Fi} module was mounted on a Gateworks GW$6200$ \gls{sbc} platform. Four antennas were connected to the Wi-Fi module.}
    \label{fig:transmitter}
\end{figure}

We evaluate the effectiveness of \FW using off-the-shelf devices and through extensive experimental evaluation. To this end, we set up a Wi-Fi network consisting of one \gls{ap} (beamformer) and two \glspl{sta} (beamformees). The \gls{ap} was implemented through a Gateworks \mbox{GW6200} \gls{sbc} equipped with a Compex \mbox{WLE1216v5-23} IEEE~802.11ac module, as shown in \figurename~\ref{fig:transmitter}. 
Two Netgear Nighthawk X4S \mbox{AC2600} routers, with $N \in$~\{1, 2\} out of 4 antennas enabled, acted as \glspl{sta} (beamformees). At the \gls{ap}, $M=$~3 antennas were used to sound the channel for \gls{dl} \gls{mum} transmission mode and the \glspl{sta} were served with $N_{\rm SS} \in$~\{1, 2\} spatial streams each. Note that implementation specific constraints prevent the use of $M=$~4 for \gls{dl} \gls{mum}. For the data transmission between the \gls{ap} and the \glspl{sta}, we used channel 42, i.e., $f_c=$~5.21~GHz with 80~MHz bandwidth. The number of \gls{ofdm} sub-channels sounded is $K=$~234 as the mechanism does not consider the 14 control sub-channels and the 8 pilot ones. The AP uses the quantization parameters $b_{\phi}=$~9 and $b_{\psi}=$~7 for $\phi$ and $\psi$ feedback angles, respectively. We generated UDP traffic in the \gls{dl} direction to induce the \gls{ap} to trigger the channel sounding mechanism, and collected the angles $(\phi,\psi)$ that were sent back by the beamformees using the Wireshark network analyzed toolkit~\cite{orebaugh2006wireshark} running on an off-the-shelf laptop equipped with an IEEE~802.11ac Wi-Fi card. This allows retrieving the $\mathbf{\Tilde{V}}$ matrices associated with each sounding operation, and computing the beamformer fingerprint (see Section~\ref{subsec:processing}).

\begin{figure}[t]
    \centering
    \includegraphics[width=0.85\columnwidth]{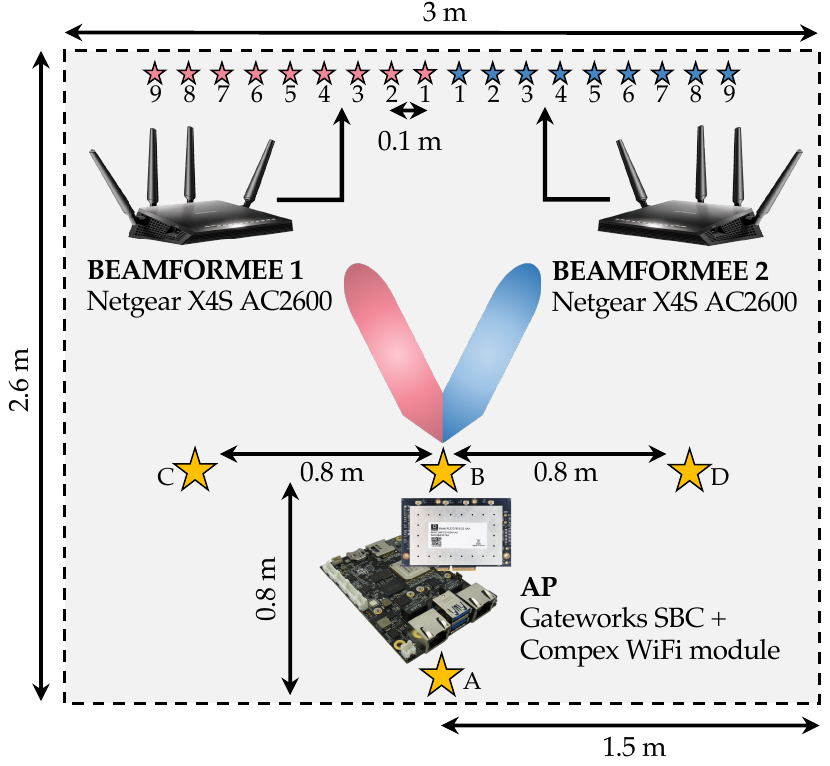}
    \setlength\abovecaptionskip{-.02cm}
    \setlength\belowcaptionskip{-.6cm}
    \caption{Indoor environment configuration. For dataset \texttt{D1}, the position of the \gls{ap} remains the same for all the acquisitions (yellow star A). The beamformees are first placed in front of the \gls{ap} and next, for each new experiment, beamformees 1 and 2 are respectively moved 10~cm to the left and 10~cm to the right. The subsequent positions of the beamformees are marked with red and blue stars respectively and labeled with a number $\in \{1, \dots, 9\}$. For the dynamic dataset \texttt{D2}, the beamformees remain fixed in position 3 while the \gls{ap} moves following the path described by the yellow stars A-B-C-D-B-A.}
    \label{fig:scenario}
\end{figure}

Two datasets -- namely \texttt{D1} and \texttt{D2} -- were collected. As for the former, the \glspl{sta} were deployed at different positions as depicted in \figurename~\ref{fig:scenario} to generate different beam patterns and different \gls{snr} regimes. The number of enabled antennas is $N=$~2 for each beamformee and each of them is served with $N_{\rm SS}=$~2 spatial streams.
Dataset \texttt{D1} allows evaluating the performance of \FW in different static conditions. The purpose of dataset \texttt{D2} is to evaluate the impact of mobility in the beamformer identification. The data were collected while the AP was manually moved following the path described in \figurename~\ref{fig:scenario} by the yellow stars A-B-C-D-B-A, entailing both vertical and horizontal movements. Here, $N=N_{\rm SS}=$~1 for the first beamformee and $N=N_{\rm SS}=$~2 for the second. The datasets were collected in two different indoor environments reproducing the same configuration depicted in \figurename~\ref{fig:scenario}. This allows evaluating the general applicability of the developed algorithm in recognizing the beamformer in the wild.

\noindent\textbf{We pledge to share our datasets with the community for reproducibility and benchmarking purposes~\cite{dataset-code}.}

\subsection{Datasets Structure}\label{subsec:dataset}

The datasets consist of the beamforming feedback angles associated with $N_{\rm modules}=$~10 different Compex Wi-Fi modules, which are the target of the proposed fingerprinting mechanism. They are collected in two indoor environments where the three entities constituting the experimental Wi-Fi network are placed as shown in \figurename~\ref{fig:scenario} and no obstacles are present between the \gls{ap} and the \glspl{sta}.
At the \gls{ap}, the \gls{sbc}, the antennas and the coaxial cables remain the same across all the considered network setups, by only changing the Compex Wi-Fi module. This ensures that the fingerprint procedure only relies on the hardware imperfections of the \mbox{Wi-Fi} module.

For the static dataset \texttt{D1}, we collected 9 different measurements for each Compex module by keeping it fixed in position A and changing the positions of the \glspl{sta}. Specifically, the beamformees are first placed in front of the beamformer, i.e., with an \gls{aoa} for the direct path of nearly zero degrees, and next moved of multiples of 10~cm respectively to the left and to the right with respect to their initial position (see the colored stars in \figurename~\ref{fig:scenario}). The positions of the \glspl{sta} are maintained fixed for the entire duration of each measurement. These configurations allow obtaining data associated with different beam shaping for the ongoing \gls{dl} \gls{mum} transmissions. Overall, we collected 90 traces, i.e., 9 traces for each of the 10 Compex \mbox{Wi-Fi} modules. 

As for the dynamic dataset \texttt{D2}, we collected 11 measurements for each Compex module. Four measurements are collected with the \gls{ap} fixed in position A. The remaining seven traces are collected while moving the \gls{ap} following the path described above, i.e., first, the \gls{ap} is moved 80~cm from position A toward the beamformees reaching position B, next is shifted 80~cm to the left and subsequently 160~cm to the right -- up to positions C and D respectively -- and finally it is brought back in position A passing from B. The beamformees are kept fixed in position 3. This dataset allows evaluating the performance of \FW in the presence of beamformer mobility. Overall, it consists of 11 traces for each of the 10 Compex \mbox{Wi-Fi} modules for a total of 110 traces.

Each trace contains the feedback angles sent by the two beamformees during two minutes of transmission. Such feedbacks can be promptly grouped based on the beamformee identifier by applying a filter on the packets source address. 

\subsection{\FW Training and Testing Procedure}\label{subsec:training}

The \FW classifier (see \figurename~\ref{fig:neural_network}) was trained using different \gls{phy} configurations, to evaluate its robustness in correctly identifying the beamformer device (the \gls{ap}) as the position of the beamformees change -- dataset \texttt{D1} -- and when the beamformer moves within the environment -- dataset \texttt{D2}. 

Table~\ref{tab:configs} summarizes the different training/testing sets that were considered for dataset \texttt{D1}, where the beamformees positions are depicted in \figurename~\ref{fig:scenario}. When the same positions are considered in the training and testing phase, the first 80\% of the collected data is used for training and validating the model, while the remaining 20\% serves as test data. In all cases, the last 20\% of training data is used for model validation. As part of our evaluation, we also assess the performance of \FW on $\mathbf{\Tilde{V}}$ \mbox{sub-matrices}. This makes it possible to evaluate the impact of using (i) different groups of transmitter antennas and spatial streams, and (ii) different portions of the radio spectrum. For (i), we vary $N_{\rm ch}$ and $N_{\rm row}$. For (ii), we pick a subset of the $K$ available \mbox{sub-channels}.
\begin{table}[t!]
\centering
\resizebox{0.49\textwidth}{!}{
\begin{tabular}{>{\raggedleft\arraybackslash}p{0.28cm}||*{9}{>{\small\centering\arraybackslash}p{0.1cm}} || *{9}{>{\small\centering\arraybackslash}p{0.1cm}}}
\toprule
\tabcolsep=0.11cm
\multirow{3}{*}{set} & \multicolumn{18}{c}{beamformees position}  \\
& \multicolumn{9}{c}{training} & \multicolumn{9}{c}{testing} \\ 
& 1 & 2 & 3 & 4 & 5 & 6 & 7 & 8 & 9 & 1 & 2 & 3 & 4 & 5 & 6 & 7 & 8 & 9 \\
\midrule
\texttt{S1} & \cellcolor{gray!40} & \cellcolor{gray!40} & \cellcolor{gray!40} & \cellcolor{gray!40} & \cellcolor{gray!40} & \cellcolor{gray!40} & \cellcolor{gray!40} & \cellcolor{gray!40} & \cellcolor{gray!40} & \cellcolor{gray!40} & \cellcolor{gray!40} & \cellcolor{gray!40} & \cellcolor{gray!40} & \cellcolor{gray!40} & \cellcolor{gray!40} & \cellcolor{gray!40} & \cellcolor{gray!40} & \cellcolor{gray!40} \\
\texttt{S2} & \cellcolor{gray!40} & & \cellcolor{gray!40} & & \cellcolor{gray!40} & & \cellcolor{gray!40} & & \cellcolor{gray!40} & & \cellcolor{gray!40} & & \cellcolor{gray!40} & & \cellcolor{gray!40} & & \cellcolor{gray!40} & \\
\texttt{S3} & \cellcolor{gray!40} & \cellcolor{gray!40} & \cellcolor{gray!40} & \cellcolor{gray!40} & \cellcolor{gray!40} & & & & & & & & & & \cellcolor{gray!40} & \cellcolor{gray!40} & \cellcolor{gray!40} & \cellcolor{gray!40}   \\
\bottomrule
\end{tabular}
}
\setlength\belowcaptionskip{-.2cm}
\caption{Dataset \texttt{D1}, training/testing sets to assess the \FW performance when varying the beamformees position in \{1, $\dots$, 9\} (see \figurename~\ref{fig:scenario}).\label{tab:configs}}
\end{table}

The training/test sets considered for dataset \texttt{D2} are detailed in Table~\ref{tab:configs_d2}.
For ease of readability, we combined the eleven traces composing the dataset into four groups. `Fix1' and `fix2' collect the four traces -- two traces each -- acquired keeping fixed the position of the \gls{ap}. The mobility traces -- i.e., collected while the \gls{ap} is manually moved in the environment -- are grouped in `mob1' and `mob2', where the first group contains four measurements while the remaining three traces compose the second group. Note that the mobility traces encode variations associated with the manual movement of the \gls{ap}. This implies that the positions taken by the \gls{ap} during the acquisition of the traces are \textit{approximately} the same due to slight variations in the movements. Moreover, a person is always present in the proximity of the \gls{ap} to perform the operation, introducing additional variability.
\begin{table}[t!]
\centering
\resizebox{0.49\textwidth}{!}{
\begin{tabular}{>{\raggedleft\arraybackslash}p{0.28cm}||*{4}{>{\small\raggedleft\arraybackslash}p{0.7cm}} || *{4}{>{\small\raggedleft\arraybackslash}p{0.7cm}}}
\toprule
\tabcolsep=0.11cm
\multirow{3}{*}{set} & \multicolumn{8}{c}{measurement group identifier}  \\
& \multicolumn{4}{c}{training} & \multicolumn{4}{c}{testing} \\ 
& fix1 & fix2 & mob1 & mob2 & fix1 & fix2 & mob1 & mob2  \\
\midrule
\texttt{S4} & & & \cellcolor{gray!40} & & & & & \cellcolor{gray!40} \\
\texttt{S5} & \cellcolor{gray!40} & \cellcolor{gray!40} & & & & & \cellcolor{gray!40} & \cellcolor{gray!40} \\
\texttt{S6} & & & \cellcolor{gray!40} & \cellcolor{gray!40} & \cellcolor{gray!40} & \cellcolor{gray!40} & & \\
\bottomrule
\end{tabular}
}
\setlength\belowcaptionskip{-.6cm}
\caption{Dataset \texttt{D2}, training/testing sets. `Fix1' and `fix2' group two static traces each, i.e., the \gls{ap} is fixed in position A (see \figurename~\ref{fig:scenario}). `Mob1' and `mob2' contain respectively four and three mobility traces, i.e., collected while the \gls{ap} is manually moved following the path detailed in \figurename~\ref{fig:scenario}. \label{tab:configs_d2}}
\end{table}

For each configuration, \FW is independently trained on the feedbacks from the two beamformees, obtaining one model for each of them. In this way, we evaluate a realistic usage scenario where each beamformee authenticates the beamformer based on local information, without relying on some other, possibly malicious, entities. The results considering both the beamformees are also reported for completeness.

\section{Experimental Results}\label{sec:results} 

\FW was experimentally evaluated on the Wi-Fi network setups of Tables~\ref{tab:configs} and \ref{tab:configs_d2}, assessing the effectiveness of the extracted beamformer fingerprint for different beamformer and beamformees configurations. 
We first briefly discuss the \gls{dnn} hyper parameters selection process and then present the \FW performance by varying the \gls{phy} parameters of the \gls{mum} transmission mode. In the first part, the \FW performance are assessed on dataset \texttt{D1}, evaluating the effect of the beamformees' positions. Dataset \texttt{D2} is considered in the second part to analyze the impact of the beamformer mobility on the device identification accuracy.

\begin{figure}[t]
    \centering
    \begin{subfigure}[t]{0.48\columnwidth}
    \centering
    \includegraphics[width=0.85\columnwidth]{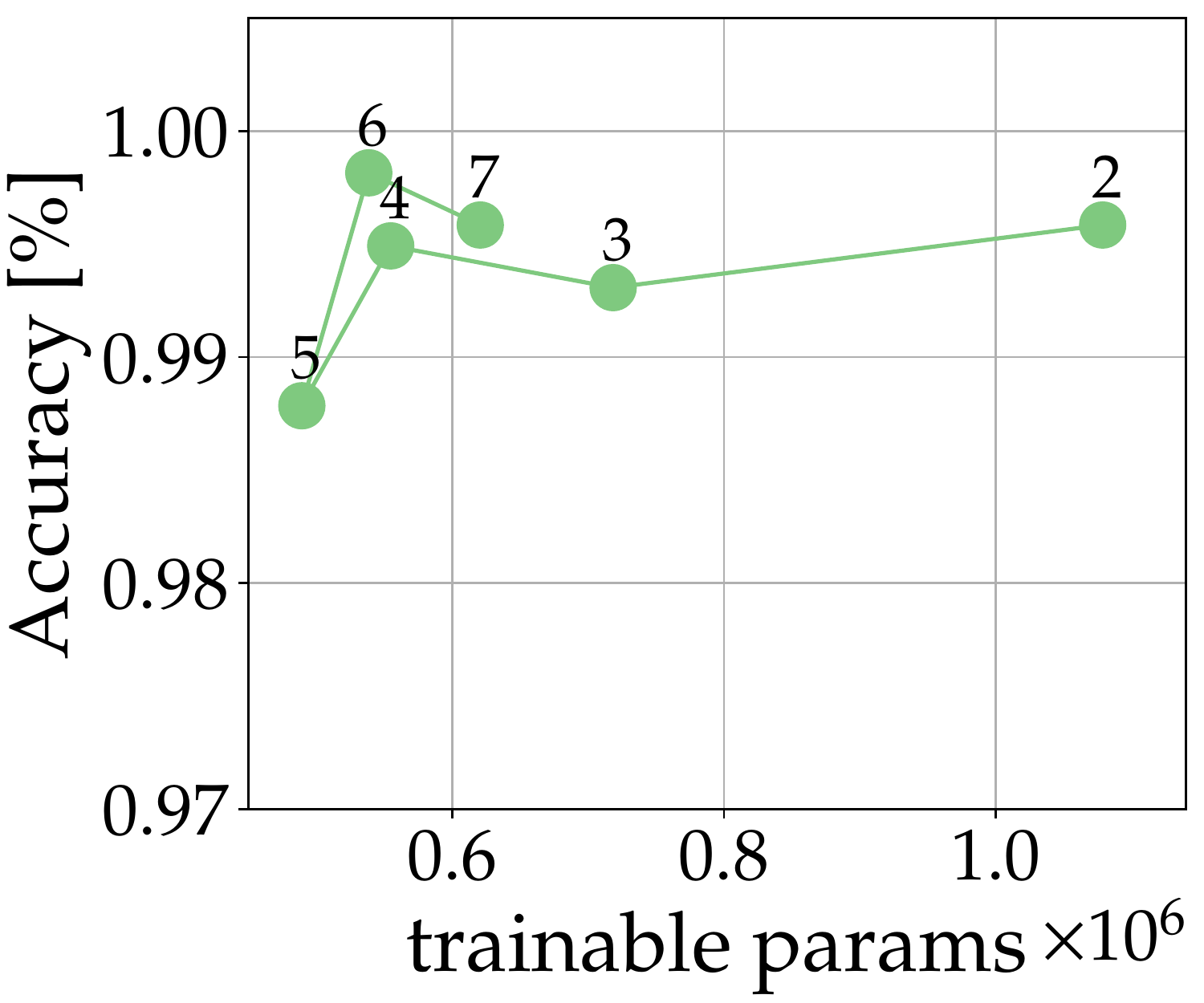}
    \caption{\FW accuracy by varying the number of convolutional layers, with 128 filters each, from 2 to 7.}
    \label{fig:hyperparam_layers}
    \end{subfigure}
    \hfill
    \begin{subfigure}[t]{0.48\columnwidth}
    \centering
    \includegraphics[width=0.85\columnwidth]{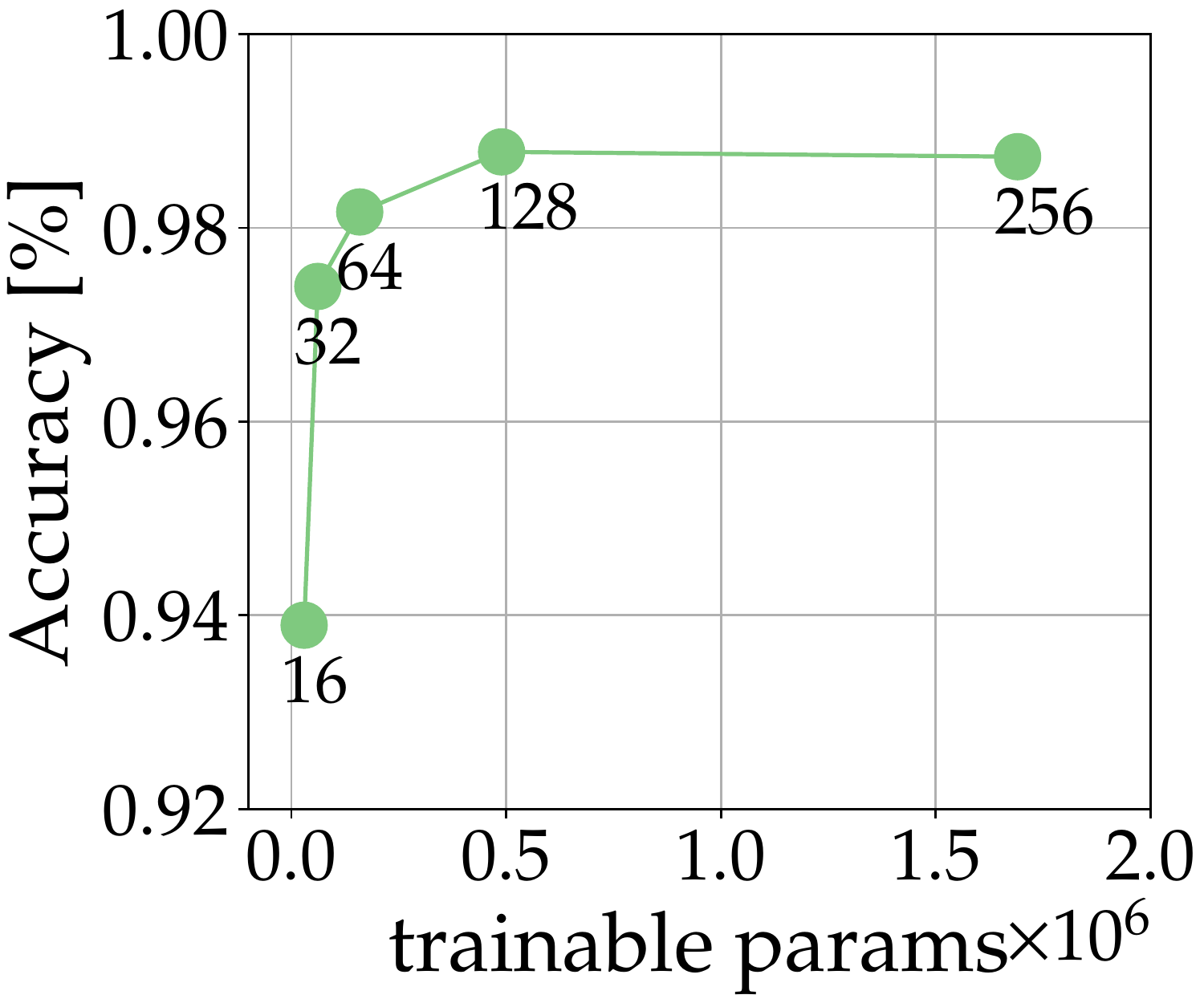}
    \caption{\FW accuracy by using 5 conv. layers and varying the no. of filters in each layer, from 16 to 256.}
    \label{fig:hyperparam_filters}
    \end{subfigure}
    \setlength\abovecaptionskip{0.1cm}
    \setlength\belowcaptionskip{-.35cm}
    \caption{\FW accuracy for beamformer 1, on \texttt{S1} validation data, by varying the \gls{dnn} parameters.}
    \label{fig:hyperparams}
\end{figure}

\smallskip
\noindent\textbf{\FW hyper parameters selection.}
\figurename~\ref{fig:hyperparam_layers} and \figurename~\ref{fig:hyperparam_filters} respectively evaluate the effect of tuning the number of convolutional layers and filters for the \gls{dnn} presented in Section~\ref{subsec:processing}. Noticeably, the accuracy remains almost constant when varying the number of layers. Also, the accuracy increases with an increasing number of filters, at a cost of a higher network complexity (i.e., more trainable parameters). As a trade-off between accuracy and complexity, we selected $N_{\rm conv}=$~5 convolutional layers with 128 filters each and kernel sizes of (1, 7) for the first three layers, (1,5) for the fourth and (1,3) for the last one by using the elbow method~\cite{ketchen-1996}. The max-pooling kernels are set to (1, 2) and the alpha-dropout between the three dense layers is applied with retain probability of 0.5 and 0.2, respectively.

\smallskip
\noindent\textbf{\FW performance using different beamformees configurations.}
\begin{figure}[t]
    \centering
    \begin{subfigure}[t]{0.32\columnwidth}
    \centering
    \setlength\fwidth{.75\columnwidth}
    \setlength\fheight{0.5\columnwidth}
    \begin{tikzpicture}
\pgfplotsset{every tick label/.append style={font=\tiny}}

\begin{axis}[
enlargelimits=false,
colorbar,
colormap/Purples,
width=\fwidth,
height=\fheight,
at={(0\fwidth,0\fheight)},
scale only axis,
tick align=inside,
xlabel={Predicted ID},
xmin=-0.5,
xmax=9.5,
xtick style={draw=none},
xlabel style={font=\scriptsize\color{white!15!black}},
ylabel style={font=\scriptsize\color{white!15!black}},
ylabel={Actual ID},
ymin=-0.5,
ymax=9.5,
xlabel shift=-5pt,
ylabel shift=-5pt,
ytick style={draw=none},
axis background/.style={fill=white},
colorbar horizontal,
colorbar style={
at={(0,1.05)},               % <-- (changed)
anchor=below south west,    % <-- (changed)
% change the width of the colorbar relative to the main `axis' environment
width=\pgfkeysvalueof{/pgfplots/parent axis width},
xtick={0, 0.5, 1},
xmin=0,
xmax=1,
axis x line*=top,
xticklabel shift=-1pt,
point meta min=0,
point meta max=1.05,
},
colorbar/width=2mm,
]
\addplot [matrix plot,point meta=explicit]
 coordinates {
(0,0) [1.0] (0,1) [0.0] (0,2) [0.0] (0,3) [0.0] (0,4) [0.0] (0,5) [0.0] (0,6) [0.0] (0,7) [0.0] (0,8) [0.0] (0,9) [0.0] 

(1,0) [0.0] (1,1) [0.9924812030075187] (1,2) [0.0] (1,3) [0.0037593984962406013] (1,4) [0.0] (1,5) [0.0037593984962406013] (1,6) [0.0] (1,7) [0.0] (1,8) [0.0] (1,9) [0.0] 

(2,0) [0.0] (2,1) [0.0] (2,2) [1.0] (2,3) [0.0] (2,4) [0.0] (2,5) [0.0] (2,6) [0.0] (2,7) [0.0] (2,8) [0.0] (2,9) [0.0] 

(3,0) [0.0] (3,1) [0.0] (3,2) [0.0] (3,3) [0.9961977186311787] (3,4) [0.0] (3,5) [0.0038022813688212928] (3,6) [0.0] (3,7) [0.0] (3,8) [0.0] (3,9) [0.0] 

(4,0) [0.0] (4,1) [0.0] (4,2) [0.0] (4,3) [0.0] (4,4) [1.0] (4,5) [0.0] (4,6) [0.0] (4,7) [0.0] (4,8) [0.0] (4,9) [0.0] 

(5,0) [0.0] (5,1) [0.0] (5,2) [0.0] (5,3) [0.0] (5,4) [0.0] (5,5) [1.0] (5,6) [0.0] (5,7) [0.0] (5,8) [0.0] (5,9) [0.0] 

(6,0) [0.0] (6,1) [0.0] (6,2) [0.0] (6,3) [0.0] (6,4) [0.0] (6,5) [0.0] (6,6) [1.0] (6,7) [0.0] (6,8) [0.0] (6,9) [0.0] 

(7,0) [0.0] (7,1) [0.0] (7,2) [0.0] (7,3) [0.0] (7,4) [0.0] (7,5) [0.0] (7,6) [0.0] (7,7) [0.8297101449275363] (7,8) [0.08695652173913043] (7,9) [0.08333333333333333] 

(8,0) [0.0] (8,1) [0.0] (8,2) [0.0] (8,3) [0.0] (8,4) [0.0] (8,5) [0.0] (8,6) [0.0] (8,7) [0.0] (8,8) [1.0] (8,9) [0.0] 

(9,0) [0.0] (9,1) [0.0] (9,2) [0.0] (9,3) [0.0] (9,4) [0.0] (9,5) [0.0] (9,6) [0.0] (9,7) [0.015384615384615385] (9,8) [0.0] (9,9) [0.9846153846153847]

};
\end{axis}
\end{tikzpicture}
    \setlength\abovecaptionskip{-.2cm}
    \caption{\texttt{S1}. Acc. 98.02\%}
    \label{fig:train_test_s1_1_ss_0}
    \end{subfigure}
    \hfill
    \begin{subfigure}[t]{0.32\columnwidth}
    \centering
    \setlength\fwidth{.75\columnwidth}
    \setlength\fheight{0.5\columnwidth}
    \begin{tikzpicture}
\pgfplotsset{every tick label/.append style={font=\tiny}}

\begin{axis}[
enlargelimits=false,
colorbar,
colormap/Purples,
width=\fwidth,
height=\fheight,
at={(0\fwidth,0\fheight)},
scale only axis,
tick align=inside,
xlabel={Predicted ID},
xmin=-0.5,
xmax=9.5,
xtick style={draw=none},
xlabel style={font=\scriptsize\color{white!15!black}},
ylabel style={font=\scriptsize\color{white!15!black}},
ylabel={Actual ID},
ymin=-0.5,
ymax=9.5,
xlabel shift=-5pt,
ylabel shift=-5pt,
ytick style={draw=none},
axis background/.style={fill=white},
colorbar horizontal,
colorbar style={
at={(0,1.05)},               % <-- (changed)
anchor=below south west,    % <-- (changed)
% change the width of the colorbar relative to the main `axis' environment
width=\pgfkeysvalueof{/pgfplots/parent axis width},
xtick={0, 0.5, 1},
xmin=0,
xmax=1,
axis x line*=top,
xticklabel shift=-1pt,
point meta min=0,
point meta max=1.05,
},
colorbar/width=2mm,
]
\addplot [matrix plot,point meta=explicit]
 coordinates {
(0,0) [0.9625212947189097] (0,1) [0.0] (0,2) [0.0] (0,3) [0.0] (0,4) [0.0] (0,5) [0.0] (0,6) [0.0] (0,7) [0.03747870528109029] (0,8) [0.0] (0,9) [0.0] 

(1,0) [0.10559006211180125] (1,1) [0.717391304347826] (1,2) [0.09472049689440994] (1,3) [0.004658385093167702] (1,4) [0.06366459627329192] (1,5) [0.006211180124223602] (1,6) [0.003105590062111801] (1,7) [0.0] (1,8) [0.0] (1,9) [0.004658385093167702] 

(2,0) [0.004559270516717325] (2,1) [0.0] (2,2) [0.9574468085106383] (2,3) [0.0] (2,4) [0.0243161094224924] (2,5) [0.0] (2,6) [0.00303951367781155] (2,7) [0.0060790273556231] (2,8) [0.004559270516717325] (2,9) [0.0] 

(3,0) [0.012131715771230503] (3,1) [0.0] (3,2) [0.0] (3,3) [0.9878682842287695] (3,4) [0.0] (3,5) [0.0] (3,6) [0.0] (3,7) [0.0] (3,8) [0.0] (3,9) [0.0] 

(4,0) [0.0018867924528301887] (4,1) [0.0018867924528301887] (4,2) [0.0] (4,3) [0.0] (4,4) [0.9320754716981132] (4,5) [0.03207547169811321] (4,6) [0.020754716981132074] (4,7) [0.0018867924528301887] (4,8) [0.005660377358490566] (4,9) [0.0037735849056603774] 

(5,0) [0.030911901081916538] (5,1) [0.00927357032457496] (5,2) [0.0] (5,3) [0.0030911901081916537] (5,4) [0.0] (5,5) [0.740340030911901] (5,6) [0.0] (5,7) [0.0] (5,8) [0.21483771251931993] (5,9) [0.0015455950540958269] 

(6,0) [0.0] (6,1) [0.0016366612111292963] (6,2) [0.01800327332242226] (6,3) [0.0] (6,4) [0.024549918166939442] (6,5) [0.0] (6,6) [0.9525368248772504] (6,7) [0.0] (6,8) [0.0032733224222585926] (6,9) [0.0] 

(7,0) [0.0] (7,1) [0.016574585635359115] (7,2) [0.27808471454880296] (7,3) [0.001841620626151013] (7,4) [0.055248618784530384] (7,5) [0.08839779005524862] (7,6) [0.014732965009208104] (7,7) [0.40515653775322286] (7,8) [0.0] (7,9) [0.13996316758747698] 

(8,0) [0.0] (8,1) [0.027649769585253458] (8,2) [0.1966205837173579] (8,3) [0.021505376344086023] (8,4) [0.0] (8,5) [0.0] (8,6) [0.10138248847926268] (8,7) [0.043010752688172046] (8,8) [0.6082949308755761] (8,9) [0.0015360983102918587] 

(9,0) [0.23971377459749552] (9,1) [0.1771019677996422] (9,2) [0.04830053667262969] (9,3) [0.0] (9,4) [0.017889087656529516] (9,5) [0.0017889087656529517] (9,6) [0.0] (9,7) [0.259391771019678] (9,8) [0.01967799642218247] (9,9) [0.23613595706618962]

};
\end{axis}
\end{tikzpicture}
    \setlength\abovecaptionskip{-.2cm}
    \caption{\texttt{S2}. Acc. 75.41\%}
    \label{fig:train_test_s2_1_ss_0}
    \end{subfigure}
    \hfill
    \begin{subfigure}[t]{0.32\columnwidth}
    \centering
    \setlength\fwidth{.75\columnwidth}
    \setlength\fheight{0.5\columnwidth}
    \begin{tikzpicture}
\pgfplotsset{every tick label/.append style={font=\tiny}}

\begin{axis}[
enlargelimits=false,
colorbar,
colormap/Purples,
width=\fwidth,
height=\fheight,
at={(0\fwidth,0\fheight)},
scale only axis,
tick align=inside,
xlabel={Predicted ID},
xmin=-0.5,
xmax=9.5,
xtick style={draw=none},
xlabel style={font=\scriptsize\color{white!15!black}},
ylabel style={font=\scriptsize\color{white!15!black}},
ylabel={Actual ID},
ymin=-0.5,
ymax=9.5,
xlabel shift=-5pt,
ylabel shift=-5pt,
ytick style={draw=none},
axis background/.style={fill=white},
colorbar horizontal,
colorbar style={
at={(0,1.05)},               % <-- (changed)
anchor=below south west,    % <-- (changed)
% change the width of the colorbar relative to the main `axis' environment
width=\pgfkeysvalueof{/pgfplots/parent axis width},
xtick={0, 0.5, 1},
xmin=0,
xmax=1,
axis x line*=top,
xticklabel shift=-1pt,
point meta min=0,
point meta max=1.05,
},
colorbar/width=2mm,
]
\addplot [matrix plot,point meta=explicit]
 coordinates {
(0,0) [0.8726415094339622] (0,1) [0.036163522012578615] (0,2) [0.06446540880503145] (0,3) [0.0] (0,4) [0.026729559748427674] (0,5) [0.0] (0,6) [0.0] (0,7) [0.0] (0,8) [0.0] (0,9) [0.0] 

(1,0) [0.07678244972577697] (1,1) [0.45886654478976235] (1,2) [0.21206581352833637] (1,3) [0.11517367458866545] (1,4) [0.10054844606946983] (1,5) [0.02010968921389397] (1,6) [0.012797074954296161] (1,7) [0.0] (1,8) [0.0] (1,9) [0.003656307129798903] 

(2,0) [0.001890359168241966] (2,1) [0.017013232514177693] (2,2) [0.7088846880907372] (2,3) [0.0] (2,4) [0.0] (2,5) [0.0] (2,6) [0.27032136105860116] (2,7) [0.0] (2,8) [0.0] (2,9) [0.001890359168241966] 

(3,0) [0.0] (3,1) [0.2814159292035398] (3,2) [0.01415929203539823] (3,3) [0.6] (3,4) [0.0] (3,5) [0.005309734513274336] (3,6) [0.0] (3,7) [0.0] (3,8) [0.09911504424778761] (3,9) [0.0] 

(4,0) [0.006734006734006734] (4,1) [0.23063973063973064] (4,2) [0.0] (4,3) [0.0] (4,4) [0.51010101010101] (4,5) [0.04377104377104377] (4,6) [0.15656565656565657] (4,7) [0.04040404040404041] (4,8) [0.010101010101010102] (4,9) [0.0016835016835016834] 

(5,0) [0.49133858267716535] (5,1) [0.10866141732283464] (5,2) [0.0] (5,3) [0.007874015748031496] (5,4) [0.0015748031496062992] (5,5) [0.38110236220472443] (5,6) [0.0] (5,7) [0.0] (5,8) [0.0015748031496062992] (5,9) [0.007874015748031496] 

(6,0) [0.0] (6,1) [0.1574539363484087] (6,2) [0.2814070351758794] (6,3) [0.0016750418760469012] (6,4) [0.005025125628140704] (6,5) [0.0] (6,6) [0.5544388609715243] (6,7) [0.0] (6,8) [0.0] (6,9) [0.0] 

(7,0) [0.028481012658227847] (7,1) [0.0] (7,2) [0.08227848101265822] (7,3) [0.28322784810126583] (7,4) [0.23575949367088608] (7,5) [0.14082278481012658] (7,6) [0.12974683544303797] (7,7) [0.00949367088607595] (7,8) [0.0] (7,9) [0.09018987341772151] 

(8,0) [0.0] (8,1) [0.24358974358974358] (8,2) [0.1778846153846154] (8,3) [0.1362179487179487] (8,4) [0.0] (8,5) [0.08333333333333333] (8,6) [0.04487179487179487] (8,7) [0.08974358974358974] (8,8) [0.22275641025641027] (8,9) [0.0016025641025641025] 

(9,0) [0.00718132854578097] (9,1) [0.22800718132854578] (9,2) [0.0017953321364452424] (9,3) [0.02154398563734291] (9,4) [0.44703770197486536] (9,5) [0.2926391382405745] (9,6) [0.0] (9,7) [0.0] (9,8) [0.0] (9,9) [0.0017953321364452424]

};
\end{axis}
\end{tikzpicture}
    \setlength\abovecaptionskip{-.2cm}
    \caption{\texttt{S3}. Acc. 42.97\%}
    \label{fig:train_test_s3_1_ss_0}
    \end{subfigure}
    
    \setlength\abovecaptionskip{0.1cm}
    \setlength\belowcaptionskip{-.6cm}
    \caption{Confusion matrices for beamformee 1, 3 TX antennas, spatial stream 0. ID in this and in the following plots refers to the \gls{ap} Wi-Fi module identifier.}
    \label{fig:train_test_s1_s2_s3_beamformee1}
\end{figure}
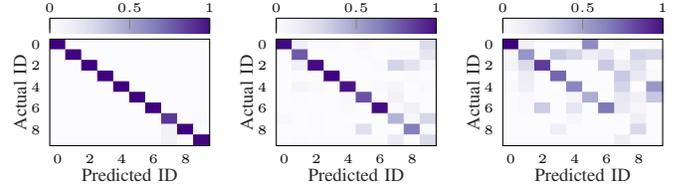
Fig.~\ref{fig:train_test_s1_s2_s3_beamformee1} shows the accuracy of \FW in correctly identifying the beamformer among the 10 Compex \mbox{Wi-Fi} modules in the dataset. The results were obtained using the beamforming feedback angles from a single beamformee. The confusion matrices are reported for each of the three training/testing configurations in Table~\ref{tab:configs}, where \gls{id} refers to the \gls{ap} module identifier. We notice that the accuracy increases with more spatial diversity in the training data, reaching 98.02\% when all the configurations are used at training time (see Fig.~\ref{fig:train_test_s1_1_ss_0} for set \texttt{S1}). With sets \texttt{S2} and \texttt{S3}, the beamformee positions at training and testing times differ. The lowest accuracy is obtained with \texttt{S3} (\mbox{worst-case} configuration). This is because \texttt{S3} is the set with the largest difference between training and testing positions. 
The performance improves when going from \texttt{S3} to \texttt{S2}, as the latter provides \FW with a more balanced set of positions during training, allowing the classifier to fill the knowledge gaps by ``interpolating'' the patterns learned from adjacent positions. 
The network reuses information from similar beam patterns leading to an identification accuracy of 75\%, even when the beamformee is at a position that was not contained in the training set (see Fig.~\ref{fig:train_test_s2_1_ss_0}). 
The same applies to \figurename~\ref{fig:train_test_s1_s2_s3_beamformees12}, where the beamforming feedback angles of both beamformees are used to build the training set. This allows to slightly increase the \FW accuracy on sets \texttt{S2} and \texttt{S3}. However, using this technique in \mbox{real-world} scenarios poses security concerns associated with the reciprocal trustworthiness of the beamformees in a \mbox{Wi-Fi} network. 
\begin{figure}[t]
    \centering
    \begin{subfigure}[t]{0.32\columnwidth}
    \centering
    \setlength\fwidth{.75\columnwidth}
    \setlength\fheight{0.5\columnwidth}
    \begin{tikzpicture}
\pgfplotsset{every tick label/.append style={font=\tiny}}

\begin{axis}[
enlargelimits=false,
colorbar,
colormap/Purples,
width=\fwidth,
height=\fheight,
at={(0\fwidth,0\fheight)},
scale only axis,
tick align=inside,
xlabel={Predicted ID},
xmin=-0.5,
xmax=9.5,
xtick style={draw=none},
xlabel style={font=\scriptsize\color{white!15!black}},
ylabel style={font=\scriptsize\color{white!15!black}},
ylabel={Actual ID},
ymin=-0.5,
ymax=9.5,
xlabel shift=-5pt,
ylabel shift=-5pt,
ytick style={draw=none},
axis background/.style={fill=white},
colorbar horizontal,
colorbar style={
at={(0,1.05)},               % <-- (changed)
anchor=below south west,    % <-- (changed)
% change the width of the colorbar relative to the main `axis' environment
width=\pgfkeysvalueof{/pgfplots/parent axis width},
xtick={0, 0.5, 1},
xmin=0,
xmax=1,
axis x line*=top,
xticklabel shift=-1pt,
point meta min=0,
point meta max=1.05,
},
colorbar/width=2mm,
]
\addplot [matrix plot,point meta=explicit]
 coordinates {
(0,0) [0.9699842022116903] (0,1) [0.001579778830963665] (0,2) [0.0] (0,3) [0.02843601895734597] (0,4) [0.0] (0,5) [0.0] (0,6) [0.0] (0,7) [0.0] (0,8) [0.0] (0,9) [0.0] 

(1,0) [0.0033277870216306157] (1,1) [0.8951747088186356] (1,2) [0.0] (1,3) [0.08652246256239601] (1,4) [0.0] (1,5) [0.0066555740432612314] (1,6) [0.0] (1,7) [0.008319467554076539] (1,8) [0.0] (1,9) [0.0] 

(2,0) [0.0] (2,1) [0.0] (2,2) [1.0] (2,3) [0.0] (2,4) [0.0] (2,5) [0.0] (2,6) [0.0] (2,7) [0.0] (2,8) [0.0] (2,9) [0.0] 

(3,0) [0.0] (3,1) [0.0] (3,2) [0.0] (3,3) [1.0] (3,4) [0.0] (3,5) [0.0] (3,6) [0.0] (3,7) [0.0] (3,8) [0.0] (3,9) [0.0] 

(4,0) [0.0] (4,1) [0.0] (4,2) [0.0] (4,3) [0.0] (4,4) [1.0] (4,5) [0.0] (4,6) [0.0] (4,7) [0.0] (4,8) [0.0] (4,9) [0.0] 

(5,0) [0.0] (5,1) [0.0] (5,2) [0.0] (5,3) [0.0] (5,4) [0.0] (5,5) [1.0] (5,6) [0.0] (5,7) [0.0] (5,8) [0.0] (5,9) [0.0] 

(6,0) [0.0] (6,1) [0.0] (6,2) [0.0] (6,3) [0.0] (6,4) [0.0] (6,5) [0.0] (6,6) [1.0] (6,7) [0.0] (6,8) [0.0] (6,9) [0.0] 

(7,0) [0.0] (7,1) [0.0] (7,2) [0.0] (7,3) [0.0] (7,4) [0.05536912751677853] (7,5) [0.0] (7,6) [0.0016778523489932886] (7,7) [0.9060402684563759] (7,8) [0.0] (7,9) [0.03691275167785235] 

(8,0) [0.0] (8,1) [0.0] (8,2) [0.0] (8,3) [0.0] (8,4) [0.0] (8,5) [0.0] (8,6) [0.0] (8,7) [0.0] (8,8) [1.0] (8,9) [0.0] 

(9,0) [0.0] (9,1) [0.0] (9,2) [0.0] (9,3) [0.0] (9,4) [0.0] (9,5) [0.0] (9,6) [0.0] (9,7) [0.005128205128205128] (9,8) [0.0] (9,9) [0.9948717948717949]

};
\end{axis}
\end{tikzpicture}
    \setlength\abovecaptionskip{-.4cm}
    \caption{\texttt{S1}. Acc. 97.62\%}
    \label{fig:train_test_s1_12_ss_0}
    \end{subfigure}
    \hfill
    \begin{subfigure}[t]{0.32\columnwidth}
    \centering
    \setlength\fwidth{.75\columnwidth}
    \setlength\fheight{0.5\columnwidth}
    \begin{tikzpicture}
\pgfplotsset{every tick label/.append style={font=\tiny}}

\begin{axis}[
enlargelimits=false,
colorbar,
colormap/Purples,
width=\fwidth,
height=\fheight,
at={(0\fwidth,0\fheight)},
scale only axis,
tick align=inside,
xlabel={Predicted ID},
xmin=-0.5,
xmax=9.5,
xtick style={draw=none},
xlabel style={font=\scriptsize\color{white!15!black}},
ylabel style={font=\scriptsize\color{white!15!black}},
ylabel={Actual ID},
ymin=-0.5,
ymax=9.5,
xlabel shift=-5pt,
ylabel shift=-5pt,
ytick style={draw=none},
axis background/.style={fill=white},
colorbar horizontal,
colorbar style={
at={(0,1.05)},               % <-- (changed)
anchor=below south west,    % <-- (changed)
% change the width of the colorbar relative to the main `axis' environment
width=\pgfkeysvalueof{/pgfplots/parent axis width},
xtick={0, 0.5, 1},
xmin=0,
xmax=1,
axis x line*=top,
xticklabel shift=-1pt,
point meta min=0,
point meta max=1.05,
},
colorbar/width=2mm,
]
\addplot [matrix plot,point meta=explicit]
 coordinates {
(0,0) [0.9845320959010054] (0,1) [0.0030935808197989174] (0,2) [0.0030935808197989174] (0,3) [0.0] (0,4) [0.0] (0,5) [0.0] (0,6) [0.0] (0,7) [0.008507347254447023] (0,8) [0.0] (0,9) [0.0007733952049497294] 

(1,0) [0.06164383561643835] (1,1) [0.8112633181126332] (1,2) [0.0228310502283105] (1,3) [0.0624048706240487] (1,4) [0.0197869101978691] (1,5) [0.00684931506849315] (1,6) [0.0] (1,7) [0.0] (1,8) [0.0] (1,9) [0.015220700152207] 

(2,0) [0.0] (2,1) [0.0] (2,2) [0.9817073170731707] (2,3) [0.0] (2,4) [0.0007621951219512195] (2,5) [0.0] (2,6) [0.001524390243902439] (2,7) [0.010670731707317074] (2,8) [0.001524390243902439] (2,9) [0.0038109756097560975] 

(3,0) [0.0076045627376425855] (3,1) [0.08136882129277566] (3,2) [0.0] (3,3) [0.7604562737642585] (3,4) [0.0] (3,5) [0.045627376425855515] (3,6) [0.045627376425855515] (3,7) [0.003041825095057034] (3,8) [0.05627376425855513] (3,9) [0.0] 

(4,0) [0.0033305578684429643] (4,1) [0.0008326394671107411] (4,2) [0.0008326394671107411] (4,3) [0.0] (4,4) [0.7885095753538718] (4,5) [0.0] (4,6) [0.10741049125728559] (4,7) [0.0008326394671107411] (4,8) [0.005828476269775187] (4,9) [0.09242298084929226] 

(5,0) [0.021423635107118175] (5,1) [0.03178991015894955] (5,2) [0.0] (5,3) [0.10642709053213545] (5,4) [0.0] (5,5) [0.6855563234277816] (5,6) [0.0] (5,7) [0.1354526606772633] (5,8) [0.018659295093296474] (5,9) [0.000691085003455425] 

(6,0) [0.0] (6,1) [0.0007886435331230284] (6,2) [0.012618296529968454] (6,3) [0.0] (6,4) [0.007886435331230283] (6,5) [0.0] (6,6) [0.9787066246056783] (6,7) [0.0] (6,8) [0.0] (6,9) [0.0] 

(7,0) [0.0] (7,1) [0.013127413127413128] (7,2) [0.12432432432432433] (7,3) [0.008494208494208495] (7,4) [0.021621621621621623] (7,5) [0.05791505791505792] (7,6) [0.05637065637065637] (7,7) [0.5629343629343629] (7,8) [0.0] (7,9) [0.1552123552123552] 

(8,0) [0.0] (8,1) [0.06521739130434782] (8,2) [0.01424287856071964] (8,3) [0.15667166416791603] (8,4) [0.0] (8,5) [0.0] (8,6) [0.05697151424287856] (8,7) [0.03673163418290855] (8,8) [0.6701649175412294] (8,9) [0.0] 

(9,0) [0.09481361426256078] (9,1) [0.008103727714748784] (9,2) [0.02188006482982172] (9,3) [0.0] (9,4) [0.09562398703403566] (9,5) [0.0] (9,6) [0.0] (9,7) [0.26094003241491087] (9,8) [0.0] (9,9) [0.5186385737439222]

};
\end{axis}
\end{tikzpicture}
    \setlength\abovecaptionskip{-.4cm}
    \caption{\texttt{S2}. Acc. 77.38\%}
    \label{fig:train_test_s2_12_ss_0}
    \end{subfigure}
    \hfill
    \begin{subfigure}[t]{0.32\columnwidth}
    \centering
    \setlength\fwidth{.75\columnwidth}
    \setlength\fheight{0.5\columnwidth}
    \begin{tikzpicture}
\pgfplotsset{every tick label/.append style={font=\tiny}}

\begin{axis}[
enlargelimits=false,
colorbar,
colormap/Purples,
width=\fwidth,
height=\fheight,
at={(0\fwidth,0\fheight)},
scale only axis,
tick align=inside,
xlabel={Predicted ID},
xmin=-0.5,
xmax=9.5,
xtick style={draw=none},
xlabel style={font=\scriptsize\color{white!15!black}},
ylabel style={font=\scriptsize\color{white!15!black}},
ylabel={Actual ID},
ymin=-0.5,
ymax=9.5,
xlabel shift=-5pt,
ylabel shift=-5pt,
ytick style={draw=none},
axis background/.style={fill=white},
colorbar horizontal,
colorbar style={
at={(0,1.05)},               % <-- (changed)
anchor=below south west,    % <-- (changed)
% change the width of the colorbar relative to the main `axis' environment
width=\pgfkeysvalueof{/pgfplots/parent axis width},
xtick={0, 0.5, 1},
xmin=0,
xmax=1,
axis x line*=top,
xticklabel shift=-1pt,
point meta min=0,
point meta max=1.05,
},
colorbar/width=2mm,
]
\addplot [matrix plot,point meta=explicit]
 coordinates {
(0,0) [0.45736994219653176] (0,1) [0.028901734104046242] (0,2) [0.11271676300578035] (0,3) [0.0] (0,4) [0.03684971098265896] (0,5) [0.08526011560693642] (0,6) [0.0050578034682080926] (0,7) [0.27239884393063585] (0,8) [0.001445086705202312] (0,9) [0.0] 

(1,0) [0.0906958561376075] (1,1) [0.3057075840500391] (1,2) [0.07818608287724785] (1,3) [0.053166536356528536] (1,4) [0.040656763096168884] (1,5) [0.001563721657544957] (1,6) [0.012509773260359656] (1,7) [0.41360437842064113] (1,8) [0.0] (1,9) [0.003909304143862392] 

(2,0) [0.045304777594728174] (2,1) [0.0008237232289950577] (2,2) [0.41762767710049425] (2,3) [0.002471169686985173] (2,4) [0.0016474464579901153] (2,5) [0.023064250411861616] (2,6) [0.028830313014827018] (2,7) [0.4744645799011532] (2,8) [0.0] (2,9) [0.005766062602965404] 

(3,0) [0.0029784065524944155] (3,1) [0.07892777364110201] (3,2) [0.0] (3,3) [0.37379002233804914] (3,4) [0.03425167535368578] (3,5) [0.16232315711094564] (3,6) [0.039463886820551006] (3,7) [0.09307520476545049] (3,8) [0.21518987341772153] (3,9) [0.0] 

(4,0) [0.036431226765799254] (4,1) [0.046096654275092935] (4,2) [0.0] (4,3) [0.0] (4,4) [0.8111524163568773] (4,5) [0.010408921933085501] (4,6) [0.016356877323420074] (4,7) [0.07732342007434945] (4,8) [0.0007434944237918215] (4,9) [0.001486988847583643] 

(5,0) [0.30493273542600896] (5,1) [0.013452914798206279] (5,2) [0.0] (5,3) [0.0037369207772795215] (5,4) [0.0] (5,5) [0.6771300448430493] (5,6) [0.0] (5,7) [0.0] (5,8) [0.0007473841554559044] (5,9) [0.0] 

(6,0) [0.0] (6,1) [0.03975993998499625] (6,2) [0.12003000750187547] (6,3) [0.0007501875468867217] (6,4) [0.17704426106526633] (6,5) [0.0] (6,6) [0.6579144786196549] (6,7) [0.0] (6,8) [0.00450112528132033] (6,9) [0.0] 

(7,0) [0.005054151624548736] (7,1) [0.034657039711191336] (7,2) [0.10180505415162455] (7,3) [0.01588447653429603] (7,4) [0.0736462093862816] (7,5) [0.19350180505415163] (7,6) [0.09025270758122744] (7,7) [0.4584837545126354] (7,8) [0.009386281588447653] (7,9) [0.017328519855595668] 

(8,0) [0.0007942811755361397] (8,1) [0.1938046068308181] (8,2) [0.050833995234312944] (8,3) [0.059571088165210485] (8,4) [0.0698967434471803] (8,5) [0.056393963463065924] (8,6) [0.0007942811755361397] (8,7) [0.17315329626687848] (8,8) [0.3931691818903892] (8,9) [0.0015885623510722795] 

(9,0) [0.11522965350523771] (9,1) [0.10394842868654311] (9,2) [0.0008058017727639] (9,3) [0.031426269137792104] (9,4) [0.0821917808219178] (9,5) [0.009669621273166801] (9,6) [0.10233682514101532] (9,7) [0.41982272360999195] (9,8) [0.0] (9,9) [0.13456889605157132]

};
\end{axis}
\end{tikzpicture}
    \setlength\abovecaptionskip{-.4cm}
    \caption{\texttt{S3}. Acc. 47.28\%}
    \label{fig:train_test_s3_12_ss_0}
    \end{subfigure}
    \setlength\abovecaptionskip{-0.05cm}
    \setlength\belowcaptionskip{-.3cm}
    \caption{Confusion matrices, mixed beamformees, 3 TX ant., spatial stream 0.}
    \label{fig:train_test_s1_s2_s3_beamformees12}
\end{figure}
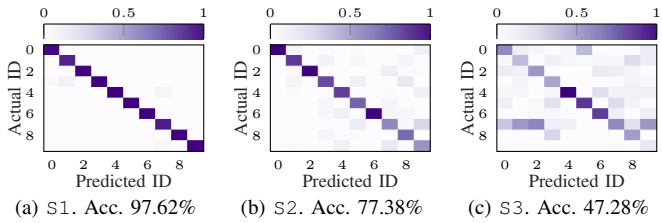
\begin{figure}[t]
    \centering
    \includegraphics[width=0.9\columnwidth]{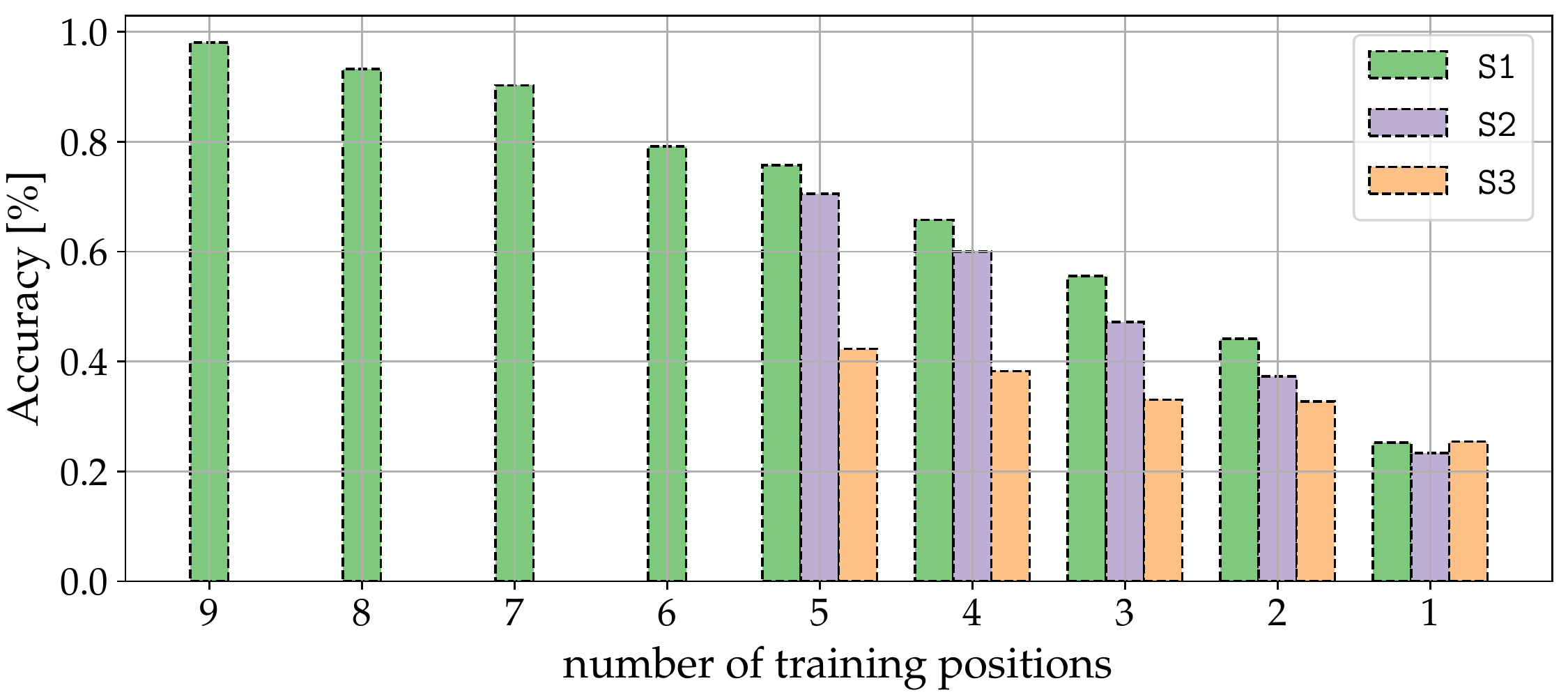}
    \setlength\abovecaptionskip{-0.02cm}
    \setlength\belowcaptionskip{-.6cm}
    \caption{\FW accuracy by varying the number of training positions from the considered set (see Table~\ref{tab:configs}). Set \texttt{S1} is trained on a maximum of 9 beamformee positions while \texttt{S2} and \texttt{S3} on 5.}
    \label{fig:comparison_train_pos}
\end{figure}
The impact of the number of beamformee training positions is evaluated in \figurename~\ref{fig:comparison_train_pos}. We report the accuracy obtained by increasing the number of positions used at training time from 1 to 9 for set \texttt{S1} and from 1 to 5 for sets \texttt{S2} and \texttt{S3}, according to Table~\ref{tab:configs}. In all the cases, the accuracy increases with more beamformee positions in the training data, which confirms that the fingerprint is more effective when high spatial diversity is present in the training data.
In \figurename~\ref{fig:train_one_test_another}, we evaluate the effect of swapping the beamformees used at training and testing times for the same network configuration. We trained \FW with data from a given beamformee and used the trained \gls{dnn} model to identify the \gls{ap} module from the $\mathbf{\Tilde{V}}$ matrices computed by a different beamformee (for the same \gls{ap} module). 
The learned fingerprint in this case performs poorly as matrix $\mathbf{\Tilde{V}}$ captures hardware inaccuracies of both devices, i.e., the beamformer (the \gls{ap}) {\it and} the beamformee. While a well designed learning architecture can identify with high accuracy the beamformer when the beamformee remains the same at training and testing times, it hardly succeeds when these devices differ. We reasonably believe that in a \mbox{real-world} scenario the impact of this will be even stronger, as the beamformees can be from different vendors and have different hardware configurations.

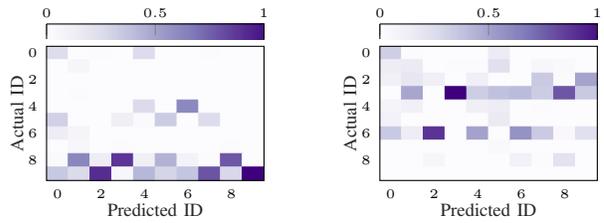
\begin{figure}[t]
    \centering
    \begin{subfigure}[t]{0.48\columnwidth}
    \centering
    \setlength\fwidth{.68\columnwidth}
    \setlength\fheight{0.42\columnwidth}
    \begin{tikzpicture}
\pgfplotsset{every tick label/.append style={font=\tiny}}

\begin{axis}[
enlargelimits=false,
colorbar,
colormap/Purples,
width=\fwidth,
height=\fheight,
at={(0\fwidth,0\fheight)},
scale only axis,
tick align=inside,
xlabel={Predicted ID},
xmin=-0.5,
xmax=9.5,
xtick style={draw=none},
xlabel style={font=\scriptsize\color{white!15!black}},
ylabel style={font=\scriptsize\color{white!15!black}},
ylabel={Actual ID},
ymin=-0.5,
ymax=9.5,
xlabel shift=-5pt,
ylabel shift=-5pt,
ytick style={draw=none},
axis background/.style={fill=white},
colorbar horizontal,
colorbar style={
at={(0,1.05)},               % <-- (changed)
anchor=below south west,    % <-- (changed)
% change the width of the colorbar relative to the main `axis' environment
width=\pgfkeysvalueof{/pgfplots/parent axis width},
xtick={0, 0.5, 1},
xmin=0,
xmax=1,
axis x line*=top,
xticklabel shift=-1pt,
point meta min=0,
point meta max=1.05,
},
colorbar/width=2mm,
]
\addplot [matrix plot,point meta=explicit]
 coordinates {
(0,0) [0.23225806451612904] (0,1) [0.0] (0,2) [0.0] (0,3) [0.0] (0,4) [0.0] (0,5) [0.29354838709677417] (0,6) [0.12903225806451613] (0,7) [0.0] (0,8) [0.016129032258064516] (0,9) [0.32903225806451614] 

(1,0) [0.0] (1,1) [0.055970149253731345] (1,2) [0.0] (1,3) [0.007462686567164179] (1,4) [0.0] (1,5) [0.0] (1,6) [0.05970149253731343] (1,7) [0.048507462686567165] (1,8) [0.582089552238806] (1,9) [0.2462686567164179] 

(2,0) [0.007017543859649123] (2,1) [0.0] (2,2) [0.0] (2,3) [0.0] (2,4) [0.0] (2,5) [0.0] (2,6) [0.0] (2,7) [0.0] (2,8) [0.13333333333333333] (2,9) [0.8596491228070176] 

(3,0) [0.0] (3,1) [0.0] (3,2) [0.0] (3,3) [0.0] (3,4) [0.0] (3,5) [0.10894941634241245] (3,6) [0.0] (3,7) [0.011673151750972763] (3,8) [0.8326848249027238] (3,9) [0.04669260700389105] 

(4,0) [0.23828125] (4,1) [0.0] (4,2) [0.0] (4,3) [0.0] (4,4) [0.24609375] (4,5) [0.015625] (4,6) [0.0] (4,7) [0.0] (4,8) [0.10546875] (4,9) [0.39453125] 

(5,0) [0.0] (5,1) [0.0] (5,2) [0.0] (5,3) [0.0] (5,4) [0.0] (5,5) [0.3111888111888112] (5,6) [0.0] (5,7) [0.0] (5,8) [0.4160839160839161] (5,9) [0.2727272727272727] 

(6,0) [0.0] (6,1) [0.0] (6,2) [0.0] (6,3) [0.0] (6,4) [0.5741444866920152] (6,5) [0.0] (6,6) [0.0] (6,7) [0.0] (6,8) [0.08365019011406843] (6,9) [0.34220532319391633] 

(7,0) [0.010869565217391304] (7,1) [0.0] (7,2) [0.0] (7,3) [0.0] (7,4) [0.0] (7,5) [0.23550724637681159] (7,6) [0.0] (7,7) [0.0] (7,8) [0.010869565217391304] (7,9) [0.7427536231884058] 

(8,0) [0.0] (8,1) [0.0] (8,2) [0.0] (8,3) [0.0] (8,4) [0.0] (8,5) [0.0] (8,6) [0.0] (8,7) [0.006993006993006993] (8,8) [0.7342657342657343] (8,9) [0.25874125874125875] 

(9,0) [0.0] (9,1) [0.0] (9,2) [0.0] (9,3) [0.0] (9,4) [0.0] (9,5) [0.0] (9,6) [0.0] (9,7) [0.0] (9,8) [0.0] (9,9) [1.0]

};
\end{axis}
\end{tikzpicture}
    \setlength\abovecaptionskip{-.02cm}
    \caption{Train on beamformee 1 and testing on beamformee 2. Acc. 25.86\%}
    \label{fig:train_1_test_2}
    \end{subfigure}
    \hfill
    \begin{subfigure}[t]{0.48\columnwidth}
    \centering
    \setlength\fwidth{.68\columnwidth}
    \setlength\fheight{0.42\columnwidth}
    \setlength\abovecaptionskip{-.02cm}
    \begin{tikzpicture}
\pgfplotsset{every tick label/.append style={font=\tiny}}

\begin{axis}[
enlargelimits=false,
colorbar,
colormap/Purples,
width=\fwidth,
height=\fheight,
at={(0\fwidth,0\fheight)},
scale only axis,
tick align=inside,
xlabel={Predicted ID},
xmin=-0.5,
xmax=9.5,
xtick style={draw=none},
xlabel style={font=\scriptsize\color{white!15!black}},
ylabel style={font=\scriptsize\color{white!15!black}},
ylabel={Actual ID},
ymin=-0.5,
ymax=9.5,
xlabel shift=-5pt,
ylabel shift=-5pt,
ytick style={draw=none},
axis background/.style={fill=white},
colorbar horizontal,
colorbar style={
at={(0,1.05)},               % <-- (changed)
anchor=below south west,    % <-- (changed)
% change the width of the colorbar relative to the main `axis' environment
width=\pgfkeysvalueof{/pgfplots/parent axis width},
xtick={0, 0.5, 1},
xmin=0,
xmax=1,
axis x line*=top,
xticklabel shift=-1pt,
point meta min=0,
point meta max=1.05,
},
colorbar/width=2mm,
]
\addplot [matrix plot,point meta=explicit]
 coordinates {
(0,0) [0.29900332225913623] (0,1) [0.12956810631229235] (0,2) [0.09966777408637874] (0,3) [0.0] (0,4) [0.0664451827242525] (0,5) [0.08305647840531562] (0,6) [0.3222591362126246] (0,7) [0.0] (0,8) [0.0] (0,9) [0.0] 

(1,0) [0.011278195488721804] (1,1) [0.13909774436090225] (1,2) [0.17293233082706766] (1,3) [0.4548872180451128] (1,4) [0.09398496240601503] (1,5) [0.0] (1,6) [0.12781954887218044] (1,7) [0.0] (1,8) [0.0] (1,9) [0.0] 

(2,0) [0.0] (2,1) [0.0] (2,2) [0.11029411764705882] (2,3) [0.003676470588235294] (2,4) [0.0] (2,5) [0.0] (2,6) [0.8345588235294118] (2,7) [0.0] (2,8) [0.051470588235294115] (2,9) [0.0] 

(3,0) [0.0] (3,1) [0.0076045627376425855] (3,2) [0.0] (3,3) [0.9923954372623575] (3,4) [0.0] (3,5) [0.0] (3,6) [0.0] (3,7) [0.0] (3,8) [0.0] (3,9) [0.0] 

(4,0) [0.0] (4,1) [0.007722007722007722] (4,2) [0.11196911196911197] (4,3) [0.30115830115830117] (4,4) [0.10810810810810811] (4,5) [0.0] (4,6) [0.47104247104247104] (4,7) [0.0] (4,8) [0.0] (4,9) [0.0] 

(5,0) [0.1387900355871886] (5,1) [0.20640569395017794] (5,2) [0.0] (5,3) [0.35587188612099646] (5,4) [0.12811387900355872] (5,5) [0.14590747330960854] (5,6) [0.0] (5,7) [0.02491103202846975] (5,8) [0.0] (5,9) [0.0] 

(6,0) [0.0] (6,1) [0.0] (6,2) [0.0] (6,3) [0.3745173745173745] (6,4) [0.0] (6,5) [0.0] (6,6) [0.5328185328185329] (6,7) [0.0] (6,8) [0.09266409266409266] (6,9) [0.0] 

(7,0) [0.0] (7,1) [0.043478260869565216] (7,2) [0.3188405797101449] (7,3) [0.30434782608695654] (7,4) [0.010869565217391304] (7,5) [0.0] (7,6) [0.31521739130434784] (7,7) [0.0] (7,8) [0.007246376811594203] (7,9) [0.0] 

(8,0) [0.0] (8,1) [0.02422145328719723] (8,2) [0.01730103806228374] (8,3) [0.7301038062283737] (8,4) [0.0] (8,5) [0.0] (8,6) [0.031141868512110725] (8,7) [0.0] (8,8) [0.1972318339100346] (8,9) [0.0] 

(9,0) [0.0] (9,1) [0.0038461538461538464] (9,2) [0.46923076923076923] (9,3) [0.3192307692307692] (9,4) [0.0] (9,5) [0.0] (9,6) [0.2076923076923077] (9,7) [0.0] (9,8) [0.0] (9,9) [0.0]

};
\end{axis}
\end{tikzpicture}
    \caption{Train on beamformee 2 and testing on beamformee 1. Acc. 25.02\%}
    \label{fig:train_2_test_1}
    \end{subfigure}
    \setlength\abovecaptionskip{0.1cm}
    \setlength\belowcaptionskip{-.3cm}
    \caption{Confusion matrices for set \texttt{S1}, training on one beamformee and testing on the other,  3 TX antennas, spatial stream 0.}
    \label{fig:train_one_test_another}
\end{figure}

\begin{figure}[t]
    \centering
    \begin{subfigure}[t]{0.48\columnwidth}
    \centering
    \includegraphics[width=0.86\columnwidth]{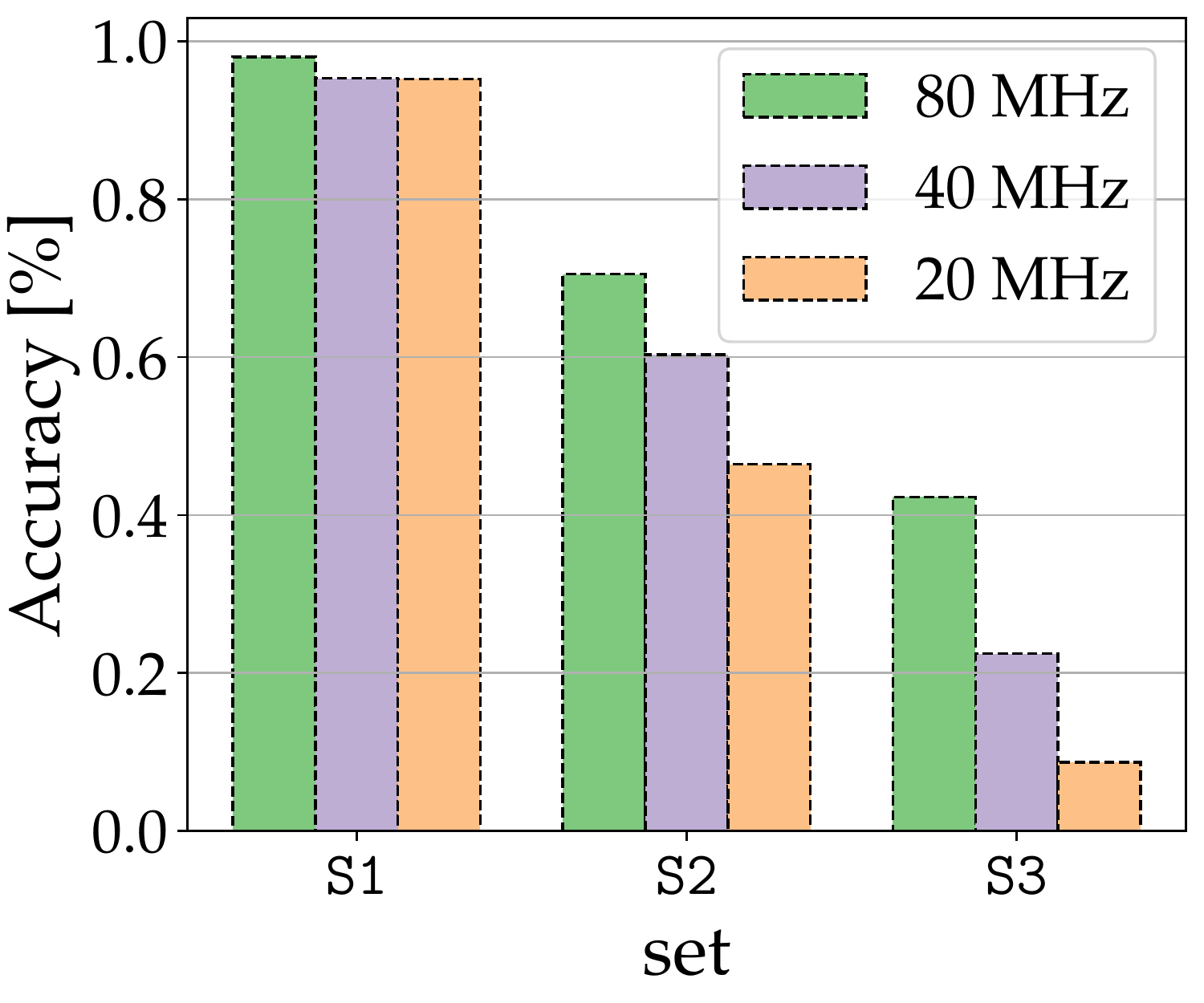}
    \setlength\abovecaptionskip{-0.02cm}
    \caption{Accuracy by varying the channel bandwidth, i.e., selecting respectively $N_{\rm col}=$~234, 110, 54, out of the $K=$~234 \gls{ofdm} sub-channels.}
    \label{fig:comparison_bandwidth}
    \end{subfigure}
    \hfill
    \begin{subfigure}[t]{0.48\columnwidth}
    \centering
    \includegraphics[width=0.86\columnwidth]{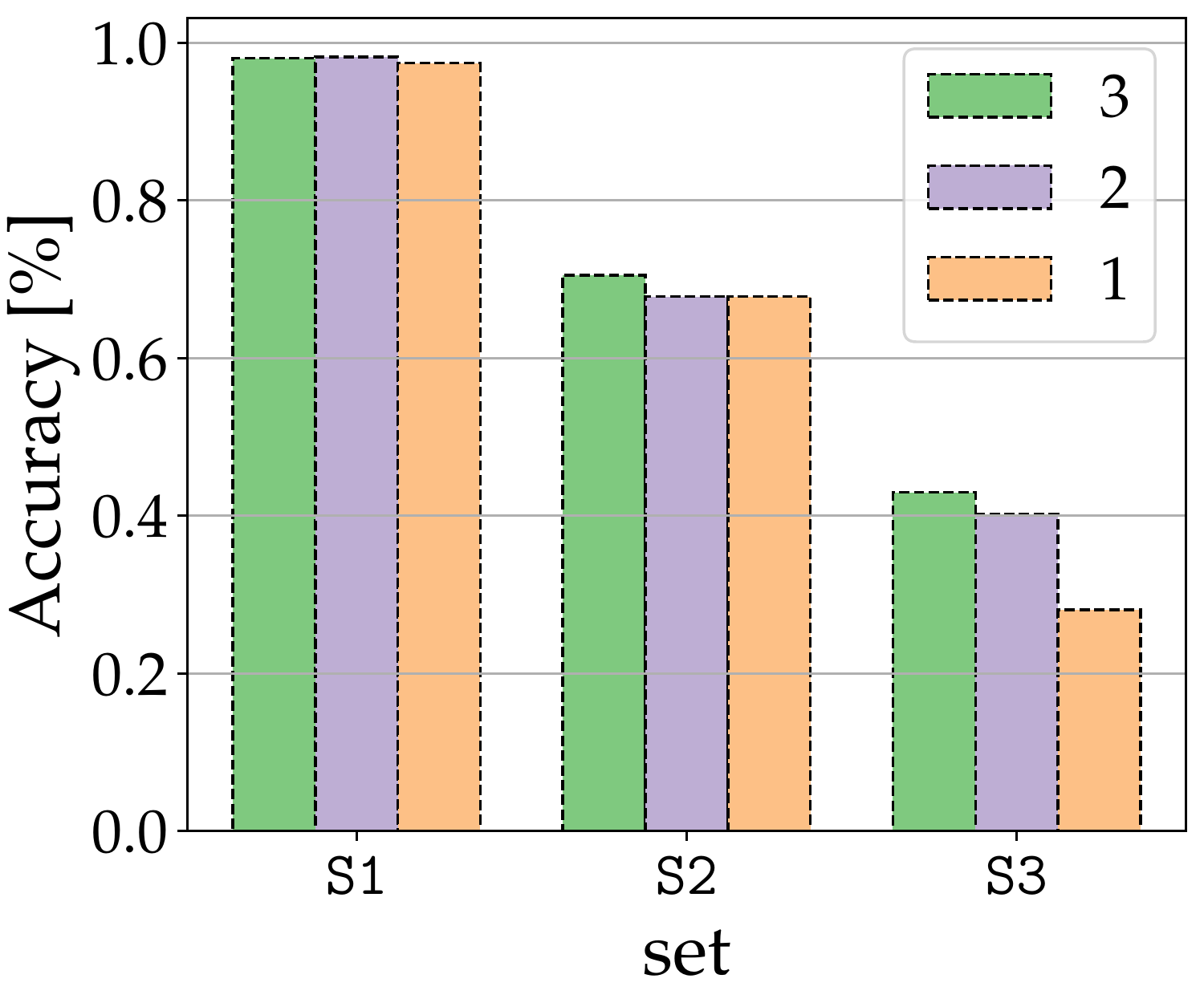}
    \setlength\abovecaptionskip{-0.02cm}
    \caption{Accuracy by varying the no. of transmitter antennas, i.e., selecting respectively $N_{\rm ch}=$~3, 2, 1 rows of the beamforming feedback matrix $\mathbf{\Tilde{V}}$.}
    \label{fig:comparison_tx_ant}
    \end{subfigure}
    \setlength\belowcaptionskip{-.6cm}
    \caption{\FW accuracy by varying the channel bandwidth and the number of transmitter antennas, using spatial stream 0.}
    \label{fig:comparisons}
\end{figure}

\smallskip
\noindent\textbf{\FW performance when varying the beamformer transmission parameters.}
In \figurename~\ref{fig:comparison_bandwidth} we compare the accuracy of \FW when considering different portions of the radio spectrum. 
According to the IEEE~802.11ac \gls{ofdm} channels specifications~\cite{IEEE-80211ac}, from the 234 sub-channels on an 80~MHz channel, we can group sub-channels belonging to two 40~MHz and four 20~MHz channels. Therefore, from the data collected on the IEEE channel 42 at 80~MHz, we extracted 110 sub-channels for the 40~MHz channel 38 and 54 sub-channels for the 20~MHz channel 36, and assessed the performance of \FW on these subsets. These results prove that the accuracy increases with a larger bandwidth, especially when considering the most challenging configurations \texttt{S2} and \texttt{S3}. \figurename~\ref{fig:comparison_tx_ant} evaluates the impact of increasing from 1 to 3 the number of transmitter antennas used to compute the fingerprint. Note that the accuracy mainly depends on the number of selected antennas and only weakly depends on their IDs. Thus, we only show results for a single selection pattern out of the possible ones for each number of antennas. The \FW performance remains almost constant on set \texttt{S1}, while the accuracy increases on \texttt{S2} and \texttt{S3} going from 1 to 3 transmitter antennas. These results confirm that exploiting to the maximum extent the spatial diversity at the beamformer -- by considering all the \gls{ofdm} sub-channels and transmitter antennas -- is key to designing robust RFP algorithms.

\begin{figure}[t]
    \centering
    \begin{subfigure}[t]{0.48\columnwidth}
    \centering
    \includegraphics[width=0.95\columnwidth]{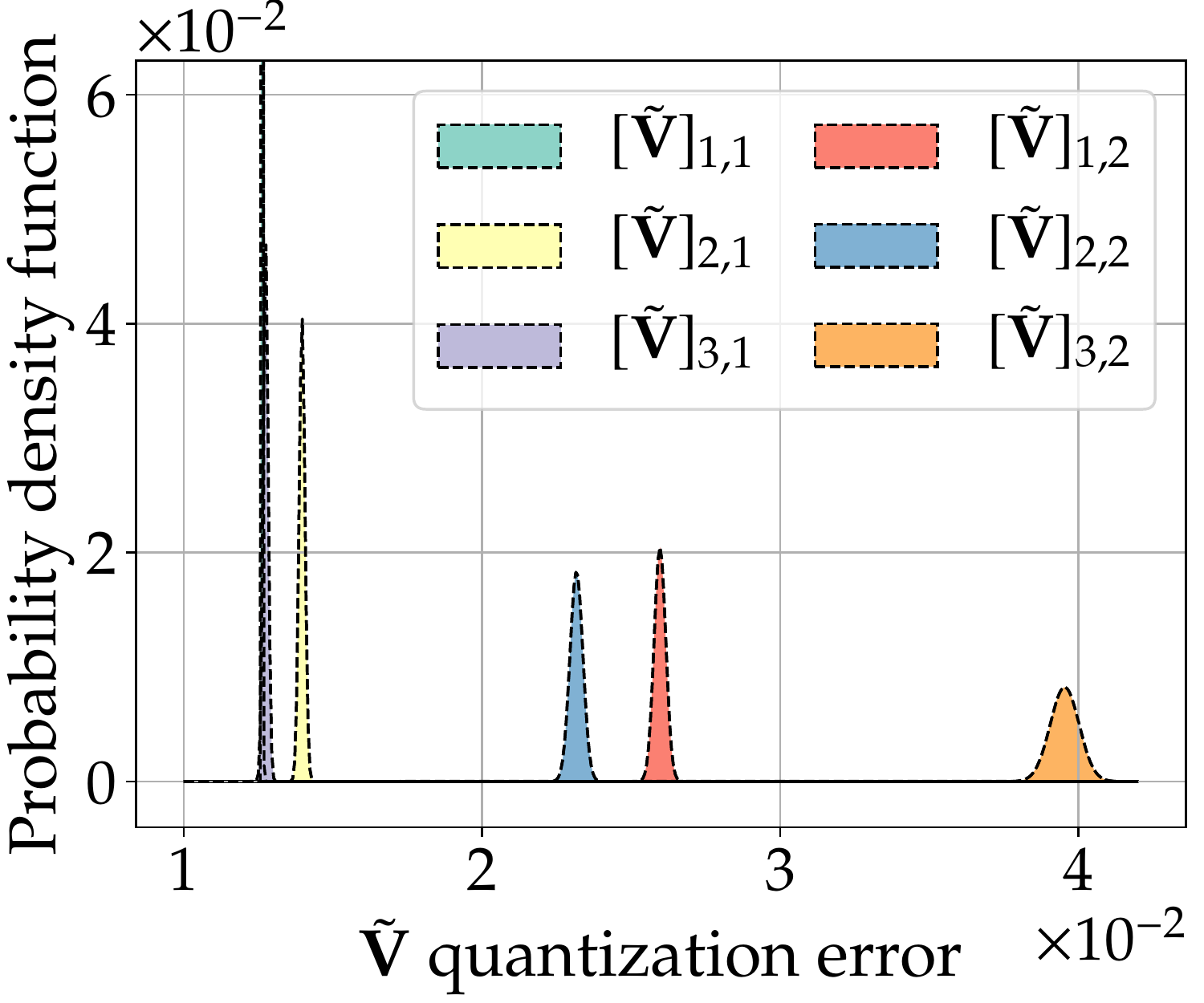}
    \caption{$b_{\psi}=5$ and $b_{\phi}=7$ bit.}
    \label{fig:pdf_error_psi5}
    \end{subfigure}
    \hfill
    \begin{subfigure}[t]{0.48\columnwidth}
    \centering
    \includegraphics[width=0.95\columnwidth]{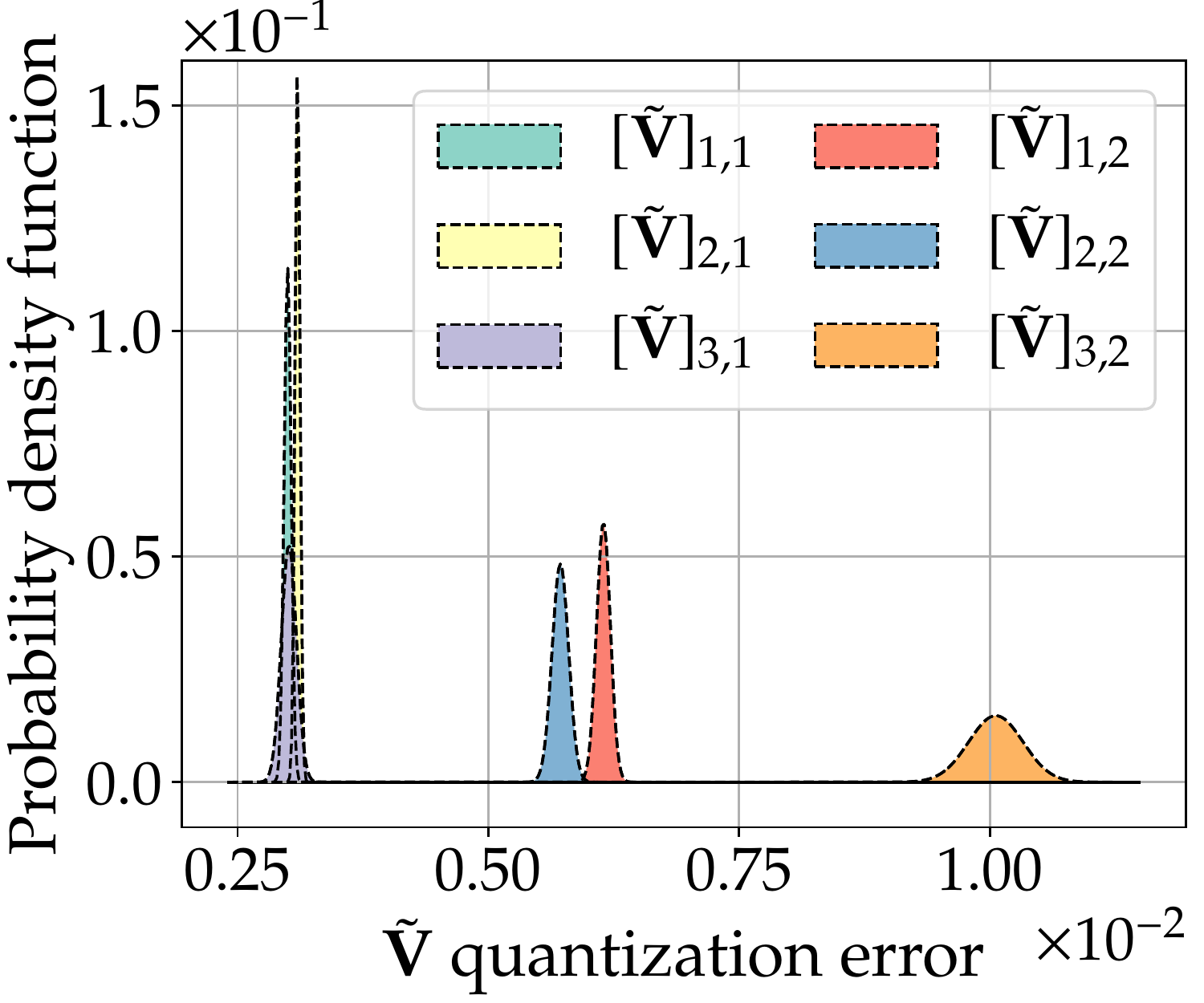}
    \caption{$b_{\psi}=7$ and $b_{\phi}=9$ bit.}
    \label{fig:pdf_error_psi7}
    \end{subfigure}
    \setlength\belowcaptionskip{-.4cm}
    \caption{PDF of the $\mathbf{\Tilde{V}}$ quantization error using the two standard-compliant sets of values for the beamforming feedback angles quantization bits.}
    \label{fig:pdf_error}
\end{figure}

\begin{figure}[t]
    \centering
    \begin{subfigure}[t]{0.32\columnwidth}
    \centering
    \setlength\fwidth{.75\columnwidth}
    \setlength\fheight{0.5\columnwidth}
    \begin{tikzpicture}
\pgfplotsset{every tick label/.append style={font=\tiny}}

\begin{axis}[
enlargelimits=false,
colorbar,
colormap/Purples,
width=\fwidth,
height=\fheight,
at={(0\fwidth,0\fheight)},
scale only axis,
tick align=inside,
xlabel={Predicted ID},
xmin=-0.5,
xmax=9.5,
xtick style={draw=none},
xlabel style={font=\scriptsize\color{white!15!black}},
ylabel style={font=\scriptsize\color{white!15!black}},
ylabel={Actual ID},
ymin=-0.5,
ymax=9.5,
xlabel shift=-5pt,
ylabel shift=-5pt,
ytick style={draw=none},
axis background/.style={fill=white},
colorbar horizontal,
colorbar style={
at={(0,1.05)},               % <-- (changed)
anchor=below south west,    % <-- (changed)
% change the width of the colorbar relative to the main `axis' environment
width=\pgfkeysvalueof{/pgfplots/parent axis width},
xtick={0, 0.5, 1},
xmin=0,
xmax=1,
axis x line*=top,
xticklabel shift=-1pt,
point meta min=0,
point meta max=1.05,
},
colorbar/width=2mm,
]
\addplot [matrix plot,point meta=explicit]
 coordinates {
(0,0) [0.9302325581395349] (0,1) [0.0] (0,2) [0.0] (0,3) [0.06976744186046512] (0,4) [0.0] (0,5) [0.0] (0,6) [0.0] (0,7) [0.0] (0,8) [0.0] (0,9) [0.0] 

(1,0) [0.0] (1,1) [0.9849624060150376] (1,2) [0.0] (1,3) [0.011278195488721804] (1,4) [0.0] (1,5) [0.0] (1,6) [0.0] (1,7) [0.0037593984962406013] (1,8) [0.0] (1,9) [0.0] 

(2,0) [0.014705882352941176] (2,1) [0.0] (2,2) [0.9852941176470589] (2,3) [0.0] (2,4) [0.0] (2,5) [0.0] (2,6) [0.0] (2,7) [0.0] (2,8) [0.0] (2,9) [0.0] 

(3,0) [0.0] (3,1) [0.011406844106463879] (3,2) [0.0] (3,3) [0.9885931558935361] (3,4) [0.0] (3,5) [0.0] (3,6) [0.0] (3,7) [0.0] (3,8) [0.0] (3,9) [0.0] 

(4,0) [0.0] (4,1) [0.0] (4,2) [0.0] (4,3) [0.0] (4,4) [1.0] (4,5) [0.0] (4,6) [0.0] (4,7) [0.0] (4,8) [0.0] (4,9) [0.0] 

(5,0) [0.0] (5,1) [0.0] (5,2) [0.0] (5,3) [0.0] (5,4) [0.0] (5,5) [1.0] (5,6) [0.0] (5,7) [0.0] (5,8) [0.0] (5,9) [0.0] 

(6,0) [0.0] (6,1) [0.0] (6,2) [0.0] (6,3) [0.0] (6,4) [0.0] (6,5) [0.0] (6,6) [1.0] (6,7) [0.0] (6,8) [0.0] (6,9) [0.0] 

(7,0) [0.0036231884057971015] (7,1) [0.007246376811594203] (7,2) [0.0] (7,3) [0.0] (7,4) [0.0] (7,5) [0.0] (7,6) [0.0036231884057971015] (7,7) [0.927536231884058] (7,8) [0.0] (7,9) [0.057971014492753624] 

(8,0) [0.0] (8,1) [0.0] (8,2) [0.0] (8,3) [0.0] (8,4) [0.0] (8,5) [0.0] (8,6) [0.09342560553633218] (8,7) [0.006920415224913495] (8,8) [0.8996539792387543] (8,9) [0.0] 

(9,0) [0.0] (9,1) [0.0] (9,2) [0.0] (9,3) [0.0] (9,4) [0.0] (9,5) [0.0] (9,6) [0.0] (9,7) [0.0] (9,8) [0.0] (9,9) [1.0]

};
\end{axis}
\end{tikzpicture}
    \setlength\abovecaptionskip{-.4cm}
    \caption{\texttt{S1}. Acc. 97.03\%}
    \label{fig:train_test_s1_1_ss_1}
    \end{subfigure}
    \hfill
    \begin{subfigure}[t]{0.32\columnwidth}
    \centering
    \setlength\fwidth{.75\columnwidth}
    \setlength\fheight{0.5\columnwidth}
    \begin{tikzpicture}
\pgfplotsset{every tick label/.append style={font=\tiny}}

\begin{axis}[
enlargelimits=false,
colorbar,
colormap/Purples,
width=\fwidth,
height=\fheight,
at={(0\fwidth,0\fheight)},
scale only axis,
tick align=inside,
xlabel={Predicted ID},
xmin=-0.5,
xmax=9.5,
xtick style={draw=none},
xlabel style={font=\scriptsize\color{white!15!black}},
ylabel style={font=\scriptsize\color{white!15!black}},
ylabel={Actual ID},
ymin=-0.5,
ymax=9.5,
xlabel shift=-5pt,
ylabel shift=-5pt,
ytick style={draw=none},
axis background/.style={fill=white},
colorbar horizontal,
colorbar style={
at={(0,1.05)},               % <-- (changed)
anchor=below south west,    % <-- (changed)
% change the width of the colorbar relative to the main `axis' environment
width=\pgfkeysvalueof{/pgfplots/parent axis width},
xtick={0, 0.5, 1},
xmin=0,
xmax=1,
axis x line*=top,
xticklabel shift=-1pt,
point meta min=0,
point meta max=1.05,
},
colorbar/width=2mm,
]
\addplot [matrix plot,point meta=explicit]
 coordinates {
(0,0) [0.24361158432708688] (0,1) [0.42248722316865417] (0,2) [0.018739352640545145] (0,3) [0.030664395229982964] (0,4) [0.23679727427597955] (0,5) [0.0] (0,6) [0.0034071550255536627] (0,7) [0.022146507666098807] (0,8) [0.0068143100511073255] (0,9) [0.015332197614991482] 

(1,0) [0.024844720496894408] (1,1) [0.38198757763975155] (1,2) [0.17701863354037267] (1,3) [0.0] (1,4) [0.055900621118012424] (1,5) [0.03260869565217391] (1,6) [0.003105590062111801] (1,7) [0.2857142857142857] (1,8) [0.0] (1,9) [0.03881987577639751] 

(2,0) [0.05623100303951368] (2,1) [0.02127659574468085] (2,2) [0.3677811550151976] (2,3) [0.425531914893617] (2,4) [0.00303951367781155] (2,5) [0.0] (2,6) [0.00303951367781155] (2,7) [0.00303951367781155] (2,8) [0.004559270516717325] (2,9) [0.11550151975683891] 

(3,0) [0.024263431542461005] (3,1) [0.0] (3,2) [0.2322357019064125] (3,3) [0.14038128249566725] (3,4) [0.1923743500866551] (3,5) [0.0] (3,6) [0.0017331022530329288] (3,7) [0.10051993067590988] (3,8) [0.30675909878682844] (3,9) [0.0017331022530329288] 

(4,0) [0.17169811320754716] (4,1) [0.3886792452830189] (4,2) [0.14528301886792452] (4,3) [0.045283018867924525] (4,4) [0.05849056603773585] (4,5) [0.0] (4,6) [0.0] (4,7) [0.07169811320754717] (4,8) [0.10943396226415095] (4,9) [0.009433962264150943] 

(5,0) [0.017001545595054096] (5,1) [0.7063369397217929] (5,2) [0.0] (5,3) [0.2689335394126739] (5,4) [0.00463678516228748] (5,5) [0.0] (5,6) [0.0] (5,7) [0.0030911901081916537] (5,8) [0.0] (5,9) [0.0] 

(6,0) [0.1751227495908347] (6,1) [0.03273322422258593] (6,2) [0.1260229132569558] (6,3) [0.265139116202946] (6,4) [0.07855973813420622] (6,5) [0.0] (6,6) [0.0] (6,7) [0.27495908346972175] (6,8) [0.019639934533551555] (6,9) [0.027823240589198037] 

(7,0) [0.029465930018416207] (7,1) [0.5580110497237569] (7,2) [0.053406998158379376] (7,3) [0.0] (7,4) [0.001841620626151013] (7,5) [0.0] (7,6) [0.289134438305709] (7,7) [0.009208103130755065] (7,8) [0.04604051565377532] (7,9) [0.01289134438305709] 

(8,0) [0.009216589861751152] (8,1) [0.46697388632872505] (8,2) [0.07987711213517665] (8,3) [0.10599078341013825] (8,4) [0.006144393241167435] (8,5) [0.0] (8,6) [0.030721966205837174] (8,7) [0.19201228878648233] (8,8) [0.0] (8,9) [0.10906298003072197] 

(9,0) [0.13774597495527727] (9,1) [0.4293381037567084] (9,2) [0.008944543828264758] (9,3) [0.05008944543828265] (9,4) [0.0] (9,5) [0.0] (9,6) [0.005366726296958855] (9,7) [0.007155635062611807] (9,8) [0.26833631484794274] (9,9) [0.09302325581395349]

};
\end{axis}
\end{tikzpicture}
    \setlength\abovecaptionskip{-.4cm}
    \caption{\texttt{S2}. Acc. 13.32\%}
    \label{fig:train_test_s2_1_ss_1}
    \end{subfigure}
    \hfill
    \begin{subfigure}[t]{0.32\columnwidth}
    \centering
    \setlength\fwidth{.75\columnwidth}
    \setlength\fheight{0.5\columnwidth}
    \begin{tikzpicture}
\pgfplotsset{every tick label/.append style={font=\tiny}}

\begin{axis}[
enlargelimits=false,
colorbar,
colormap/Purples,
width=\fwidth,
height=\fheight,
at={(0\fwidth,0\fheight)},
scale only axis,
tick align=inside,
xlabel={Predicted ID},
xmin=-0.5,
xmax=9.5,
xtick style={draw=none},
xlabel style={font=\scriptsize\color{white!15!black}},
ylabel style={font=\scriptsize\color{white!15!black}},
ylabel={Actual ID},
ymin=-0.5,
ymax=9.5,
xlabel shift=-5pt,
ylabel shift=-5pt,
ytick style={draw=none},
axis background/.style={fill=white},
colorbar horizontal,
colorbar style={
at={(0,1.05)},               % <-- (changed)
anchor=below south west,    % <-- (changed)
% change the width of the colorbar relative to the main `axis' environment
width=\pgfkeysvalueof{/pgfplots/parent axis width},
xtick={0, 0.5, 1},
xmin=0,
xmax=1,
axis x line*=top,
xticklabel shift=-1pt,
point meta min=0,
point meta max=1.05,
},
colorbar/width=2mm,
]
\addplot [matrix plot,point meta=explicit]
 coordinates {
(0,0) [0.0047169811320754715] (0,1) [0.009433962264150943] (0,2) [0.0031446540880503146] (0,3) [0.0] (0,4) [0.0220125786163522] (0,5) [0.4858490566037736] (0,6) [0.0] (0,7) [0.27358490566037735] (0,8) [0.20125786163522014] (0,9) [0.0] 

(1,0) [0.08226691042047532] (1,1) [0.16270566727605118] (1,2) [0.3619744058500914] (1,3) [0.054844606946983544] (1,4) [0.0] (1,5) [0.05850091407678245] (1,6) [0.009140767824497258] (1,7) [0.21206581352833637] (1,8) [0.031078610603290677] (1,9) [0.027422303473491772] 

(2,0) [0.0] (2,1) [0.05103969754253308] (2,2) [0.03969754253308128] (2,3) [0.001890359168241966] (2,4) [0.0] (2,5) [0.13799621928166353] (2,6) [0.22873345935727787] (2,7) [0.003780718336483932] (2,8) [0.21739130434782608] (2,9) [0.31947069943289225] 

(3,0) [0.0] (3,1) [0.184070796460177] (3,2) [0.1929203539823009] (3,3) [0.0] (3,4) [0.0] (3,5) [0.02831858407079646] (3,6) [0.0] (3,7) [0.007079646017699115] (3,8) [0.5575221238938053] (3,9) [0.03008849557522124] 

(4,0) [0.003367003367003367] (4,1) [0.003367003367003367] (4,2) [0.12457912457912458] (4,3) [0.0] (4,4) [0.2222222222222222] (4,5) [0.03198653198653199] (4,6) [0.0] (4,7) [0.2760942760942761] (4,8) [0.2845117845117845] (4,9) [0.05387205387205387] 

(5,0) [0.03622047244094488] (5,1) [0.1732283464566929] (5,2) [0.0031496062992125984] (5,3) [0.07086614173228346] (5,4) [0.2551181102362205] (5,5) [0.048818897637795275] (5,6) [0.0] (5,7) [0.1952755905511811] (5,8) [0.2173228346456693] (5,9) [0.0] 

(6,0) [0.0] (6,1) [0.1574539363484087] (6,2) [0.10720268006700168] (6,3) [0.010050251256281407] (6,4) [0.0] (6,5) [0.3969849246231156] (6,6) [0.008375209380234505] (6,7) [0.2747068676716918] (6,8) [0.03350083752093802] (6,9) [0.011725293132328308] 

(7,0) [0.004746835443037975] (7,1) [0.25949367088607594] (7,2) [0.0] (7,3) [0.2674050632911392] (7,4) [0.011075949367088608] (7,5) [0.32436708860759494] (7,6) [0.011075949367088608] (7,7) [0.06645569620253164] (7,8) [0.030063291139240507] (7,9) [0.02531645569620253] 

(8,0) [0.0] (8,1) [0.296474358974359] (8,2) [0.23076923076923078] (8,3) [0.01282051282051282] (8,4) [0.0] (8,5) [0.296474358974359] (8,6) [0.0673076923076923] (8,7) [0.07692307692307693] (8,8) [0.016025641025641024] (8,9) [0.003205128205128205] 

(9,0) [0.00718132854578097] (9,1) [0.06463195691202872] (9,2) [0.14542190305206462] (9,3) [0.01436265709156194] (9,4) [0.18312387791741472] (9,5) [0.03770197486535009] (9,6) [0.00718132854578097] (9,7) [0.20287253141831238] (9,8) [0.3375224416517056] (9,9) [0.0]

};
\end{axis}
\end{tikzpicture}
    \setlength\abovecaptionskip{-.4cm}
    \caption{\texttt{S3}. Acc. 5.63\%}
    \label{fig:train_test_s3_1_ss_1}
    \end{subfigure}
    \setlength\belowcaptionskip{-.45cm}
    \caption{Confusion matrices, beamformee 1, 3 TX antennas, spatial stream 1.}
    \label{fig:train_test_s1_s2_s3_ss1}
\end{figure}
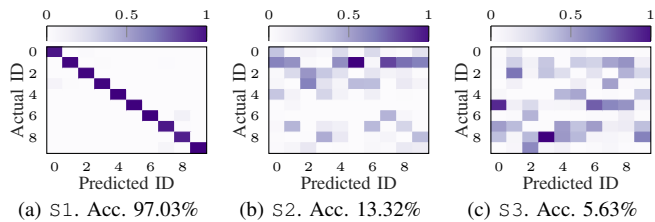

\smallskip
\noindent\textbf{\FW performance when changing the reference beamformee spatial stream.}
To evaluate the effect of changing the \gls{dnn} input spatial stream on the beamformer fingerprinting accuracy, we consider the impact of the beamforming feedback angles quantization on the columns of $\mathbf{\Tilde{V}}$, representing the spatial streams dimensions. From Algorithm~\ref{alg:beamf_feedback}, it follows that the impact of the quantization error increases going from the first to the last reconstructed stream. We verified this fact by simulating an \gls{ofdm} \gls{mum} channel, considering the ray tracing model of~\cite{TGac}. We obtained the channel matrix $\mathbf{H}$ for 100,000 transmissions in \gls{mum} mode, and we derived $\mathbf{\Tilde{V}}$ via \gls{svd}. Hence, we computed the $q_{\phi}$ and $q_{\psi}$ quantized angles following Algorithm~\ref{alg:beamf_feedback} and using the quantization parameters defined in the standards~\cite{IEEE-80211ac, IEEE-80211ax}. These operations are the same performed by the beamformees to generate the feedback. Next, we reconstructed $\mathbf{\Tilde{V}}$ from the quantized angles and evaluated the reconstruction error on each combination of transmitter antennas and spatial streams. 
We plot the probability density functions (PDFs) of the quantization error using ($b_{\psi}=$~5, $b_{\phi}=$~7) and ($b_{\psi}=$~7, $b_{\psi}=$~9) bits for quantization in Figs.~\ref{fig:pdf_error_psi5} and \ref{fig:pdf_error_psi7}. 
We notice the reconstruction of the second column of $\mathbf{\Tilde{V}}$, i.e., the second stream, is less accurate than the reconstruction of the first, for all the three transmitter antennas. This is intrinsically related to the construction of the $\mathbf{D}_{k, i}$ and $\mathbf{G}_{k, \ell, i}$ matrices from the quantized angles, and to their combination for the computation of $\mathbf{\Tilde{V}}$ (see \eq{eq:v_matrix}). Indeed, the algorithm has a recursive structure by which the quantization error on the first stream propagates to the next ones, leading to worse approximations for the higher order columns of matrix $\mathbf{\Tilde{V}}$. The quantization error can also be visualized from our empirical measurements. 
\begin{figure}[t]
    \centering
    \includegraphics[width=0.85\columnwidth]{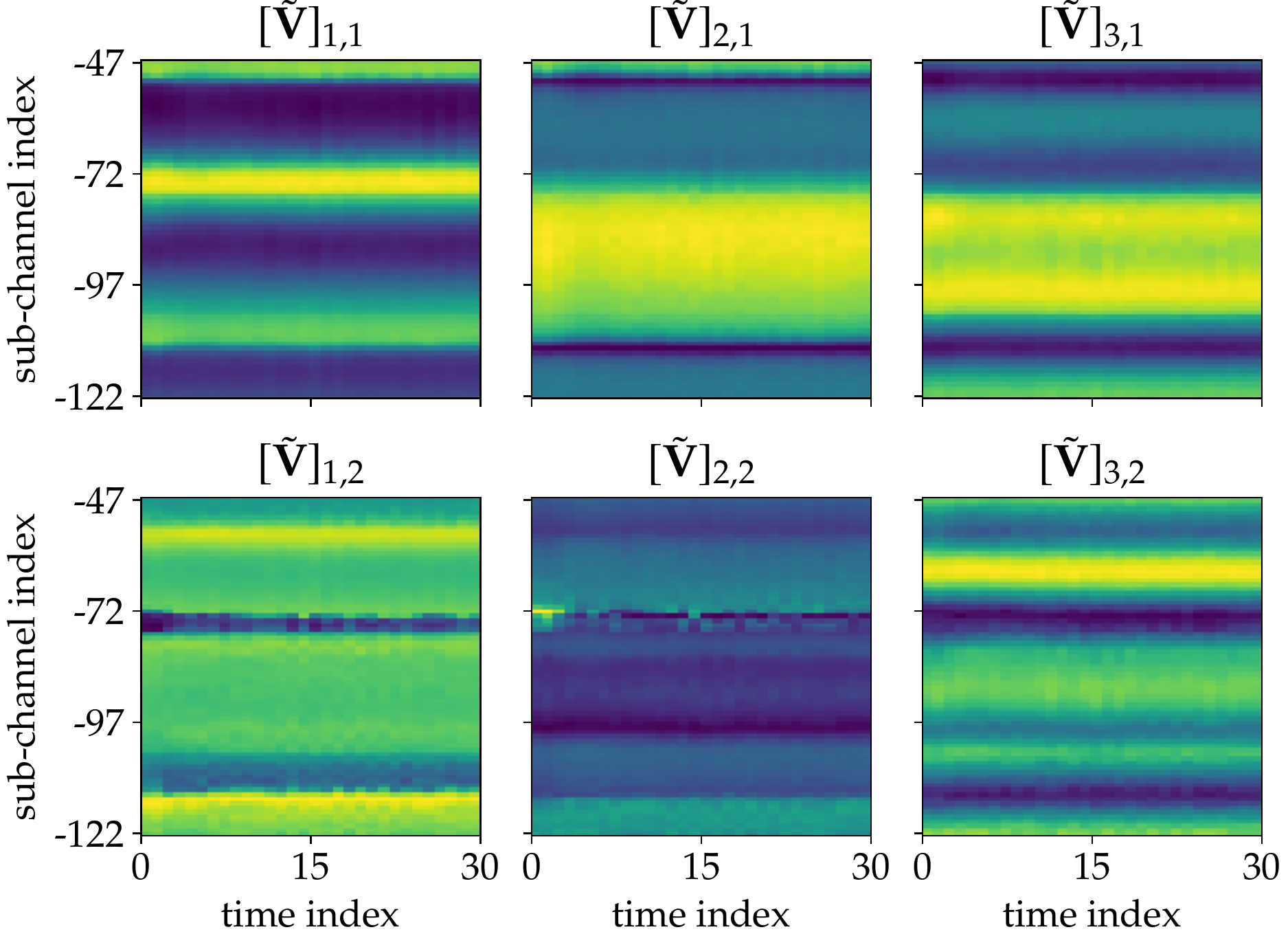}
    \setlength\abovecaptionskip{-0.04cm}
    \setlength\belowcaptionskip{-.5cm}
    \caption{Time evolution of $\mathbf{\Tilde{V}}$ for the first 75 \gls{ofdm} sub-channels, in static conditions. The columns refer to the transmit antennas while the rows to the spatial streams.}
    \label{fig:heatmap_v}
\end{figure}
In \figurename~\ref{fig:heatmap_v}, we plot an excerpt of the $\mathbf{\Tilde{V}}$ matrix reconstructed by \FW from the quantized angles obtained at the beamformee side in static conditions. The quantization error is clearly visible for the second spatial stream (column 2 of matrix $\mathbf{\Tilde{V}}$). Thus, the performance of \FW decreases when considering as \gls{dnn} input the data associated with the second spatial stream (\figurename~\ref{fig:train_test_s1_s2_s3_ss1}) instead of the first one (\figurename~\ref{fig:train_test_s1_s2_s3_beamformee1}). While on set \texttt{S1} the beamformer can still be identified with high accuracy using data from the second spatial stream, when considering sets \texttt{S2} and \texttt{S3} -- thus reducing the number of training positions -- the beamformer fingerprint can no longer be effectively extracted, leading to a considerable drop in the classification accuracy.

\begin{figure}[t]
    \centering
    \begin{subfigure}[t]{0.48\columnwidth}
    \centering
    \includegraphics[width=0.85\columnwidth]{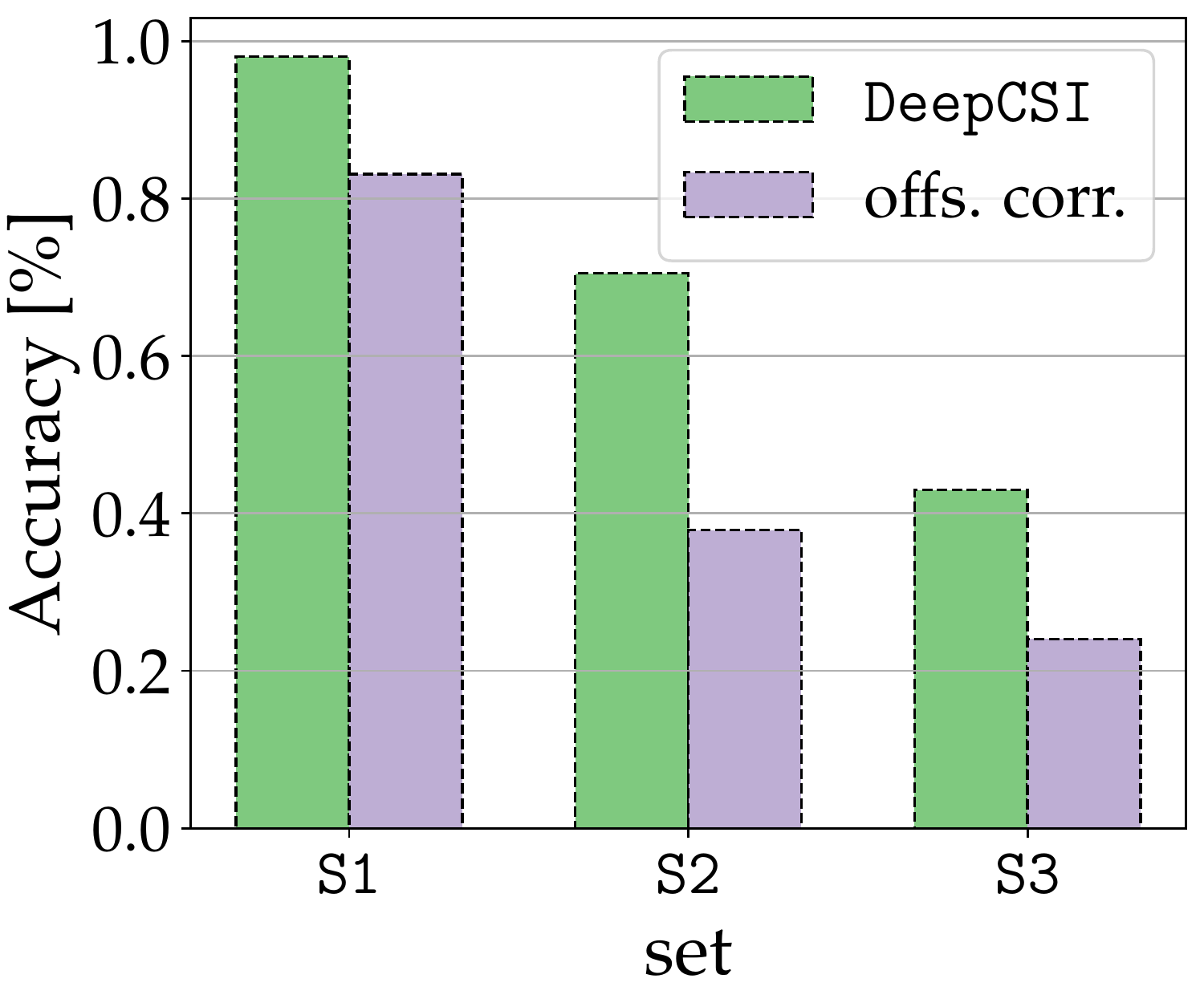}
    \setlength\abovecaptionskip{-0.02cm}
    \caption{Accuracy compared with the one obtained using the processed input.}
    \label{fig:comparison_phase_correction}
    \end{subfigure}
    \hfill
    \begin{subfigure}[t]{0.48\columnwidth}
    \centering
    \setlength\fwidth{.68\columnwidth}
    \setlength\fheight{.42\columnwidth}
    \begin{tikzpicture}
\pgfplotsset{every tick label/.append style={font=\tiny}}

\begin{axis}[
enlargelimits=false,
colorbar,
colormap/Purples,
width=\fwidth,
height=\fheight,
at={(0\fwidth,0\fheight)},
scale only axis,
tick align=inside,
xlabel={Predicted ID},
xmin=-0.5,
xmax=9.5,
xtick style={draw=none},
xlabel style={font=\scriptsize\color{white!15!black}},
ylabel style={font=\scriptsize\color{white!15!black}},
ylabel={Actual ID},
ymin=-0.5,
ymax=9.5,
xlabel shift=-5pt,
ylabel shift=-5pt,
ytick style={draw=none},
axis background/.style={fill=white},
colorbar horizontal,
colorbar style={
at={(0,1.05)},               % <-- (changed)
anchor=below south west,    % <-- (changed)
% change the width of the colorbar relative to the main `axis' environment
width=\pgfkeysvalueof{/pgfplots/parent axis width},
xtick={0, 0.5, 1},
xmin=0,
xmax=1,
axis x line*=top,
xticklabel shift=-1pt,
point meta min=0,
point meta max=1.1,
},
colorbar/width=2mm,
]
\addplot [matrix plot,point meta=explicit]
 coordinates {
(0,0) [0.8981481481481481] (0,1) [0.0030864197530864196] (0,2) [0.0030864197530864196] (0,3) [0.0] (0,4) [0.0] (0,5) [0.08024691358024691] (0,6) [0.0030864197530864196] (0,7) [0.0030864197530864196] (0,8) [0.0] (0,9) [0.009259259259259259] 

(1,0) [0.03571428571428571] (1,1) [0.8214285714285714] (1,2) [0.0] (1,3) [0.03571428571428571] (1,4) [0.017857142857142856] (1,5) [0.03869047619047619] (1,6) [0.0] (1,7) [0.03869047619047619] (1,8) [0.005952380952380952] (1,9) [0.005952380952380952] 

(2,0) [0.0] (2,1) [0.0] (2,2) [0.9644012944983819] (2,3) [0.0] (2,4) [0.003236245954692557] (2,5) [0.0] (2,6) [0.02912621359223301] (2,7) [0.0] (2,8) [0.0] (2,9) [0.003236245954692557] 

(3,0) [0.07861635220125786] (3,1) [0.0440251572327044] (3,2) [0.0] (3,3) [0.6163522012578616] (3,4) [0.0440251572327044] (3,5) [0.08490566037735849] (3,6) [0.025157232704402517] (3,7) [0.07547169811320754] (3,8) [0.02830188679245283] (3,9) [0.0031446540880503146] 

(4,0) [0.03536977491961415] (4,1) [0.0] (4,2) [0.0] (4,3) [0.0] (4,4) [0.9131832797427653] (4,5) [0.006430868167202572] (4,6) [0.003215434083601286] (4,7) [0.01929260450160772] (4,8) [0.0] (4,9) [0.022508038585209004] 

(5,0) [0.038011695906432746] (5,1) [0.05555555555555555] (5,2) [0.0] (5,3) [0.017543859649122806] (5,4) [0.005847953216374269] (5,5) [0.8362573099415205] (5,6) [0.0] (5,7) [0.011695906432748537] (5,8) [0.0] (5,9) [0.03508771929824561] 

(6,0) [0.0] (6,1) [0.1390728476821192] (6,2) [0.0033112582781456954] (6,3) [0.0] (6,4) [0.023178807947019868] (6,5) [0.009933774834437087] (6,6) [0.7814569536423841] (6,7) [0.023178807947019868] (6,8) [0.016556291390728478] (6,9) [0.0033112582781456954] 

(7,0) [0.009404388714733543] (7,1) [0.0] (7,2) [0.003134796238244514] (7,3) [0.0] (7,4) [0.03761755485893417] (7,5) [0.006269592476489028] (7,6) [0.0219435736677116] (7,7) [0.8934169278996865] (7,8) [0.018808777429467086] (7,9) [0.009404388714733543] 

(8,0) [0.006557377049180328] (8,1) [0.0] (8,2) [0.003278688524590164] (8,3) [0.003278688524590164] (8,4) [0.03278688524590164] (8,5) [0.09508196721311475] (8,6) [0.08196721311475409] (8,7) [0.0] (8,8) [0.7540983606557377] (8,9) [0.022950819672131147] 

(9,0) [0.07430340557275542] (9,1) [0.0] (9,2) [0.0] (9,3) [0.0030959752321981426] (9,4) [0.009287925696594427] (9,5) [0.0030959752321981426] (9,6) [0.006191950464396285] (9,7) [0.07120743034055728] (9,8) [0.0030959752321981426] (9,9) [0.8297213622291022]

};
\end{axis}
\end{tikzpicture}
    \setlength\abovecaptionskip{-0.02cm}
    \caption{Confusion matrix for \texttt{S1} after offset correction. Accuracy: 83.10\%}
    \label{fig:comparison_phase_correction_cm}
    \end{subfigure}
    \setlength\abovecaptionskip{0.1cm}
    \setlength\belowcaptionskip{-.5cm}
    \caption{Comparison with the accuracy obtained by learning the fingerprints from the processed version of $\mathbf{\Tilde{V}}$, i.e., after applying the offsets correction (offs. corr.) in~\cite{meneghello2021}. Beamformee 1, 3 TX antennas, spatial stream 0.}
    \label{fig:comparisons_phase}
\end{figure}
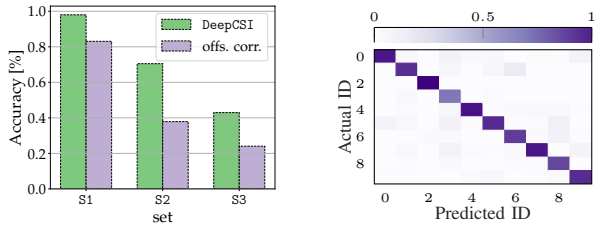

\smallskip
\noindent\textbf{\FW performance compared with learning from a processed input.}
\FW learns beamformer-specific features directly from the I/Q samples of matrix $\mathbf{\Tilde{V}}$. As an alternative approach, we evaluated the effect of pre-processing such I/Q data before using it as input for the \gls{dnn}. Specifically, we applied to the beamforming feedback matrices the data cleaning algorithm presented in~\cite{meneghello2021}. 
The \gls{cfr} estimated at the beamformee on the \gls{ndp} -- and from which $\mathbf{\Tilde{V}}$ is derived -- slightly deviates from the theoretical model in \eq{eq:h_subcarrier} due to hardware imperfections causing undesired phase offsets~\cite{Zhu2018}. Among these imperfections, the most significant are:
(i) the \gls{cfo}, which originates from the difference between the carrier frequency at transmitter and receiver sides;
(ii) the \gls{sfo}, which is due to clocks synchronization error;
(iii) the \gls{pdd}, i.e., the receiver decoding time;
(iv) the \gls{ppo}, which is associated with the random generation of the initial signal phase by the phase-locked loop module;
and (v) the \gls{pa}, which accounts for the phase difference (multiples of $\pi$) among the signals at the transmitter antennas.
By considering these contributions, the overall phase offset, $\theta_{{\rm offs}, k, m, n}$, can be formulated as 
\beq \label{eq:phase_offset}
	\theta_{{\rm offs}, k, m, n} = \theta_{\rm CFO} 
	- 2\pi k(\tau_{{\rm SFO}} + \tau_{{\rm PDD}})/T + \theta_{{\rm PPO}} + \theta_{{\rm PA}},
\eeq
and, in turn, the \gls{cfr} estimated at the beamformee during the channel sounding procedure becomes
\beq \label{eq:h_subcarrier_offset}
	 \hat{H}_{k, m, n} = H_{k, m, n} e^{j\theta_{{\rm offs}, k, m, n}}.
\eeq
Besides the \gls{pdd}, all the other contributions to \eq{eq:phase_offset} are associated with imperfections at the transmitter device, that in our case is the target of our fingerprinting technique, i.e., the \gls{ap}. Our key intuition is that also the beamforming feedback matrix $\mathbf{\Tilde{V}}$ -- derived from $\mathbf{H}$ as discussed in Section~\ref{subsec:beamforming_feedback} -- would be affected by the phase offsets (i)-(v). Thus, we may use the offsets cleaning algorithm of~\cite{meneghello2021} to improve its quality. Along this line of reasoning, we evaluate in \figurename~\ref{fig:comparisons_phase} the impact of a preliminary offset cleaning phase on matrix $\mathbf{\Tilde{V}}$ on the fingerprinting accuracy. \FW (with no offsets cleaning) outperforms its version with the described offset correction capability across all the training/testing sets. In other words, the offsets introduced by the beamformer hardware imperfections are strategic to reliably recognize the device, and any offset cleaning may result in their partial removal, affecting the fingerprinting quality.

\smallskip
\noindent\textbf{\FW performance in the presence of beamformer mobility.}
The robustness of \FW on beamformer's mobility is evaluated through dataset \texttt{D2}. 
In Figs.~\ref{fig:train_test_mobility_1}-\ref{fig:train_test_mobility_2} we report the performance of \FW on set \texttt{S4} (see Table \ref{tab:configs_d2}). Specifically, in \figurename~\ref{fig:train_test_mobility_1} the entire mobility path for both the training and testing sets is considered. We remind that even if the theoretical path is the same for all the measurements, the operation is performed manually and, in turn, the actual shifts undergo uncontrolled variations that reflect on the collected traces. The results show that \FW is able to effectively learn a fingerprint of the \gls{ap} from the MU-MIMO beamforming feedback matrices even when the \gls{ap} moves, reaching an accuracy above 80\%. The proposed learning architecture allows compensating for the slight variations introduced by the manual shifts of the \gls{ap} and the presence of the person moving in the vicinity.
The fingerprint is less effective when the environmental conditions sharply depart from the training ones. The high-scale modifications on the beamforming feedback matrices -- associated with the channel variations -- prevent the neural network from effectively capturing the small-scale variations that descend from the hardware imperfections. We show this behaviour in \figurename~\ref{fig:train_test_mobility_2}, where \FW is trained and tested in different mobility conditions. The training and validation phases are performed on the first half of the traces in `mob1', i.e., the portions related to the sub-path A-B-C-B. The test is executed on the fraction of the traces in `mob2' collected while the \gls{ap} spans the segments B-D-B. Overall, the results in Figs.~\ref{fig:train_test_mobility_1}-\ref{fig:train_test_mobility_2} indicate that the higher the variability in the training set, the more likely \FW will learn fingerprints that are robust to changing radio channel conditions. The variability in the training set allows the learning algorithm to properly detect the elements that are in common to the traces and, in turn, identify the hardware-related features.
In Figs.~\ref{fig:train_test_mobility_3}-\ref{fig:train_test_mobility_4} we report the performance of \FW on sets \texttt{S5} and \texttt{S6}. When \FW is trained on the sole static traces -- set \texttt{S5} -- the learned fingerprint is not effective in recognizing the beamformer when it moves in the environment. On the other hand, once trained on the dynamic traces, \FW is able to correctly identify the \gls{ap} in static conditions (about 88\% of accuracy on set \texttt{S6}). These last results confirm that the diversity in the training set is desirable to obtain a robust algorithm able to generalize over different conditions.
\begin{figure}[t]
    \centering
    \begin{subfigure}[t]{0.48\columnwidth}
    \centering
    \setlength\fwidth{.68\columnwidth}
    \setlength\fheight{0.42\columnwidth}
    \begin{tikzpicture}
\pgfplotsset{every tick label/.append style={font=\tiny}}

\begin{axis}[
enlargelimits=false,
colorbar,
colormap/Purples,
width=\fwidth,
height=\fheight,
at={(0\fwidth,0\fheight)},
scale only axis,
tick align=inside,
xlabel={Predicted ID},
xmin=-0.5,
xmax=9.5,
xtick style={draw=none},
xlabel style={font=\scriptsize\color{white!15!black}},
ylabel style={font=\scriptsize\color{white!15!black}},
ylabel={Actual ID},
ymin=-0.5,
ymax=9.5,
xlabel shift=-5pt,
ylabel shift=-5pt,
ytick style={draw=none},
axis background/.style={fill=white},
colorbar horizontal,
colorbar style={
at={(0,1.05)},               % <-- (changed)
anchor=below south west,    % <-- (changed)
% change the width of the colorbar relative to the main `axis' environment
width=\pgfkeysvalueof{/pgfplots/parent axis width},
xtick={0, 0.5, 1},
xmin=0,
xmax=1,
axis x line*=top,
xticklabel shift=-1pt,
point meta min=0,
point meta max=1.05,
},
colorbar/width=2mm,
]
\addplot [matrix plot,point meta=explicit]
 coordinates {
(0,0) [0.6977917981072556] (0,1) [0.017665615141955835] (0,2) [0.007570977917981073] (0,3) [0.0056782334384858045] (0,4) [0.006309148264984227] (0,5) [0.025236593059936908] (0,6) [0.0031545741324921135] (0,7) [0.138801261829653] (0,8) [0.0031545741324921135] (0,9) [0.0946372239747634] 

(1,0) [0.005564830272676683] (1,1) [0.7518085698386199] (1,2) [0.027267668336115748] (1,3) [0.015025041736227046] (1,4) [0.016138007790762382] (1,5) [0.004451864218141347] (1,6) [0.09237618252643294] (1,7) [0.008903728436282694] (1,8) [0.05286588759042849] (1,9) [0.025598219254312743] 

(2,0) [0.0034129692832764505] (2,1) [0.015602145294978059] (2,2) [0.8868844466114091] (2,3) [0.004388103364212579] (2,4) [0.008288639687957094] (2,5) [0.00048756704046806434] (2,6) [0.01462701121404193] (2,7) [0.0048756704046806435] (2,8) [0.025353486104339348] (2,9) [0.03607996099463676] 

(3,0) [0.0028555111364934323] (3,1) [0.0028555111364934323] (3,2) [0.00799543118218161] (3,3) [0.8812107367218732] (3,4) [0.0] (3,5) [0.023415191319246145] (3,6) [0.002284408909194746] (3,7) [0.017133066818960593] (3,8) [0.06225014277555682] (3,9) [0.0] 

(4,0) [0.008095238095238095] (4,1) [0.015238095238095238] (4,2) [0.011904761904761904] (4,3) [0.0] (4,4) [0.9104761904761904] (4,5) [0.0] (4,6) [0.01] (4,7) [0.006666666666666667] (4,8) [0.0004761904761904762] (4,9) [0.037142857142857144] 

(5,0) [0.040280210157618214] (5,1) [0.005253940455341506] (5,2) [0.0035026269702276708] (5,3) [0.03561004086398132] (5,4) [0.0005837711617046118] (5,5) [0.854057209573847] (5,6) [0.0011675423234092236] (5,7) [0.021015761821366025] (5,8) [0.03210741389375365] (5,9) [0.00642148277875073] 

(6,0) [0.0034856700232378003] (6,1) [0.025561580170410533] (6,2) [0.02943454686289698] (6,3) [0.0019364833462432224] (6,4) [0.01471727343144849] (6,5) [0.0003872966692486445] (6,6) [0.9109217660728117] (6,7) [0.0] (6,8) [0.0003872966692486445] (6,9) [0.013168086754453912] 

(7,0) [0.05454545454545454] (7,1) [0.004422604422604423] (7,2) [0.03095823095823096] (7,3) [0.05847665847665848] (7,4) [0.010319410319410319] (7,5) [0.015724815724815724] (7,6) [0.0004914004914004914] (7,7) [0.744963144963145] (7,8) [0.033906633906633905] (7,9) [0.046191646191646195] 

(8,0) [0.0020999580008399833] (8,1) [0.04745905081898362] (8,2) [0.0125997480050399] (8,3) [0.015539689206215875] (8,4) [0.0] (8,5) [0.023519529609407813] (8,6) [0.0] (8,7) [0.021839563208735827] (8,8) [0.8416631667366653] (8,9) [0.03527929441411172] 

(9,0) [0.07717629846378932] (9,1) [0.03182150694952451] (9,2) [0.0032918800292611556] (9,3) [0.0] (9,4) [0.1075347476225311] (9,5) [0.0021945866861741038] (9,6) [0.006949524506217996] (9,7) [0.02011704462326262] (9,8) [0.0018288222384784199] (9,9) [0.7490855888807608]

};
\end{axis}
\end{tikzpicture}
    \setlength\abovecaptionskip{-0.02cm}
    \caption{\texttt{S4}. Train and test on the complete \gls{ap} mobility path. Acc.82.56\%}
    \label{fig:train_test_mobility_1}
    \end{subfigure}
    \hfill
    \begin{subfigure}[t]{0.48\columnwidth}
    \centering
    \setlength\fwidth{.68\columnwidth}
    \setlength\fheight{0.42\columnwidth}
    \begin{tikzpicture}
\pgfplotsset{every tick label/.append style={font=\tiny}}

\begin{axis}[
enlargelimits=false,
colorbar,
colormap/Purples,
width=\fwidth,
height=\fheight,
at={(0\fwidth,0\fheight)},
scale only axis,
tick align=inside,
xlabel={Predicted ID},
xmin=-0.5,
xmax=9.5,
xtick style={draw=none},
xlabel style={font=\scriptsize\color{white!15!black}},
ylabel style={font=\scriptsize\color{white!15!black}},
ylabel={Actual ID},
ymin=-0.5,
ymax=9.5,
xlabel shift=-5pt,
ylabel shift=-5pt,
ytick style={draw=none},
axis background/.style={fill=white},
colorbar horizontal,
colorbar style={
at={(0,1.05)},               % <-- (changed)
anchor=below south west,    % <-- (changed)
% change the width of the colorbar relative to the main `axis' environment
width=\pgfkeysvalueof{/pgfplots/parent axis width},
xtick={0, 0.5, 1},
xmin=0,
xmax=1,
axis x line*=top,
xticklabel shift=-1pt,
point meta min=0,
point meta max=1.05,
},
colorbar/width=2mm,
]
\addplot [matrix plot,point meta=explicit]
 coordinates {
(0,0) [0.006309148264984227] (0,1) [0.138801261829653] (0,2) [0.0] (0,3) [0.0031545741324921135] (0,4) [0.006309148264984227] (0,5) [0.0473186119873817] (0,6) [0.0] (0,7) [0.7066246056782335] (0,8) [0.04100946372239748] (0,9) [0.050473186119873815] 

(1,0) [0.002785515320334262] (1,1) [0.025069637883008356] (1,2) [0.002785515320334262] (1,3) [0.005571030640668524] (1,4) [0.4428969359331476] (1,5) [0.0] (1,6) [0.1615598885793872] (1,7) [0.3342618384401114] (1,8) [0.016713091922005572] (1,9) [0.008356545961002786] 

(2,0) [0.0] (2,1) [0.0] (2,2) [0.3] (2,3) [0.0] (2,4) [0.07317073170731707] (2,5) [0.0] (2,6) [0.0] (2,7) [0.6268292682926829] (2,8) [0.0] (2,9) [0.0] 

(3,0) [0.002857142857142857] (3,1) [0.0] (3,2) [0.0] (3,3) [0.8428571428571429] (3,4) [0.04] (3,5) [0.0] (3,6) [0.0] (3,7) [0.11428571428571428] (3,8) [0.0] (3,9) [0.0] 

(4,0) [0.0] (4,1) [0.0] (4,2) [0.0] (4,3) [0.007142857142857143] (4,4) [0.780952380952381] (4,5) [0.0] (4,6) [0.007142857142857143] (4,7) [0.20476190476190476] (4,8) [0.0] (4,9) [0.0] 

(5,0) [0.011661807580174927] (5,1) [0.0] (5,2) [0.0029154518950437317] (5,3) [0.06705539358600583] (5,4) [0.008746355685131196] (5,5) [0.3935860058309038] (5,6) [0.0] (5,7) [0.5160349854227405] (5,8) [0.0] (5,9) [0.0] 

(6,0) [0.0] (6,1) [0.0] (6,2) [0.0] (6,3) [0.029069767441860465] (6,4) [0.5116279069767442] (6,5) [0.0] (6,6) [0.4108527131782946] (6,7) [0.040697674418604654] (6,8) [0.0] (6,9) [0.007751937984496124] 

(7,0) [0.0] (7,1) [0.0] (7,2) [0.0] (7,3) [0.0] (7,4) [0.07371007371007371] (7,5) [0.0] (7,6) [0.0] (7,7) [0.918918918918919] (7,8) [0.0] (7,9) [0.007371007371007371] 

(8,0) [0.0] (8,1) [0.0] (8,2) [0.0] (8,3) [0.0041928721174004195] (8,4) [0.0020964360587002098] (8,5) [0.0020964360587002098] (8,6) [0.0] (8,7) [0.710691823899371] (8,8) [0.27882599580712786] (8,9) [0.0020964360587002098] 

(9,0) [0.0] (9,1) [0.003656307129798903] (9,2) [0.0] (9,3) [0.0] (9,4) [0.3546617915904936] (9,5) [0.0] (9,6) [0.0] (9,7) [0.4680073126142596] (9,8) [0.0] (9,9) [0.1736745886654479]

};
\end{axis}
\end{tikzpicture}
    \setlength\abovecaptionskip{-0.02cm}
    \caption{\texttt{S4}. Train and test on different \gls{ap} mobility sub-paths. Acc. 41.15\%}
    \label{fig:train_test_mobility_2}
    \end{subfigure}
    \hfill
     \begin{subfigure}[t]{0.48\columnwidth}
    \centering
    \setlength\fwidth{.68\columnwidth}
    \setlength\fheight{0.42\columnwidth}
    \begin{tikzpicture}
\pgfplotsset{every tick label/.append style={font=\tiny}}

\begin{axis}[
enlargelimits=false,
colorbar,
colormap/Purples,
width=\fwidth,
height=\fheight,
at={(0\fwidth,0\fheight)},
scale only axis,
tick align=inside,
xlabel={Predicted ID},
xmin=-0.5,
xmax=9.5,
xtick style={draw=none},
xlabel style={font=\scriptsize\color{white!15!black}},
ylabel style={font=\scriptsize\color{white!15!black}},
ylabel={Actual ID},
ymin=-0.5,
ymax=9.5,
xlabel shift=-5pt,
ylabel shift=-5pt,
ytick style={draw=none},
axis background/.style={fill=white},
colorbar horizontal,
colorbar style={
at={(0,1.05)},               % <-- (changed)
anchor=below south west,    % <-- (changed)
% change the width of the colorbar relative to the main `axis' environment
width=\pgfkeysvalueof{/pgfplots/parent axis width},
xtick={0, 0.5, 1},
xmin=0,
xmax=1,
axis x line*=top,
xticklabel shift=-1pt,
point meta min=0,
point meta max=1.05,
},
colorbar/width=2mm,
]
\addplot [matrix plot,point meta=explicit]
 coordinates {
(0,0) [0.018316831683168316] (0,1) [0.023514851485148515] (0,2) [0.13193069306930694] (0,3) [0.0034653465346534654] (0,4) [0.010148514851485149] (0,5) [0.2527227722772277] (0,6) [0.04405940594059406] (0,7) [0.4] (0,8) [0.09207920792079208] (0,9) [0.023762376237623763] 

(1,0) [0.0007511266900350526] (1,1) [0.08487731597396093] (1,2) [0.3202303455182774] (1,3) [0.024036054081121683] (1,4) [0.049323985978968456] (1,5) [0.1484727090635954] (1,6) [0.10565848773159739] (1,7) [0.18502754131196794] (1,8) [0.030045067601402103] (1,9) [0.05157736604907361] 

(2,0) [0.0] (2,1) [0.01818181818181818] (2,2) [0.4857142857142857] (2,3) [0.011471861471861472] (2,4) [0.001948051948051948] (2,5) [0.14307359307359307] (2,6) [0.08701298701298701] (2,7) [0.19437229437229436] (2,8) [0.052597402597402594] (2,9) [0.005627705627705628] 

(3,0) [0.0020552637588490525] (3,1) [0.01918246174925782] (3,2) [0.0890614295501256] (3,3) [0.08380908883306691] (3,4) [0.0020552637588490525] (3,5) [0.3601278830783284] (3,6) [0.06987896780086778] (3,7) [0.3039506736697876] (3,8) [0.06599680292304179] (3,9) [0.0038821648778259877] 

(4,0) [0.004580152671755725] (4,1) [0.07459105779716467] (4,2) [0.07066521264994548] (4,3) [0.011559432933478735] (4,4) [0.04798255179934569] (4,5) [0.17251908396946564] (4,6) [0.08745910577971647] (4,7) [0.4772082878953108] (4,8) [0.014830970556161395] (4,9) [0.0386041439476554] 

(5,0) [0.0004880429477794046] (5,1) [0.01903367496339678] (5,2) [0.07906295754026355] (5,3) [0.014641288433382138] (5,4) [0.00951683748169839] (5,5) [0.3194241093216203] (5,6) [0.030258662762323085] (5,7) [0.4253294289897511] (5,8) [0.09419228892142509] (5,9) [0.008052708638360176] 

(6,0) [0.005728011825572801] (6,1) [0.04804138950480414] (6,2) [0.2682926829268293] (6,3) [0.0336289726533629] (6,4) [0.009053954175905396] (6,5) [0.07150776053215077] (6,6) [0.3501478196600148] (6,7) [0.18163340724316335] (6,8) [0.011271249076127124] (6,9) [0.020694752402069475] 

(7,0) [0.0022467320261437907] (7,1) [0.03982843137254902] (7,2) [0.10375816993464053] (7,3) [0.012459150326797386] (7,4) [0.00959967320261438] (7,5) [0.20363562091503268] (7,6) [0.018790849673202614] (7,7) [0.5328839869281046] (7,8) [0.06556372549019608] (7,9) [0.011233660130718954] 

(8,0) [0.0005422013374299657] (8,1) [0.03831556117838424] (8,2) [0.400325320802458] (8,3) [0.04120730164467739] (8,4) [0.0052412795951563345] (8,5) [0.2911621181998916] (8,6) [0.020422917043195373] (8,7) [0.08819808422194109] (8,8) [0.084221941080788] (8,9) [0.030363274896078075] 

(9,0) [0.0003204101249599487] (9,1) [0.04405639218199295] (9,2) [0.13521307273309838] (9,3) [0.014258250560717719] (9,4) [0.0318808074335149] (9,5) [0.22636975328420378] (9,6) [0.027074655559115667] (9,7) [0.42790772188401155] (9,8) [0.04870233899391221] (9,9) [0.044216597244472924]

};
\end{axis}
\end{tikzpicture}
    \setlength\abovecaptionskip{-0.02cm}
    \caption{\texttt{S5}. Train on static conditions and test on mobility traces. Acc. 20.50\%}
    \label{fig:train_test_mobility_3}
    \end{subfigure}
    \hfill
    \begin{subfigure}[t]{0.48\columnwidth}
    \centering
    \setlength\fwidth{.68\columnwidth}
    \setlength\fheight{0.42\columnwidth}
    \begin{tikzpicture}
\pgfplotsset{every tick label/.append style={font=\tiny}}

\begin{axis}[
enlargelimits=false,
colorbar,
colormap/Purples,
width=\fwidth,
height=\fheight,
at={(0\fwidth,0\fheight)},
scale only axis,
tick align=inside,
xlabel={Predicted ID},
xmin=-0.5,
xmax=9.5,
xtick style={draw=none},
xlabel style={font=\scriptsize\color{white!15!black}},
ylabel style={font=\scriptsize\color{white!15!black}},
ylabel={Actual ID},
ymin=-0.5,
ymax=9.5,
xlabel shift=-5pt,
ylabel shift=-5pt,
ytick style={draw=none},
axis background/.style={fill=white},
colorbar horizontal,
colorbar style={
at={(0,1.05)},               % <-- (changed)
anchor=below south west,    % <-- (changed)
% change the width of the colorbar relative to the main `axis' environment
width=\pgfkeysvalueof{/pgfplots/parent axis width},
xtick={0, 0.5, 1},
xmin=0,
xmax=1,
axis x line*=top,
xticklabel shift=-1pt,
point meta min=0,
point meta max=1.05,
},
colorbar/width=2mm,
]
\addplot [matrix plot,point meta=explicit]
 coordinates {
(0,0) [0.584786641929499] (0,1) [0.0] (0,2) [0.0] (0,3) [0.0] (0,4) [0.0] (0,5) [0.0] (0,6) [0.0] (0,7) [0.4152133580705009] (0,8) [0.0] (0,9) [0.0] 

(1,0) [0.0] (1,1) [0.9996147919876733] (1,2) [0.0] (1,3) [0.0] (1,4) [0.0] (1,5) [0.0] (1,6) [0.0003852080123266564] (1,7) [0.0] (1,8) [0.0] (1,9) [0.0] 

(2,0) [0.01199294532627866] (2,1) [0.0] (2,2) [0.9869488536155203] (2,3) [0.0] (2,4) [0.0003527336860670194] (2,5) [0.0] (2,6) [0.0] (2,7) [0.0] (2,8) [0.0] (2,9) [0.0007054673721340388] 

(3,0) [0.0] (3,1) [0.0] (3,2) [0.0] (3,3) [0.9396607958251794] (3,4) [0.0] (3,5) [0.0009784735812133072] (3,6) [0.0] (3,7) [0.046314416177429873] (3,8) [0.01304631441617743] (3,9) [0.0] 

(4,0) [0.0] (4,1) [0.0] (4,2) [0.0] (4,3) [0.0] (4,4) [0.9958432304038005] (4,5) [0.0] (4,6) [0.0] (4,7) [0.0005938242280285036] (4,8) [0.0] (4,9) [0.0035629453681710215] 

(5,0) [0.0] (5,1) [0.0] (5,2) [0.0] (5,3) [0.08255779045331732] (5,4) [0.0] (5,5) [0.6253377364154908] (5,6) [0.0] (5,7) [0.290303212248574] (5,8) [0.0018012608826178324] (5,9) [0.0] 

(6,0) [0.0] (6,1) [0.0] (6,2) [0.0] (6,3) [0.0] (6,4) [0.0] (6,5) [0.0] (6,6) [1.0] (6,7) [0.0] (6,8) [0.0] (6,9) [0.0] 

(7,0) [0.0] (7,1) [0.0010964912280701754] (7,2) [0.0] (7,3) [0.0] (7,4) [0.28691520467836257] (7,5) [0.0] (7,6) [0.0] (7,7) [0.7006578947368421] (7,8) [0.0] (7,9) [0.011330409356725146] 

(8,0) [0.0] (8,1) [0.0] (8,2) [0.0] (8,3) [0.0] (8,4) [0.0] (8,5) [0.00039323633503735744] (8,6) [0.0] (8,7) [0.0] (8,8) [0.9996067636649626] (8,9) [0.0] 

(9,0) [0.0] (9,1) [0.0] (9,2) [0.0] (9,3) [0.0] (9,4) [0.0] (9,5) [0.0] (9,6) [0.0] (9,7) [0.0] (9,8) [0.0] (9,9) [1.0]

};
\end{axis}
\end{tikzpicture}
    \setlength\abovecaptionskip{-0.02cm}
    \caption{\texttt{S6}. Train on mobility traces and test on static conditions. Acc. 88.12\%}
    \label{fig:train_test_mobility_4}
    \end{subfigure}
    \setlength\abovecaptionskip{0.1cm}
    \setlength\belowcaptionskip{-.5cm}
    \caption{Confusion matrices, beamformee 1, 3 TX antennas, spatial stream 0.}
    \label{fig:train_test_mobility}
\end{figure}

\section{Concluding Remarks}\label{sec:conclusions} 
In this paper, we have presented a novel approach to \mbox{Wi-Fi} radio fingerprinting (RFP) which leverages IEEE 802.11-compliant steering matrices to authenticate MU-MIMO \mbox{Wi-Fi} devices.
The present work makes the following key advances:

$\bullet$ \textbf{We have demonstrated for the first time the feasibility of \gls{rfp} for MU-MIMO Wi-Fi}. To this end, \FW leverages the beamforming feedback matrices computed by any of the beamformees and transmitted in clear to the beamformer. \textbf{We have verified that the matrices are affected by the beamformer hardware imperfections and, in turn, can be used to identify the device}. Moreover, the feedback is not affected by inter-stream and inter-user interference, thus increasing robustness. \FW is independent of the number of beamformees associated with the target beamformer: different beamformer's fingerprints can be computed, one from each beamformee. Conversely from prior work, \FW does not require direct CSI computation and, in turn, can be run on any Wi-Fi device without requiring SDRs.

$\bullet$ \textbf{We have performed a massive data collection campaign with off-the-shelf Wi-Fi equipment}, where 10 \mbox{Wi-Fi} radios emit MU-MIMO signals in different positions. Experimental results indicate that \FW is able to correctly identify the transmitter with accuracy above 98\%. We have evaluated \FW fingerprinting accuracy by differentiating the set of positions for the devices at training and testing times. Our technique achieves accuracy of 73\% when training is performed on a more balanced set of spatial points, which allows the classifier to interpolate the training patterns for the missing points, using those from adjacent training positions. 

$\bullet$ \textbf{For the first time, we evaluated the proposed \gls{rfp} technique with moving Wi-Fi devices.} \FW reaches an accuracy above 82\%, showing the robustness of the learned fingerprint to changing radio channel conditions. \textbf{Our results show that the higher the variability in the traffic traces used for the training phase, the higher is the accuracy when the algorithm is used at run-time to identify the devices}. This indicates the need for extensive datasets to train effective RFP algorithms. In this vision, \textbf{we pledge to make our contribution by sharing our datasets}~\cite{dataset-code}.

As part of ongoing work, we plan to further investigate the effect of the beam patterns and of the positions of the receivers on the fingerprinting performance, also in the presence of interference from overlapping channels. As a further extension, we intend to investigate lifelong machine learning techniques to accumulate knowledge and improve the device's fingerprint as it moves in the environment.%\vspace{-0.3cm}

\section*{Acknowledgement}

This material is based upon work supported in part by the National Science Foundation (NSF) under Grant No. CNS-2134973 and CNS-2120447, and by the Italian Ministry of Education, University and Research (MIUR) through the initiative ``Departments of Excellence'' (Law 232/2016). The views and opinions are those of the authors and do not necessarily reflect those of the funding institutions.%\vspace{-0.1cm}

\footnotesize
\bibliographystyle{IEEEtran}
\bibliography{mybib} 
% that's all folks
\end{document}